\documentclass[aps,prc,showpacs,showkeys,nofootinbib]{revtex4}

\usepackage{graphicx}
\usepackage{dcolumn}
\usepackage{amsthm}
\usepackage{bm}

\newcommand{\sigmabf}{\mbox{\boldmath $\sigma$}}
\newcommand{\xibf}{\mbox{\boldmath $\xi$}}

\begin{document}

\title{Model of the bremsstrahlung emission accompanying interactions between protons and nuclei from low up to intermediate energies:
role of magnetic emission}



\author{Sergei~P.~Maydanyuk\thanks{\emph{e-mail:} maidan@kinr.kiev.ua}}

\affiliation{Institute for Nuclear Research, National Academy of Science of Ukraine, Kiev, 03680, Ukraine}

\date{\small\today}


\begin{abstract}
A new model of the bremsstrahlung emission which accompanies proton decay and collisions of protons off nuclei in the energy region from the lowest up to intermediate, has been developed. This model includes spin formalism, potential approach for description of interaction between protons and nuclei, and operator of emission includes component of the magnetic emission (defined on the basis of Pauli equation). In the problem of the bremsstrahlung during the proton decay in the first time a role of the magnetic emission is studied using such a model. For the studied $^{146}{\rm Tm}$ nucleus it has been studied the following:
(1) How much does the magnetic emission change the full bremsstrahlung spectrum?
(2) At which angle is the magnetic emission the most intensive relatively electric one?
(3) Is there some space region where the magnetic emission increases strongly relatively electric one?
(4) How intensive is the magnetic emission in the tunneling region?
(5) Which values has the probability at its maximum and at zero energy limit of the emitted photons?
It is demonstrated that the model is able to describe enough well experimental data of the bremsstrahlung emission
which accompanies collisions of protons off
the $^{9}{\rm C}$, $^{64}{\rm Cu}$ and $^{107}{\rm Ag}$ nuclei at the incident energy $T_{\rm lab}=72$~MeV (at the photon energy up to 60~MeV),
the $^{9}{\rm Be}$, $^{12}{\rm C}$ and $^{208}{\rm Pb}$ nuclei at the incident energy $T_{\rm lab}=140$~MeV (at the photon energy up to 120~MeV).
\end{abstract}


\maketitle

\section{Introduction
\label{sec.Introduction}}

According to theory of collisions of protons off nuclei, interactions between two nucleons play important role, which become leading at increasing of energy. By such a way, interaction between two nucleons (i.e. nucleon-nucleon, or two-nucleons interaction) is putted into the basis of relativistic models of collisions, with further application of formalism of Feynman's diagrams. But, on the other side, consideration of nucleus as medium allows to include space distribution of all nucleons into the model. Such a way takes into account non-locality of quantum mechanics, one of its fundamental aspects. Comparing these two different considerations, a question arises: what is more fundamental, interaction between different point-like nucleons of the studied nuclear system or quantum effects of non-locality in it?

How important non-local effect in study of many-nucleons interactions? How much are they small? Results of \cite{Maydanyuk.2011.JMP} give some answer on this question: it was shown that fully quantum consideration of the boundary and initial conditions in the problem of proton decay has essential influence on the calculated half-life (for example, half-lives calculated in \cite{Gurvitz.1987.PRL,Buck.1992.PRC,Aberg_Nazarewicz.1997.PRC,Esbensen.2000.PRC,Hagino.2001.PRC,Gurvitz.2004.PRA,Delion.2006.PRL,Dong.2009.PRC,Delion.2009.PRC}
can be changed up to 200 times after taking such conditions into account, while assumed error is only some percents in that models). This estimation indicates that non-local effects are not so small and their inclusion into calculations is able to essentially change results sometimes.

Another aspect is collective motion. Models with nucleon-nucleon interaction should be the most accurate, if the collective effects caused by interactions between nucleons of the complete nuclear system were very small. However, we know that this is not so at low energies. One can assume that many-nucleons interactions disappear at increasing of energy of interacting nucleons. If to analyze bremsstrahlung emission, which accompanies collisions of protons off nuclei, then there are indications that two-nucleons interactions give the largest intensity of emission. But, we find that many-nucleons effects should arise at increasing of energy of the emitted photons\footnote{For example, in the problem of $\alpha$-decay at increasing of energy of the emitted photon for obtaining stable value of the emission probability, it needs to continuously increase external boundary of space region of integration. In the task of fission this problem is essentially more difficult (see \cite{Maydanyuk.2010.PRC}).}.
We find confirmation about essential influence of many-nucleons interactions on the process of emission and all importance of its study in literature (for example, see \cite{Kopitin.1997.YF}; in particular, two-nucleons approaches do not give positive explanation of nature of hard photons).

Properties of the bremsstrahlung accompanying scattering of protons off nuclei have been studied enough well (for example, see review \cite{Pluiko.1987.PEPAN}, also \cite{Kamanin.1989.PEPAN} for emission in collisions between heavy ions). As a rule, as emitter of photons in nuclear system, both the nucleus as medium, and different nucleons in it were considered. The process of emission is studied as result of deacceleration of motion of nucleons in the averaged field of nucleus or in consequence of nucleon-nucleon collisions.
At the same time, it was pointed out (for example, see \cite{Kopitin.1997.YF}) that properties of the nuclear bremsstrahlung emission accompanying nucleon-nucleus and nucleus-nucleus collisions (especially, in region of intermediate energies up to 150~MeV / nucleon) have been studied worst of all.
This causes our interest in use of the optical model potentials \cite{Becchetti.1969.PR} and folding potentials \cite{Khoa.2000.NPA} for investigations of the bremsstrahlung emission, which accompanies interactions protons with nuclei. It could be interesting to obtain the model, which allows to describe the spectra in energy region from minimal up to intermediate. Possibility to take quantum non-local properties into account in description of such interactions reinforces our interest in such potential approach.

However, in investigations of the bremsstrahlung emission, which accompanies $\alpha$-decay of nuclei
\cite{Batkin.1986.SJNCA,D'Arrigo.1994.PHLTA,Dyakonov.1996.PRLTA,Kasagi.1997.JPHGB,Kasagi.1997.PRLTA,%
Papenbrock.1998.PRLTA,Dyakonov.1999.PHRVA,Bertulani.1999.PHRVA,Takigawa.1999.PHRVA,Flambaum.1999.PRLTA,%
Tkalya.1999.JETP,Tkalya.1999.PHRVA,So_Kim.2000.JKPS,Misicu.2001.JPHGB,Dijk.2003.FBSSE,Maydanyuk.2003.PTP,%
Maydanyuk.2006.EPJA,Ohtsuki.2006.CzJP,Amusia.2007.JETP,Boie.2007.PRL,Jentschura.2008.PRC,Maydanyuk.2008.EPJA,%
Giardina.2008.MPLA,Maydanyuk.2009.NPA,Maydanyuk.2009.TONPPJ,Maydanyuk.2009.JPS},
spontaneous fission of nuclei \cite{Sobel.1973.PRC,Dietrich.1974.PRC,Kasagi.1989.JPSJS,Luke.1991.PRC,Ploeg.1992.PRL,%
Hofman.1993.PRC,Ploeg.1995.PRC,Varlachev.2007.IRAN,Maydanyuk.2010.PRC,Eremin.2010.IJMPE,Pandit.2010.PLB},
ternary fission of nuclei \cite{Maydanyuk.2011.JPCS}, and also collisions of nucleons off nuclei \cite{Kursunoglu.1957.PR,D'Arrigo.1992.NPA,D'Arrigo.1993.NPA,Kopitin.1997.YF}, ions and nuclei off nuclei at non-relativistic energies \cite{Kamanin.1989.PEPAN}, the emission caused by the magnetic moment of the fragment moving relatively the nucleus has not been taken into account.
Such a way could be explained, if at such energies of the emitted photons the magnetic emission is enough small and it can be neglected in calculations (for example, see~\cite{Kopitin.1997.YF}).
Microscopic models can provide a powerful formalism for study of many-nucleons interactions, where wave functions were obtained from a single-configuration resonanting group calculations. But, in particular, we see that magnetic emission was not included into such models, which were applied for description of the bremsstrahlung emission during scattering of protons on $\alpha$-particles \cite{Liu.1990.PRC}, $\alpha$-particles on $\alpha$-particles and light nuclei \cite{Baye.1985.NPA,Baye.1991.NPA}.

The magnetic emission is connected with magnetic momentum and spin of the fragment, interacting with nucleus. Attempt to take such aspects into account leads to matrix form of equations of interactions (where two-component Pauli equation is the simplest) and many-component wave function of nuclear system (for example, see \cite{Ahiezer.1981}, p.~32--35, 48--60). However, the magnetic component of emission and spin formalism are included in relativistic models of collisions of nucleons between themselves and with nuclei at intermediate energies (based on Dirac equation).
Here, I should like to note two directions of intensive investigations:
Refs.~\cite{Nakayama.1989.PRC,Herrmann.1991.PRC} and
Refs.~\cite{Liou.1987.PRC,Liou.1993.PRC,Liou.1995.PLB.v345,Liou.1995.PLB.v355,Liou.1996.PRC,Li.1998.PRC.v57,Li.1998.PRC.v58,%
Timmermans.2001.PRC,Liou.2004.PRC,Li.2005.PRC,Timmermans.2006.PRC,Li.2011.PRC}.
However, main emphasis in these papers was made on construction of correct relativistic description of interaction between two nucleons in this task, where formalism was developed in momentum representation mainly. So, it could be interesting to obtain the model, combining spin formalism of interacting fragments of nuclear system (with inclusion of magnetic momentum) and potential approach for description of interaction between themselves.

The problem of the bremsstrahlung during collisions of protons off nuclei and proton decay can be convenient in this investigation. In \cite{Maydanyuk.2011.JPG} the problem of the bremsstrahlung during proton-decay was studied (see also \cite{Kurgalin.2001.IRAN}). However, here the magnetic emission caused by the magnetic moment of proton was not taken into account (but spin-orbital component of potential was included and its influence on the spectrum was estimated). In order to clarify its role, a model with such aspect is needed.
Main aim of this paper is construction of such a model.


What interesting and new could this model give? We shall put such questions.
How much does the magnetic emission change the full bremsstrahlung spectrum?
At which angle is the magnetic emission the most intensive relatively electric one?
Is there some space region where the magnetic emission increases strongly relatively electric one?
How intensive is the magnetic emission in the tunneling region?
Which values has the probability at its maximum and at zero energy limit of the emitted photons?
We answer on such questions in this paper.



\section{Model
\label{sec.2}}

\subsection{Operator of emission of the bremsstrahlung photon
\label{sec.2.1}}

Let us consider generalization of Pauli equation for $A+1$ nucleons of the proton-nucleus system in laboratory frame
(obtained starting from eq.~(1.3.6) in \cite{Ahiezer.1981}, p.~33)
\begin{equation}
\begin{array}{lcl}
  i \hbar \displaystyle\frac{\partial \Psi}{\partial t} = \hat{H} \Psi, &
\hspace{5mm}
  \hat{H} =
  \displaystyle\sum_{i=1}^{A+1}
  \biggl\{
    \displaystyle\frac{1}{2m_{i}} \Bigl( \mathbf{p}_{i} - \displaystyle\frac{z_{i}e}{c} \mathbf{A}_{i} \Bigr)^{2} +
    z_{i}e\, A_{i,0} - \displaystyle\frac{z_{i}e \hbar}{2m_{i}c}\; \sigmabf \cdot \mathbf{rot\, A}_{i}
  \biggr\} +
  V(\mathbf{r}_{1} \ldots \mathbf{r}_{A+1}),
\end{array}
\label{eq.pauli.1}
\end{equation}
where we use for any nucleon with number $i$ (like to eq.~(1.3.4) in \cite{Ahiezer.1981} for one-particle problem)
\begin{equation}
\begin{array}{lcl}
  \chi =
  \displaystyle\frac{1}{2m_{i}c}\;
  \sigmabf\,
  \Bigl( \mathbf{p}_{i} - \displaystyle\frac{z_{i}e}{c} \mathbf{A}_{i} \Bigr)\, \psi.
\end{array}
\label{eq.pauli.1.add}
\end{equation}
Here, $\Psi = (\chi, \psi)$ is bispinor wave function of the proton-nucleus system, $m_{i}$ and $z_{i}$ are mass and charge of nucleon with number $i$, $\mathbf{A}_{i}$ is component of potential of the electromagnetic field formed by this nucleon (describing possible bremsstrahlung emission of photon caused by this nucleon), $\sigmabf$ are Puali matrixes, $A$ is mass number of nucleus, $V(\mathbf{r}_{1} \ldots \mathbf{r}_{A+1})$ is potential of (nuclear and Coulomb) interactions between all nucleons\footnote{According to~\cite{Ahiezer.1981} (see p.~32), the equation (\ref{eq.pauli.1}) is working if energy $\varepsilon_{i}$ of any nucleon with number $i$ is close to its mass $m_{i}$, i.e. $|\varepsilon_{i}-m_{i}| \ll m_{i}$ ($c=1$). From here one can obtain high energy limit for proton incident energy $\varepsilon_{p} \ll 2m_{p} \simeq 1.\,86$~GeV. By other words, inside energy region up to $\varepsilon_{\rm p}$ the equation~(\ref{eq.pauli.1}) includes all relativistic properties, which Dirac equation gives us (with application of eq.~(\ref{eq.pauli.1.add})). In particular, this limit is essentially higher intermediate energies for proton-nucleus collisions studied in this paper.}.
We pass to the center-of-masses frame, where we have distance $\mathbf{r}$ between center-of masses of proton and nucleus (for example, see Appendix~A in Ref.~\cite{Maydanyuk.2011.JPCS}, also \cite{Kopitin.1997.YF}).
Then, one can represent this hamiltonian as $\hat{H} = \hat{H}_{0} + \hat{W},$ where $\hat{W}$ combines all items of electromagnetic field, which we define as operator of emission of the bremsstrahlung photon, and $\hat{H}_{0}$ is rest of hamiltonian without the emission of photons. Neglecting by relative motion of nucleons of nucleus in calculation of $\hat{W}$,
we find:
\begin{equation}
\begin{array}{lcl}
  \vspace{2mm}
  \hat{W} & = &
  \hat{W}_{\rm el} + \hat{W}_{\rm mag}, \\

  \vspace{2mm}
  \hat{W}_{\rm el} & = &
    - Z_{\rm eff}\, \displaystyle\frac{e}{2mc} \, (\mathbf{\hat{p}A} + \mathbf{A\hat{p}}) +
    eA_{0} + Z_{\rm eff}^{2}\, \displaystyle\frac{e^{2}}{2mc^{2}}\, \mathbf{A}^{2}, \\

  \vspace{2mm}
  \hat{W}_{\rm mag} & = &
    - Z_{\rm eff}\, \displaystyle\frac{e \hbar}{2mc}\;
    \sigmabf \cdot \mathbf{rot\, A},
\end{array}
\label{eq.2.1.1}
\end{equation}
where $Z_{\rm eff}$ and $m$ are effective charge and reduced mass of the proton-nucleus system, $\mathbf{\hat{p}}$ is operator of momentum corresponding to $\mathbf{r}$. Neglecting items at $e^{2}\mathbf{A}^{2}/c^{2}$ and $A_{0}$,
the operator of emission in Coulomb gauge can be rewritten as
\begin{equation}
\begin{array}{lcl}
  \hat{W} & = &
  - Z_{\rm eff}\, \displaystyle\frac{e}{mc}\, \mathbf{A\hat{p}} -
    Z_{\rm eff}\, \displaystyle\frac{e \hbar}{2mc}\; \sigmabf\cdot \mathbf{rot\, A} =
  - Z_{\rm eff}\, \displaystyle\frac{e}{mc}\,
    \Big( \mathbf{A\hat{p}} + \displaystyle\frac{\hbar}{2}\: \sigmabf\cdot \mathbf{rot\, A} \Bigr).
\end{array}
\label{eq.2.1.2}
\end{equation}
Substituting the following form of the potential of electromagnetic field:
\begin{equation}
\begin{array}{lcl}
  \mathbf{A} & = &
  \displaystyle\sum\limits_{\alpha=1,2}
    \sqrt{\displaystyle\frac{2\pi\hbar c^{2}}{w_{\rm ph}}}\; \mathbf{e}^{(\alpha),\,*}
    e^{-i\, \mathbf{k_{\rm ph}r}},
\end{array}
\label{eq.2.1.3}
\end{equation}
we obtain
\begin{equation}
\begin{array}{lcl}
  \hat{W} & = &
  Z_{\rm eff}\, \displaystyle\frac{e}{mc}\,
  \sqrt{\displaystyle\frac{2\pi\hbar c^{2}}{w_{\rm ph}}}\;
    \displaystyle\sum\limits_{\alpha=1,2}
  e^{-i\,\mathbf{k_{\rm ph}r}}\;
  \Big(
    i\, \mathbf{e}^{(\alpha)}\, \nabla -
    \displaystyle\frac{1}{2}\: \sigmabf\cdot
      \Bigl[ \nabla \times \mathbf{e}^{(\alpha)} \Bigr] +
    i\,\displaystyle\frac{1}{2}\: \sigmabf\cdot
      \Bigl[ \mathbf{k}_{\rm ph} \times \mathbf{e}^{(\alpha)} \Bigr]
  \Bigr).
\end{array}
\label{eq.2.1.7}
\end{equation}
Here, $\mathbf{e}^{(\alpha)}$ are unit vectors of polarization of the photon emitted ($\mathbf{e}^{(\alpha), *} = \mathbf{e}^{(\alpha)}$), $\mathbf{k}_{\rm ph}$ is wave vector of the photon and $w_{\rm ph} = k_{\rm ph} c = \bigl| \mathbf{k}_{\rm ph}\bigr|c$. Vectors $\mathbf{e}^{(\alpha)}$ are perpendicular to $\mathbf{k}_{\rm ph}$ in Coulomb calibration. We have two independent polarizations $\mathbf{e}^{(1)}$ and $\mathbf{e}^{(2)}$ for the photon with impulse $\mathbf{k}_{\rm ph}$ ($\alpha=1,2$). One can develop formalism simpler in the system of units where $\hbar = 1$ and $c = 1$, but we shall write constants $\hbar$ and $c$ explicitly.
Also we have properties:
\begin{equation}
\begin{array}{lclc}
  \Bigl[ \mathbf{k}_{\rm ph} \times \mathbf{e}^{(1)} \Bigr] = k_{\rm ph}\, \mathbf{e}^{(2)}, &
  \Bigl[ \mathbf{k}_{\rm ph} \times \mathbf{e}^{(2)} \Bigr] = -\, k_{\rm ph}\, \mathbf{e}^{(1)}, &
  \Bigl[ \mathbf{k}_{\rm ph} \times \mathbf{e}^{(3)} \Bigr] = 0, &
%
  \displaystyle\sum\limits_{\alpha=1,2,3}
  \Bigl[ \mathbf{k}_{\rm ph} \times \mathbf{e}^{(\alpha)} \Bigr] =
  k_{\rm ph}\, (\mathbf{e}^{(2)} - \mathbf{e}^{(1)}).
\end{array}
\label{eq.2.1.10}
\end{equation}

\subsection{Matrix element of emission
\label{sec.2.2}}

Let us consider the matrix element in form:
\begin{equation}
\begin{array}{lcl}
  F_{fi} & \equiv &
    \Bigl< k_{f} \Bigl|\, \hat{W}\, \Bigr| \,k_{i} \Bigr> =
    \displaystyle\int
      \psi^{*}_{f}(\mathbf{r})\:
      \hat{W}\:
      \psi_{i}(\mathbf{r})\;
      \mathbf{dr},
\end{array}
\label{eq.2.2.1}
\end{equation}
where $\psi_{i}(\mathbf{r}) = |k_{i}\bigr>$ and $\psi_{f}(\mathbf{r}) = |k_{f}\bigr>$ are stationary wave functions of the proton-nucleus system in the initial $i$-state (i.e. state before emission of photon) and final $f$-state (i.e. state after emission of photon) which do not contain number of photons emitted.
Substituting the operator of emission in form~(\ref{eq.2.1.7}) into eq.~(\ref{eq.2.2.1}), we obtain:
\begin{equation}
  F_{fi} \quad = \quad
  \bigl< k_{f} \bigl|\,  \hat{W}\, \bigr| \,k_{i} \bigr> \quad = \quad
  Z_{\rm eff}\, \displaystyle\frac{e}{mc}\,
    \sqrt{\displaystyle\frac{2\pi\hbar c^{2}}{w_{\rm ph}}}\;
    \Bigl\{ p_{\rm el} + p_{\rm mag, 1} + p_{\rm mag, 2} \Bigr\},
\label{eq.2.2.3}
\end{equation}
where
\begin{equation}
\begin{array}{lcl}
  \vspace{2mm}
  p_{\rm el} & = &
  i \displaystyle\sum\limits_{\alpha=1,2}
    \mathbf{e}^{(\alpha)}\,
    \Bigl< k_{f} \Bigl|\, e^{-i\,\mathbf{k_{\rm ph}r}}\: \nabla\, \Bigr| \,k_{i} \Bigr>, \\

  \vspace{2mm}
  p_{\rm mag, 1} & = &
  \displaystyle\frac{1}{2}\:
  \displaystyle\sum\limits_{\alpha=1,2}
  \Bigl< k_{f} \Bigl|\,
    e^{-i\,\mathbf{k_{\rm ph}r}}\;
    \sigmabf\cdot \Bigl[ \mathbf{e}^{(\alpha)} \times \nabla \Bigr]\,
  \Bigr| \,k_{i} \Bigr>, \\

  p_{\rm mag, 2} & = & -
  i\,\displaystyle\frac{1}{2}\:
  \displaystyle\sum\limits_{\alpha=1,2}
  \Bigl[ \mathbf{k}_{\rm ph} \times \mathbf{e}^{(\alpha)} \Bigr]\:
  \Bigl< k_{f} \Bigl|\, e^{-i\,\mathbf{k_{\rm ph}r}}\; \sigmabf\, \Bigr| \,k_{i} \Bigr>.
\end{array}
\label{eq.2.2.4}
\end{equation}
This definition for $F_{fi}$ is in compliance with our previous formalism in
\cite{Maydanyuk.2003.PTP,Maydanyuk.2006.EPJA,Maydanyuk.2008.EPJA,Giardina.2008.MPLA,Maydanyuk.2009.NPA,%
Maydanyuk.2009.TONPPJ,Maydanyuk.2009.JPS,Maydanyuk.2010.PRC,Maydanyuk.2011.JPCS,Maydanyuk.2011.JPG}.
In particular, for square of matric element of emission we have (see eqs.~(1)--(2) in \cite{Maydanyuk.2011.JPG}):
\begin{equation}
\begin{array}{lcl}
  \vspace{2mm}
  |a_{fi}|^{2} & = &
  2\pi\, T\, |F_{fi}|^{2} \cdot \delta(w_{f} - w_{i} + w_{\rm ph}).
\end{array}
\label{eq.2.2.6}
\end{equation}

\subsection{Wave function of nuclear system and summation over spinor states
\label{sec.2.3}}

We shall define the wave function of the proton in field of the nucleus. We shall construct it in form of bilinear combination of eigenfunctions of orbital and spinor subsystems
(as eq.~(1.4.2) in \cite{Ahiezer.1981}, p.~42).
%
%
However, we shall assume that it is not possible to fix experimentally states for selected $M$ (eigenvalue of momentum operator $\hat{J}_{z}$).
So, we shall be interesting in superposition over all states with different $M$ and define the wave function so:
\begin{equation}
  \psi_{jl} (\mathbf{r}, s) =
  R\,(r)\:
  \displaystyle\sum\limits_{m=-l}^{l}
  \displaystyle\sum\limits_{\mu = \pm 1/2}
    C_{lm 1/2 \mu}^{j,M=m+\mu}\, Y_{lm}(\mathbf{n}_{\rm r})\, v_{\mu} (s),
\label{eq.2.3.1b}
\end{equation}
where $R\,(r)$ is radial scalar function (not dependent on $m$ at the same $l$), $\mathbf{n}_{\rm r} = \mathbf{r} / r$ is unit vector directed along $\mathbf{r}$, $Y_{lm}(\mathbf{n}_{\rm r})$ are spherical functions (we use definition (28,7)--(28,8), p.~119 in~\cite{Landau.v3.1989}), $C_{lm 1/2 \mu}^{jM}$ are Clebsh-Gordon coefficients, $s$ is variable of spin, $M = m + \mu$ and $l = j \pm 1/2$.
For convenience of calculations we shall use spacial wave function as
\begin{equation}
  \varphi_{lm} (\mathbf{r}) =
  R_{l}\,(r)\: Y_{lm}(\mathbf{n}_{\rm r}).
\label{eq.2.3.2}
\end{equation}

Spinor function $v_{\mu}(s)$ has two components $v_{\mu_{1}}(s)$ and $v_{\mu_{2}}(s)$, which are eigenfunctions of spin operator $\hat{s}_{z}$ having eigenvalues $\sigma_{1}$ and $\sigma_{2}$ (see~\cite{Landau.v3.1989}, p.~247).
So, we have:
\begin{equation}
\begin{array}{cc}
  v_{\mu_{1}} (s) = \delta_{\mu_{1}s}, &
  v_{\mu_{2}} (s) = \delta_{\mu_{2}s}.
\end{array}
\label{eq.2.3.3}
\end{equation}
Action of operator of spin on the wave function is given by (see eq.~(55,4) in \cite{Landau.v3.1989}, p.~248)
\begin{equation}
\begin{array}{cc}
  (\mathbf{\hat{s}}\,v_{\mu})\, (\sigma) =
  \displaystyle\sum\limits_{\sigma^{\prime}}
  s_{\sigma \sigma^{\prime}}\, v_{\mu}\, (\sigma^{\prime})
\end{array}
\label{eq.2.3.4}
\end{equation}
and we have non-zero matrix elements:
\begin{equation}
\begin{array}{lcl}
  \vspace{2mm}
  (s_{x})_{\sigma, \sigma-1} & = &
    (s_{x})_{\sigma-1, \sigma} =
    \displaystyle\frac{1}{2}\: \sqrt{(s+\sigma)\, (s-\sigma+1)}, \\
  \vspace{2mm}
  (s_{y})_{\sigma, \sigma-1} & = &
    -\, (s_{y})_{\sigma-1, \sigma} =
    -\,\displaystyle\frac{i}{2}\: \sqrt{(s+\sigma)\, (s-\sigma+1)}, \\
  \vspace{2mm}
  (s_{z})_{\sigma \sigma} & = & \sigma.
\end{array}
\label{eq.2.3.5}
\end{equation}
From eqs.~(\ref{eq.2.3.5}) (at $s=1/2$, $\sigma=\pm 1/2$) we calculate:
\begin{equation}
\begin{array}{ccl}
  \vspace{2mm}
  v_{\mu_{f}}^{*} (s_{f})\, \hat{\sigmabf}_{x}\, v_{\mu_{i}} (s_{i}) & = &
  \delta_{\mu_{f}, s_{f}}\,
  \Bigl\{
    \delta_{s_{i}, -1/2}\, \delta_{\mu_{i}, +1/2}\; +\;
    \delta_{s_{i}, +1/2}\, \delta_{\mu_{i}, -1/2}
  \Bigr\}, \\

  \vspace{2mm}
  v_{\mu_{f}}^{*} (s_{f})\, \hat{\sigmabf}_{y}\, v_{\mu_{i}} (s_{i}) & = &
  i\, \delta_{\mu_{f}, s_{f}}\,
  \Bigl\{
    \delta_{s_{i}, -1/2}\, \delta_{\mu_{i}, +1/2}\; -\;
    \delta_{s_{i}, +1/2}\, \delta_{\mu_{i}, -1/2}
  \Bigr\}, \\

  \vspace{2mm}
  v_{\mu_{f}}^{*} (s_{f})\, \hat{\sigmabf}_{z}\, v_{\mu_{i}} (s_{i}) & = &
  \delta_{\mu_{f}, s_{f}}\,
  \Bigl\{
    \delta_{s_{i}, -1/2}\, \delta_{\mu_{i}, -1/2}\; +\;
    \delta_{s_{i}, +1/2}\, \delta_{\mu_{i}, +1/2}
  \Bigr\}
\end{array}
\label{eq.2.3.9}
\end{equation}
and find summations:
\begin{equation}
\begin{array}{ccl}
  \displaystyle\sum\limits_{s_{i}, s_{f} = \pm 1/2}
  \hspace{-5mm}
    v_{\mu_{f}}^{*} (s_{f})\, \hat{\sigmabf}_{x}\, v_{\mu_{i}} (s_{i}) = 1, & 

  \displaystyle\sum\limits_{s_{i}, s_{f} = \pm 1/2}
  \hspace{-5mm}
    v_{\mu_{f}}^{*} (s_{f})\, \hat{\sigmabf}_{y}\, v_{\mu_{i}} (s_{i}) =
  i\, \Bigl\{ \delta_{\mu_{i}, +1/2}\; -\; \delta_{\mu_{i}, -1/2} \Bigr\}, & 

  \displaystyle\sum\limits_{s_{i}, s_{f} = \pm 1/2}
  \hspace{-5mm}
    v_{\mu_{f}}^{*} (s_{f})\, \hat{\sigmabf}_{z}\, v_{\mu_{i}} (s_{i}) = 1.
\end{array}
\label{eq.2.3.10}
\end{equation}
Considering vectorial form of spin operator, these formulas can be rewritten as
\begin{equation}
\begin{array}{ccl}
  \vspace{2mm}
  \displaystyle\sum\limits_{s_{i}, s_{f} = \pm 1/2}
    v_{\mu_{f}}^{*} (s_{f})\, \hat{\sigmabf}\, v_{\mu_{i}} (s_{i}) & = &
  \mathbf{e}_{\rm x}
  \displaystyle\sum\limits_{s_{i}, s_{f} = \pm 1/2}
    v_{\mu_{f}}^{*} (s_{f})\, \hat{\sigmabf}_{x}\, v_{\mu_{i}} (s_{i}) +
  \mathbf{e}_{\rm y}
  \displaystyle\sum\limits_{s_{i}, s_{f} = \pm 1/2}
    v_{\mu_{f}}^{*} (s_{f})\, \hat{\sigmabf}_{y}\, v_{\mu_{i}} (s_{i}) + \\
  & + &
  \mathbf{e}_{\rm z}
  \displaystyle\sum\limits_{s_{i}, s_{f} = \pm 1/2}
    v_{\mu_{f}}^{*} (s_{f})\, \hat{\sigmabf}_{z}\, v_{\mu_{i}} (s_{i}) =

  \mathbf{e}_{\rm x} +
  \mathbf{e}_{\rm y}\, i\, \Bigl\{ \delta_{\mu_{i}, +1/2}\; -\; \delta_{\mu_{i}, -1/2} \Bigr\} +
  \mathbf{e}_{\rm z},
\end{array}
\label{eq.2.3.11}
\end{equation}
where orthogonal unit vectors $\mathbf{e}_{\rm x}$, $\mathbf{e}_{\rm y}$, $\mathbf{e}_{\rm z}$ are used.

So, using the found eqs.~(\ref{eq.2.3.10})--(\ref{eq.2.3.11}), we perform summation in eqs.~(\ref{eq.2.2.4}) over all spinor states:
\begin{equation}
\begin{array}{ccl}
  \vspace{4mm}
  \Bigl< k_{f} \Bigl|\, e^{-i\,\mathbf{k_{\rm ph}r}}\: \nabla\, \Bigr| \,k_{i} \Bigr> & = &
  \displaystyle\sum\limits_{m_{f}, m_{i}}
  \displaystyle\sum\limits_{\mu_{i},\, \mu_{f} = \pm 1/2}
    C_{l_{f}m_{f} 1/2 \mu_{f}}^{j_{f}M_{f},\,*}\,
    C_{l_{i}m_{i} 1/2 \mu_{i}}^{j_{i}M_{i}} \cdot
    \Bigl< k_{f} \Bigl|\, e^{-i\,\mathbf{k_{\rm ph}r}}\: \nabla\, \Bigr| \,k_{i} \Bigr>_\mathbf{r}, \\

  \vspace{1mm}
  \Bigl< k_{f} \Bigl|\, e^{-i\,\mathbf{k_{\rm ph}r}}\: \sigmabf\, \Bigr| \,k_{i} \Bigr> & = &
  \displaystyle\sum\limits_{m_{f}, m_{i}}
  \displaystyle\sum\limits_{\mu_{i},\, \mu_{f} = \pm 1/2}
    C_{l_{f}m_{f} 1/2 \mu_{f}}^{j_{f}M_{f},\,*}\,
    C_{l_{i}m_{i} 1/2 \mu_{i}}^{j_{i}M_{i}} \quad \times \\
  \vspace{4mm}
  & \times &
    \Bigl[
      \mathbf{e}_{\rm x} +
      \mathbf{e}_{\rm y}\, i\, \Bigl\{ \delta_{\mu_{i}, +1/2}\; -\; \delta_{\mu_{i}, -1/2} \Bigr\} +
      \mathbf{e}_{\rm z} \Bigr]\:
    \Bigl< k_{f} \Bigl|\, e^{-i\,\mathbf{k_{\rm ph}r}}\, \Bigr| \,k_{i} \Bigr>_\mathbf{r}, \\

  \vspace{1mm}
  \Bigl< k_{f} \Bigl|\, e^{-i\,\mathbf{k_{\rm ph}r}}\: \sigmabf \cdot
    \Bigl[\mathbf{\rm e}^{(\alpha)} \times \nabla \Bigr] \Bigr| \,k_{i} \Bigr> & = &
  \displaystyle\sum\limits_{m_{f}, m_{i}}
  \displaystyle\sum\limits_{\mu_{i},\, \mu_{f} = \pm 1/2}
    C_{l_{f}m_{f} 1/2 \mu_{f}}^{j_{f}M_{f},\,*}\,
    C_{l_{i}m_{i} 1/2 \mu_{i}}^{j_{i}M_{i}} \times \\
  & \times &
    \Bigl[
      \mathbf{e}_{\rm x} +
      \mathbf{e}_{\rm y}\, i\, \Bigl\{ \delta_{\mu_{i}, +1/2}\; -\; \delta_{\mu_{i}, -1/2} \Bigr\} +
      \mathbf{e}_{\rm z} \Bigr] \cdot
    \biggl[\mathbf{\rm e}^{(\alpha)} \times
    \Bigl< k_{f} \Bigl|\, e^{-i\,\mathbf{k_{\rm ph}r}}\, \nabla \Bigr| \,k_{i} \Bigr>_\mathbf{r} \biggr],
\end{array}
\label{eq.2.3.16}
\end{equation}
where $\bigl< k_{f} \bigl|\, \ldots \bigr| \,k_{i} \bigr>_\mathbf{r}$ is one-component matrix element
\begin{equation}
\begin{array}{lcl}
  \Bigl< k_{f} \Bigl|\, \hat{f}  \Bigr| \,k_{i} \Bigr>_\mathbf{r} & \equiv &
  \displaystyle\int
    R_{f}^{*}\,(r)\:
    Y_{l_{f}m_{f}}(\mathbf{n}_{\rm r})^{*}\;
    \hat{f} \:
    R_{i}\,(r)\:
    Y_{l_{i}m_{i}}(\mathbf{n}_{\rm r})\; \mathbf{dr},
\end{array}
\label{eq.2.3.13}
\end{equation}
where integration should be performed over space coordinates only.

We orient frame vectors $\mathbf{e}_{\rm x}$, $\mathbf{e}_{\rm y}$ and $\mathbf{e}_{\rm z}$ so, that $\mathbf{e}_{\rm z}$ be directed along to $\mathbf{k}_{\rm ph}$. Then, vectors $\mathbf{e}_{\rm x}$ and $\mathbf{e}_{\rm y}$ can be directed along $\mathbf{e}^{(1)}$ and $\mathbf{e}^{(2)}$, correspondingly.
In Coulomb gauge we obtain:
\begin{equation}
\begin{array}{lcccc}
  \mathbf{e}_{\rm x} = \mathbf{e}^{(1)}, &
  \mathbf{e}_{\rm y} = \mathbf{e}^{(2)}, &
  |\mathbf{e}_{\rm x}| = |\mathbf{e}_{\rm y}| = |\mathbf{e}_{\rm z}| = 1, &
  |\mathbf{e}^{(3)}| = 0.
\end{array}
\label{eq.2.3.22}
\end{equation}
Now we perform summation in eqs.~(\ref{eq.2.2.4}) over polarization vectors and obtain:
\begin{equation}
\begin{array}{lcl}
  \vspace{2mm}
  p_{\rm el} & = &
  i\, \displaystyle\sum\limits_{m_{f}, m_{i}}
  \displaystyle\sum\limits_{\mu_{i},\, \mu_{f} = \pm 1/2}
    C_{l_{f}m_{f} 1/2 \mu_{f}}^{j_{f}M_{f},\,*}\,
    C_{l_{i}m_{i} 1/2 \mu_{i}}^{j_{i}M_{i}} \cdot
    (\mathbf{e}^{(1)} + \mathbf{e}^{(2)})\,
    \Bigl< k_{f} \Bigl|\, e^{-i\,\mathbf{k_{\rm ph}r}}\: \nabla\, \Bigr| \,k_{i} \Bigr>_\mathbf{r}, \\

  \vspace{1mm}
  p_{\rm mag,\,1} & = &
  \displaystyle\frac{1}{2}
  \displaystyle\sum\limits_{m_{f}, m_{i}}
  \displaystyle\sum\limits_{\mu_{i},\, \mu_{f} = \pm 1/2}
    C_{l_{f}m_{f} 1/2 \mu_{f}}^{j_{f}M_{f},\,*}\,
    C_{l_{i}m_{i} 1/2 \mu_{i}}^{j_{i}M_{i}} \cdot
    \Bigl[
      \mathbf{e}_{\rm x} +
      \mathbf{e}_{\rm y}\, i\, \Bigl\{ \delta_{\mu_{i}, +1/2}\; -\; \delta_{\mu_{i}, -1/2} \Bigr\} +
      \mathbf{e}_{\rm z} \Bigr] \times \\
  \vspace{2mm}
  & \times &
    \biggl[
      \displaystyle\sum\limits_{\alpha=1,2} \mathbf{\rm e}^{(\alpha)} \times
      \Bigl< k_{f} \Bigl|\, e^{-i\,\mathbf{k_{\rm ph}r}}\, \nabla \Bigr| \,k_{i} \Bigr>_\mathbf{r} \biggr], \\

  p_{\rm mag,\,2} & = &
  \displaystyle\frac{-i\,k_{\rm ph}}{2}\,
  \displaystyle\sum\limits_{m_{f}, m_{i}}
  \displaystyle\sum\limits_{\mu_{i},\, \mu_{f} = \pm 1/2}
    C_{l_{f}m_{f} 1/2 \mu_{f}}^{j_{f}M_{f},\,*}\,
    C_{l_{i}m_{i} 1/2 \mu_{i}}^{j_{i}M_{i}} \cdot
  \Bigl[ -1 + i\, \Bigl\{ \delta_{\mu_{i}, +1/2}\; -\; \delta_{\mu_{i}, -1/2} \Bigr\} \Bigr]\:
  \Bigl< k_{f} \Bigl|\, e^{-i\,\mathbf{k_{\rm ph}r}}\, \Bigr| \,k_{i} \Bigr>_\mathbf{r}.
\end{array}
\label{eq.2.3.23}
\end{equation}

\subsection{Matric elements integrated over space coordinates
\label{sec.2.4}}

We shall calculate the following matrix elements:
\begin{equation}
\begin{array}{ll}
  \Bigl< k_{f} \Bigl| \,  e^{-i\mathbf{k_{\rm ph}r}} \, \Bigr| \,k_{i} \Bigr>_\mathbf{r} =
  \displaystyle\int
    \varphi^{*}_{f}(\mathbf{r}) \:
    e^{-i\mathbf{k_{\rm ph}r}}\:
    \varphi_{i}(\mathbf{r}) \;
    \mathbf{dr}, &

  \hspace{5mm}
  \biggl< k_{f} \biggl| \,  e^{-i\mathbf{k_{\rm ph}r}} \displaystyle\frac{\partial}{\partial \mathbf{r}} \,
  \biggr| \,k_{i} \biggr>_\mathbf{r} =
  \displaystyle\int
    \varphi^{*}_{f}(\mathbf{r}) \:
    e^{-i\mathbf{k_{\rm ph}r}} \displaystyle\frac{\partial}{\partial \mathbf{r}}\:
    \varphi_{i}(\mathbf{r}) \;
    \mathbf{dr}.
\end{array}
\label{eq.2.4.1.1}
\end{equation}


\subsubsection{Expansion of the vector potential $\mathbf{A}$ by multipoles
\label{sec.2.4.3}}

Let us expand the vectorial potential $\mathbf{A}$ of electromagnetic field by multipole. According to \cite{Eisenberg.1973} (see~(2.106), p.~58),
in the spherical symmetric approximation we have:
\begin{equation}
  \mathbf{\xi}_{\mu}\, e^{i \mathbf{k_{\rm ph}r}} =
    \mu\, \sqrt{2\pi}\, \sum_{l_{\rm ph}=1}\,
    (2l_{\rm ph}+1)^{1/2}\, i^{l_{\rm ph}}\,  \cdot
    \Bigl[ \mathbf{A}_{l_{\rm ph}\mu} (\mathbf{r}, M) +
    i\mu\, \mathbf{A}_{l_{\rm ph}\mu} (\mathbf{r}, E) \Bigr],
\label{eq.2.4.3.1}
\end{equation}
where (see~\cite{Eisenberg.1973}, (2.73) in p.~49, (2.80) in p.~51)
\begin{equation}
\begin{array}{lcl}
  \vspace{2mm}
  \mathbf{A}_{l_{\rm ph}\mu}(\mathbf{r}, M) & = &
        j_{l_{\rm ph}}(k_{\rm ph}r) \: \mathbf{T}_{l_{\rm ph}l_{\rm ph},\mu} (\mathbf{n}_{\rm r}), \\
  \vspace{2mm}
  \mathbf{A}_{l_{\rm ph}\mu}(\mathbf{r}, E) & = &
        \sqrt{\displaystyle\frac{l_{\rm ph}+1}{2l_{\rm ph}+1}}\,
        j_{l_{\rm ph}-1}(k_{\rm ph}r) \: \mathbf{T}_{l_{\rm ph}l_{\rm ph}-1,\mu}(\mathbf{n}_{\rm r})\; -
        \sqrt{\displaystyle\frac{l_{\rm ph}}{2l_{\rm ph}+1}}\,
        j_{l_{\rm ph}+1}(k_{\rm ph}r) \: \mathbf{T}_{l_{\rm ph}l_{\rm ph}+1,\mu}(\mathbf{n}_{\rm r}).
\end{array}
\label{eq.2.4.3.2}
\end{equation}
Here, $\mathbf{A}_{l_{\rm ph}\mu}(\textbf{r}, M)$ and $\mathbf{A}_{l_{\rm ph}\mu}(\textbf{r}, E)$ are \emph{magnetic} and \emph{electric multipoles}, $j_{l_{\rm ph}}(k_{\rm ph}r)$ is \emph{spherical Bessel function of order $l_{\rm ph}$}, $\mathbf{T}_{l_{\rm ph}l_{\rm ph}^{\prime},\mu}(\mathbf{n}_{\rm r})$ are \emph{vector spherical harmonics}.
Eq.~(\ref{eq.2.4.3.1}) is solution of the wave equation of electromagnetic field in form of plane wave, which is presented as summation of the electrical and magnetic multipoles (for example, see p.~83--92 in~\cite{Ahiezer.1981}). Therefore, separate multipolar terms in eq.~(\ref{eq.2.4.3.1}) are solutions of this wave equation for chosen numbers $j_{\rm ph}$ and $l_{\rm ph}$ ($j_{\rm ph}$ is quantum number characterizing eigenvalue of the full momentum operator, while $l_{\rm ph}= j_{\rm ph}-1, j_{\rm ph}, j_{\rm ph}+1$ is connected with orbital momentum operator, but it defines eigenvalues of photon parity and, so, it is quantum number also).


We orient the frame so that axis $z$ be directed along the vector $\mathbf{k}_{\rm ph}$ (see~\cite{Eisenberg.1973}, (2.105) in p.~57). According to \cite{Eisenberg.1973} (see p.~45), the functions $\mathbf{T}_{l_{\rm ph}l_{\rm ph}^{\prime},\mu}(\mathbf{n}_{\rm r})$ have the following form
(${\mathbf \xi}_{0} = 0$):
\begin{equation}
  \mathbf{T}_{j_{\rm ph}l_{\rm ph},m} (\mathbf{n}_{\rm r}) =
  \sum\limits_{\mu = \pm 1} (l_{\rm ph}, 1, j_{\rm ph} \,\big| \,m-\mu, \mu, m)\;
  Y_{l_{\rm ph},m-\mu}(\mathbf{n}_{\rm r})\;
  \mathbf{\xi}_{\mu},
\label{eq.2.4.3.3}
\end{equation}
where $(l, 1, j \,\bigl| \, m-\mu, \mu, m)$ are \emph{Clebsh-Gordon coefficients},
$Y_{lm}(\theta, \varphi)$ are \emph{spherical functions} defined, according to~\cite{Landau.v3.1989} (see p.~119, (28,7)--(28,8)).
From eq.~(\ref{eq.2.4.3.1}) one can obtain such a formula (at $\mathbf{e}^{(3)}=0$):
\begin{equation}
  e^{-i \mathbf{k_{\rm ph}r}} =
  \displaystyle\frac{1}{2}\,
  \displaystyle\sum\limits_{\mu = \pm 1}
    \mathbf{\xi}_{\mu}\, \mu\, \sqrt{2\pi}\, \sum_{l_{\rm ph}=1}\,
    (2l_{\rm ph}+1)^{1/2}\, (-i)^{l_{\rm ph}}\,  \cdot
    \Bigl[ \mathbf{A}_{l_{\rm ph}\mu}^{*} (\mathbf{r}, M) -
    i\mu\, \mathbf{A}_{l_{\rm ph}\mu}^{*} (\mathbf{r}, E) \Bigr].
\label{eq.2.4.3.5}
\end{equation}

\subsubsection{Spherically symmetric decay
\label{sec.2.4.4}}

Using (\ref{eq.2.4.3.5}), for (\ref{eq.2.4.1.1}) we find:
\begin{equation}
\begin{array}{ll}
  \vspace{1mm}
  \Bigl< k_{f} \Bigl| \,  e^{-i\mathbf{k_{\rm ph}r}} \, \Bigr| \,k_{i} \Bigr>_\mathbf{r} =
  \sqrt{\displaystyle\frac{\pi}{2}}\:
  \displaystyle\sum\limits_{l_{\rm ph}=1}\,
    (-i)^{l_{\rm ph}}\, \sqrt{2l_{\rm ph}+1}\;
  \displaystyle\sum\limits_{\mu = \pm 1}
    \Bigl[ \mu\,\tilde{p}_{l_{\rm ph}\mu}^{M} - i\, \tilde{p}_{l_{\rm ph}\mu}^{E} \Bigr], \\

  \biggl< k_{f} \biggl| \,  e^{-i\mathbf{k_{\rm ph}r}} \displaystyle\frac{\partial}{\partial \mathbf{r}}\,
  \biggr| \,k_{i} \biggr>_\mathbf{r} =
  \sqrt{\displaystyle\frac{\pi}{2}}\:
  \displaystyle\sum\limits_{l_{\rm ph}=1}\,
    (-i)^{l_{\rm ph}}\, \sqrt{2l_{\rm ph}+1}\;
  \displaystyle\sum\limits_{\mu = \pm 1}
    \xibf_{\mu}\, \mu\, \times
    \Bigl[ p_{l_{\rm ph}\mu}^{M} - i\mu\: p_{l_{\rm ph}\mu}^{E} \Bigr],
\end{array}
\label{eq.2.4.4.1}
\end{equation}
where
\begin{equation}
\begin{array}{lcllcl}
  p_{l_{\rm ph}\mu}^{M} & = &
    \displaystyle\int
        \varphi^{*}_{f}(\mathbf{r}) \,
        \biggl( \displaystyle\frac{\partial}{\partial \mathbf{r}}\, \varphi_{i}(\mathbf{r}) \biggr) \,
        \mathbf{A}_{l_{\rm ph}\mu}^{*} (\mathbf{r}, M) \;
        \mathbf{dr}, &

  \hspace{7mm}
  p_{l_{\rm ph}\mu}^{E} & = &
    \displaystyle\int
        \varphi^{*}_{f}(\mathbf{r}) \,
        \biggl( \displaystyle\frac{\partial}{\partial \mathbf{r}}\, \varphi_{i}(\mathbf{r}) \biggr)\,
        \mathbf{A}_{l_{\rm ph}\mu}^{*} (\mathbf{r}, E) \;
        \mathbf{dr},
\end{array}
\label{eq.2.4.4.2}
\end{equation}
and
\begin{equation}
\begin{array}{lcllcl}
  \tilde{p}_{l_{\rm ph}\mu}^{M} & = &
    \xibf_{\mu}\,
    \displaystyle\int
      \varphi^{*}_{f}(\mathbf{r})\,
      \varphi_{i}(\mathbf{r})\;
      \mathbf{A}_{l_{\rm ph}\mu}^{*} (\mathbf{r}, M) \;
      \mathbf{dr}, &
  \hspace{7mm}
  \tilde{p}_{l_{\rm ph}\mu}^{E} & = &
    \xibf_{\mu}\,
    \displaystyle\int
      \varphi^{*}_{f}(\mathbf{r})\,
      \varphi_{i}(\mathbf{r})\;
      \mathbf{A}_{l_{\rm ph}\mu}^{*} (\mathbf{r}, E)\;
      \mathbf{dr}.
\end{array}
\label{eq.2.4.4.3}
\end{equation}

Now we shall calculate components in eqs.~(\ref{eq.2.3.23}). For the first and third items we obtain:
\begin{equation}
\begin{array}{lcl}
  \vspace{1mm}
  p_{\rm el} & = &
  i\, \sqrt{\displaystyle\frac{\pi}{2}}
  \displaystyle\sum\limits_{m_{i}, m_{f}}
  \displaystyle\sum\limits_{\mu_{i},\, \mu_{f} = \pm 1/2}
    C_{l_{f}m_{f} 1/2 \mu_{f}}^{j_{f}M_{f},\,*}\,
    C_{l_{i}m_{i} 1/2 \mu_{i}}^{j_{i}M_{i}} \cdot
    \displaystyle\sum\limits_{l_{\rm ph}=1}\,
      (-i)^{l_{\rm ph}}\, \sqrt{2l_{\rm ph}+1}\; \cdot
    \Bigl[
      p_{l_{\rm ph}}^{M} -
      i\,p_{l_{\rm ph}}^{E}
    \Bigr], \\

  \vspace{1mm}
  p_{\rm mag,\,2} & = &
  \displaystyle\frac{-i\,k_{\rm ph}}{2}\,
  \sqrt{\displaystyle\frac{\pi}{2}}\;
  \displaystyle\sum\limits_{m_{i}, m_{f}}
  \displaystyle\sum\limits_{\mu_{i},\, \mu_{f} = \pm 1/2}
    C_{l_{f}m_{f} 1/2 \mu_{f}}^{j_{f}M_{f},\,*}\,
    C_{l_{i}m_{i} 1/2 \mu_{i}}^{j_{i}M_{i}}\quad \times \\
  \vspace{1mm}
  & \times &
  \Bigl[ -1 + i\, \Bigl\{ \delta_{\mu_{i}, +1/2}\; -\; \delta_{\mu_{i}, -1/2} \Bigr\} \Bigr] \cdot
    \displaystyle\sum\limits_{l_{\rm ph}=1}\,
      (-i)^{l_{\rm ph}}\, \sqrt{2l_{\rm ph}+1} \cdot
      \Bigl[ \tilde{p}_{l_{\rm ph}}^{M} - i\, \tilde{p}_{l_{\rm ph}}^{E} \Bigr],
\end{array}
\label{eq.2.4.4.4}
\end{equation}
where
\begin{equation}
\begin{array}{cccc}
  p_{l_{\rm ph}}^{M} = \displaystyle\sum\limits_{\mu = \pm 1} h_{\mu}\, \mu\, p_{l_{\rm ph}\mu}^{M}, &
  \hspace{7mm}
  p_{l_{\rm ph}}^{E} = \displaystyle\sum\limits_{\mu = \pm 1} h_{\mu}\, p_{l_{\rm ph}\mu}^{E}, &
  \hspace{7mm}
  \tilde{p}_{l_{\rm ph}}^{M} =
    \displaystyle\sum\limits_{\mu = \pm 1} \mu\; \tilde{p}_{l_{\rm ph}\mu}^{M}, &
  \hspace{7mm}
  \tilde{p}_{l_{\rm ph}}^{E} =
    \displaystyle\sum\limits_{\mu = \pm 1} \tilde{p}_{l_{\rm ph}\mu}^{E}.
\end{array}
\label{eq.2.4.4.5}
\end{equation}
Now we shall analyze the second item in eqs.~(\ref{eq.2.3.23}) and find:
\begin{equation}
\begin{array}{ll}
  \vspace{1mm}
  p_{\rm mag,\,1} =
    \displaystyle\frac{1}{2}
  \displaystyle\sum\limits_{m_{i}, m_{f}}
  \displaystyle\sum\limits_{\mu_{i},\, \mu_{f} = \pm 1/2}
    C_{l_{f}m_{f} 1/2 \mu_{f}}^{j_{f}M_{f},\,*}\,
    C_{l_{i}m_{i} 1/2 \mu_{i}}^{j_{i}M_{i}} \cdot
    \Bigl[
      \displaystyle\frac{1}{\sqrt{2}}\, \bigl(\xibf_{-1} - \xibf_{+1}\bigr) +
      \displaystyle\frac{i}{\sqrt{2}}\, \bigl(\xibf_{-1} + \xibf_{+1}\bigr)\,
      i\, \Bigl\{ \delta_{\mu_{i}, +1/2}\; - \\

  - \quad
    \delta_{\mu_{i}, -1/2} \Bigr\} +  \mathbf{e}_{\rm z} \Bigr]\quad \times
    \biggl[
      \displaystyle\sum\limits_{\mu=\pm 1} h_{\mu} \xibf_{\mu}^{*} \times
    \sqrt{\displaystyle\frac{\pi}{2}}\:
    \displaystyle\sum\limits_{l_{\rm ph}=1}\,
      (-i)^{l_{\rm ph}}\, \sqrt{2l_{\rm ph}+1}\;
    \displaystyle\sum\limits_{\mu^{\prime} = \pm 1}
      \xibf_{\mu^{\prime}}\, \mu^{\prime}\, \times
      \Bigl[ p_{l_{\rm ph}\mu^{\prime}}^{M} - i\mu^{\prime}\: p_{l_{\rm ph}\mu^{\prime}}^{E} \Bigr]\, \biggr].
\end{array}
\label{eq.2.4.4.6}
\end{equation}
Taking properties (\ref{eq.2.4.2.7}) into account, we calculate eq.~(\ref{eq.2.4.4.6}) further and obtain:
\begin{equation}
\begin{array}{ll}
  p_{\rm mag,\,1} =
    - \displaystyle\frac{1}{2}
  \sqrt{\displaystyle\frac{\pi}{2}}\:
  \displaystyle\sum\limits_{m_{i}, m_{f}}
  \displaystyle\sum\limits_{\mu_{i},\, \mu_{f} = \pm 1/2}
    C_{l_{f}m_{f} 1/2 \mu_{f}}^{j_{f}M_{f},\,*}\,
    C_{l_{i}m_{i} 1/2 \mu_{i}}^{j_{i}M_{i}}\;
    \displaystyle\sum\limits_{l_{\rm ph}=1}\,
      (-i)^{l_{\rm ph}}\, \sqrt{2l_{\rm ph}+1}\; \cdot
    \displaystyle\sum\limits_{\mu=\pm 1}
      i\, h_{\mu}\, \mu \Bigl[\mu\, p_{l_{\rm ph}\mu}^{M} - i\, p_{l_{\rm ph}\mu}^{E} \Bigr].
\end{array}
\label{eq.2.4.4.7}
\end{equation}
So, we have found all components in (\ref{eq.2.2.4}):
\begin{equation}
\begin{array}{lcl}
  \vspace{3mm}
  p_{\rm el} & = &
  \sqrt{\displaystyle\frac{\pi}{2}}\:
  \displaystyle\sum\limits_{l_{\rm ph}=1}\,
    (-i)^{l_{\rm ph}}\, \sqrt{2l_{\rm ph}+1}\; \cdot
  \displaystyle\sum\limits_{\mu=\pm 1}
    h_{\mu} \cdot
  \displaystyle\sum\limits_{m_{i}, m_{f}}
  \displaystyle\sum\limits_{\mu_{i},\, \mu_{f} = \pm 1/2}
    C_{l_{f}m_{f} 1/2 \mu_{f}}^{j_{f}M_{f},\,*}\,
    C_{l_{i}m_{i} 1/2 \mu_{i}}^{j_{i}M_{i}}\,
    \Bigl[
      i\,\mu\, p_{l_{\rm ph}\mu}^{M m_{i} m_{f}} +
      p_{l_{\rm ph}\mu}^{E m_{i} m_{f}}
    \Bigr], \\

  \vspace{3mm}
  p_{\rm mag,1} & = &
  \displaystyle\frac{1}{2}\:
  \sqrt{\displaystyle\frac{\pi}{2}}\:
  \displaystyle\sum\limits_{l_{\rm ph}=1}\,
    (-i)^{l_{\rm ph}}\, \sqrt{2l_{\rm ph}+1}\; \cdot
  \displaystyle\sum\limits_{\mu=\pm 1}
    h_{\mu}\, \mu
  \displaystyle\sum\limits_{m_{i}, m_{f}}
  \displaystyle\sum\limits_{\mu_{i},\, \mu_{f} = \pm 1/2}
    C_{l_{f}m_{f} 1/2 \mu_{f}}^{j_{f}M_{f},\,*}\,
    C_{l_{i}m_{i} 1/2 \mu_{i}}^{j_{i}M_{i}}\,
    \Bigl[i\,\mu\, p_{l_{\rm ph}\mu}^{M m_{i} m_{f}} + p_{l_{\rm ph}\mu}^{E m_{i} m_{f}} \Bigr], \\

  \vspace{1mm}
  p_{\rm mag, 2} & = &
  \sqrt{\displaystyle\frac{\pi}{8}}\: k_{\rm ph}\;
  \displaystyle\sum\limits_{l_{\rm ph}=1}\,
    (-i)^{l_{\rm ph}}\, \sqrt{2l_{\rm ph}+1} \cdot
  \displaystyle\sum\limits_{\mu=\pm 1}
  \displaystyle\sum\limits_{m_{i}, m_{f}}
  \displaystyle\sum\limits_{\mu_{i},\, \mu_{f} = \pm 1/2}
    C_{l_{f}m_{f} 1/2 \mu_{f}}^{j_{f}M_{f},\,*}\,
    C_{l_{i}m_{i} 1/2 \mu_{i}}^{j_{i}M_{i}}\quad \times \\
  & \times &
  \Bigl[ -1 + i\, \Bigl\{ \delta_{\mu_{i}, +1/2}\; -\; \delta_{\mu_{i}, -1/2} \Bigr\} \Bigr] \cdot
    \Bigl[ i\,\mu\,\tilde{p}_{l_{\rm ph}}^{M} + \tilde{p}_{l_{\rm ph}}^{E} \Bigr].
\end{array}
\label{eq.2.4.4.8}
\end{equation}

\subsubsection{Calculations of the components $p_{l_{\rm ph}\mu}^{M}$, $p_{l_{\rm ph}\mu}^{E}$ and\,
$\tilde{p}_{l_{\rm ph}\mu}^{M}$, $\tilde{p}_{l_{\rm ph}\mu}^{E}$
\label{sec.2.4.6}}

For calculation of these components we shall use \emph{gradient formula} (see~\cite{Eisenberg.1973}, (2.56) in p.~46):
\begin{equation}
\begin{array}{lcl}
  \displaystyle\frac{\partial}{\partial \mathbf{r}}\: \varphi_{i}(\mathbf{r}) & = &
  \displaystyle\frac{\partial}{\partial \mathbf{r}}\:
    \Bigl\{ R_{i} (r)\: Y_{l_{i}m_{i}}({\mathbf n}_{\rm r}) \Bigr\} =

    \sqrt{\displaystyle\frac{l_{i}}{2l_{i}+1}}\:
    \biggl( \displaystyle\frac{dR_{i}(r)}{dr} + \displaystyle\frac{l_{i}+1}{r}\, R_{i}(r) \biggr)\,
      \mathbf{T}_{l_{i} l_{i}-1, m_{i}}({\mathbf n}_{\rm r}) - \\
  & - &
  \sqrt{\displaystyle\frac{l_{i}+1}{2l_{i}+1}}\:
    \biggl( \displaystyle\frac{dR_{i}(r)}{dr} - \displaystyle\frac{l_{i}}{r}\, R_{i}(r) \biggr)\,
      \mathbf{T}_{l_{i} l_{i}+1, m_{i}}({\mathbf n}_{\rm r}),
\end{array}
\label{eq.2.4.6.1}
\end{equation}
and obtain:
\begin{equation}
\begin{array}{lcl}
\vspace{1mm}
  p_{l_{\rm ph,\mu}}^{M} & = &
    \sqrt{\displaystyle\frac{l_{i}}{2l_{i}+1}}\:
      I_{M}(l_{i},l_{f}, l_{\rm ph}, l_{i}-1, \mu) \cdot
      \Bigl\{
        J_{1}(l_{i},l_{f},l_{\rm ph}) + (l_{i}+1) \cdot J_{2}(l_{i},l_{f},l_{\rm ph})
      \Bigr\}\; - \\
\vspace{3mm}
  & - &
    \sqrt{\displaystyle\frac{l_{i}+1}{2l_{i}+1}}\:
      I_{M}(l_{i},l_{f}, l_{\rm ph}, l_{i}+1, \mu) \cdot
      \Bigl\{
        J_{1}(l_{i},l_{f},l_{\rm ph}) - l_{i} \cdot J_{2}(l_{i},l_{f},l_{\rm ph})
      \Bigr\}, \\

\vspace{1mm}
  p_{l_{\rm ph,\mu}}^{E} & = &
    \sqrt{\displaystyle\frac{l_{i}\,(l_{\rm ph}+1)}{(2l_{i}+1)(2l_{\rm ph}+1)}} \cdot
      I_{E}(l_{i},l_{f}, l_{\rm ph}, l_{i}-1, l_{\rm ph}-1, \mu) \cdot
      \Bigl\{
        J_{1}(l_{i},l_{f},l_{\rm ph}-1)\; +
        (l_{i}+1) \cdot J_{2}(l_{i},l_{f},l_{\rm ph}-1)
      \Bigr\}\; - \\
\vspace{1mm}
    & - &
    \sqrt{\displaystyle\frac{l_{i}\,l_{\rm ph}}{(2l_{i}+1)(2l_{\rm ph}+1)}} \cdot
      I_{E} (l_{i},l_{f}, l_{\rm ph}, l_{i}-1, l_{\rm ph}+1, \mu) \cdot
      \Bigl\{
        J_{1}(l_{i},l_{f},l_{\rm ph}+1)\; +
        (l_{i}+1) \cdot J_{2}(l_{i},l_{f},l_{\rm ph}+1)
      \Bigr\}\; + \\
\vspace{1mm}
  & + &
    \sqrt{\displaystyle\frac{(l_{i}+1)(l_{\rm ph}+1)}{(2l_{i}+1)(2l_{\rm ph}+1)}} \cdot
      I_{E} (l_{i},l_{f},l_{\rm ph}, l_{i}+1, l_{\rm ph}-1, \mu) \cdot
      \Bigl\{
        J_{1}(l_{i},l_{f},l_{\rm ph}-1)\; -
        l_{i} \cdot J_{2}(l_{i},l_{f},l_{\rm ph}-1)
      \Bigr\}\; - \\
  & - &
    \sqrt{\displaystyle\frac{(l_{i}+1)\,l_{\rm ph}}{(2l_{i}+1)(2l_{\rm ph}+1)}} \cdot
      I_{E} (l_{i},l_{f}, l_{\rm ph}, l_{i}+1, l_{\rm ph}+1, \mu) \cdot
      \Bigl\{
        J_{1}(l_{i},l_{f},l_{\rm ph}+1)\; -
        l_{i} \cdot J_{2}(l_{i},l_{f},l_{\rm ph}+1)
      \Bigr\},
\end{array}
\label{eq.2.4.6.4}
\end{equation}
where
\begin{equation}
\begin{array}{ccl}
  J_{1}(l_{i},l_{f},n) & = &
  \displaystyle\int\limits^{+\infty}_{0}
    \displaystyle\frac{dR_{i}(r, l_{i})}{dr}\: R^{*}_{f}(l_{f},r)\,
    j_{n}(k_{\rm ph}r)\; r^{2} dr, \\

  J_{2}(l_{i},l_{f},n) & = &
  \displaystyle\int\limits^{+\infty}_{0}
    R_{i}(r, l_{i})\, R^{*}_{f}(l_{f},r)\: j_{n}(k_{\rm ph}r)\; r\, dr, \\

  I_{M}\, (l_{i}, l_{f}, l_{\rm ph}, l_{1}, \mu) & = &
    \displaystyle\int
      Y_{l_{f}m_{f}}^{*}(\mathbf{n}_{\rm r})\,
      \mathbf{T}_{l_{i}\, l_{1},\, m_{i}}(\mathbf{n}_{\rm r})\,
      \mathbf{T}_{l_{\rm ph}\,l_{\rm ph},\, \mu}^{*}(\mathbf{n}_{\rm r})\; d\Omega, \\

  I_{E}\, (l_{i}, l_{f}, l_{\rm ph}, l_{1}, l_{2}, \mu) & = &
    \displaystyle\int
      Y_{l_{f}m_{f}}^{*}(\mathbf{n}_{\rm r})\,
      \mathbf{T}_{l_{i} l_{1},\, m_{i}}(\mathbf{n}_{\rm r})\,
      \mathbf{T}_{l_{\rm ph} l_{2},\, \mu}^{*}(\mathbf{n}_{\rm r})\; d\Omega.
\end{array}
\label{eq.2.4.6.3}
\end{equation}
By the same way, for $\tilde{p}_{l_{\rm ph}\mu}^{M}$ и $\tilde{p}_{l_{\rm ph}\mu}^{E}$ we find:
\begin{equation}
\begin{array}{lcl}
  \tilde{p}_{l_{\rm ph}\mu}^{M} & = &
    \tilde{I}\,(l_{i},l_{f},l_{\rm ph}, l_{\rm ph}, \mu) \cdot \tilde{J}\, (l_{i},l_{f},l_{\rm ph}), \\
  \tilde{p}_{l_{\rm ph}\mu}^{E} & = &
    \sqrt{\displaystyle\frac{l_{\rm ph}+1}{2l_{\rm ph}+1}}
      \tilde{I}\,(l_{i},l_{f},l_{\rm ph},l_{\rm ph}-1,\mu) \cdot \tilde{J}\,(l_{i},l_{f},l_{\rm ph}-1) -
    \sqrt{\displaystyle\frac{l_{\rm ph}}{2l_{\rm ph}+1}}
      \tilde{I}\,(l_{i},l_{f},l_{\rm ph},l_{\rm ph}+1,\mu) \cdot \tilde{J}\,(l_{i},l_{f},l_{\rm ph}+1),
\end{array}
\label{eq.2.4.6.7}
\end{equation}
where
\begin{equation}
\begin{array}{lcl}
  \tilde{J}\,(l_{i},l_{f},n) & = &
  \displaystyle\int\limits^{+\infty}_{0}
    R_{i}(r)\, R^{*}_{f}(l,r)\, j_{n}(k_{\rm ph}r)\; r^{2} dr, \\

  \tilde{I}\,(l_{i}, l_{f}, l_{\rm ph}, n, \mu) & = &
  \xibf_{\mu} \displaystyle\int
    Y_{l_{i}m_{i}}(\mathbf{n}_{\rm r})\:
    Y_{l_{f}m_{f}}^{*}(\mathbf{n}_{\rm r})\:
    \mathbf{T}_{l_{\rm ph} n,\mu}^{*}(\mathbf{n}_{\rm r}) \: d\Omega.
\end{array}
\label{eq.2.4.6.6}
\end{equation}

\subsubsection{Differential (angular) matrix elements of emission
\label{sec.2.4.11}}

We shall be interesting in the angular emission of the bremsstrahlung photons. By such a reason let us introduce the following differential matrix elements, $dp_{l}^{M}$ and $dp_{l}^{E}$,
dependent on the angle $\theta$:
\begin{equation}
\begin{array}{lcl}
\vspace{2mm}
  \displaystyle\frac{d\, p_{l_{\rm ph}\mu}^{M}}{\sin{\theta}\,d\theta} & = &
    \sqrt{\displaystyle\frac{l_{i}}{2l_{i}+1}}\:
    \displaystyle\frac{d\, I_{M}\,(l_{i}, l_{f}, l_{\rm ph}, l_{i}-1, \mu)}{\sin{\theta}\,d\theta} \cdot
    \Bigl\{
      J_{1}(l_{i},l_{f},l_{\rm ph}) +
      (l_{i}+1) \cdot J_{2}(l_{i},l_{f},l_{\rm ph})
    \Bigr\}\; - \\
  & - &
    \sqrt{\displaystyle\frac{l_{i}+1}{2l_{i}+1}}\:
    \displaystyle\frac{d\, I_{M}\, (l_{i}, l_{f}, l_{\rm ph}, l_{i}+1, \mu)}{\sin{\theta}\,d\theta} \cdot
      \Bigl\{J_{1}(l_{i},l_{f},l_{\rm ph}) - l_{i} \cdot J_{2}(l_{i},l_{f},l_{\rm ph}) \Bigr\},
\end{array}
\label{eq.2.4.11.1}
\end{equation}
\begin{equation}
\begin{array}{lcl}
\vspace{1mm}
  \displaystyle\frac{d\, p_{l_{\rm ph}\mu}^{E}}{\sin{\theta}\,d\theta} & = &
    \sqrt{\displaystyle\frac{l_{i}\,(l_{\rm ph}+1)}{(2l_{i}+1)(2l_{\rm ph}+1)}} \cdot
    \displaystyle\frac{d\, I_{E}\,(l_{i}, l_{f}, l_{\rm ph}, l_{i}-1, l_{\rm ph}-1, \mu)}
      {\sin{\theta}\,d\theta} \cdot
      \Bigl\{
        J_{1}(l_{i},l_{f},l_{\rm ph}-1)\; +
        (l_{i}+1) \cdot J_{2}(l_{i},l_{f},l_{\rm ph}-1)
      \Bigr\}\; - \\
\vspace{1mm}
    & - &
    \sqrt{\displaystyle\frac{l_{i}\,l_{\rm ph}}{(2l_{i}+1)(2l_{\rm ph}+1)}} \cdot
    \displaystyle\frac{d\, I_{E}\,(l_{i}, l_{f}, l_{\rm ph}, l_{i}-1, l_{\rm ph}+1, \mu)}
      {\sin{\theta}\,d\theta} \cdot
      \Bigl\{
        J_{1}(l_{i},l_{f},l_{\rm ph}+1)\; +
        (l_{i}+1) \cdot J_{2}(l_{i},l_{f},l_{\rm ph}+1)
      \Bigr\}\; + \\
\vspace{1mm}
  & + &
    \sqrt{\displaystyle\frac{(l_{i}+1)(l_{\rm ph}+1)}{(2l_{i}+1)(2l_{\rm ph}+1)}} \cdot
    \displaystyle\frac{d\, I_{E}\,(l_{i}, l_{f}, l_{\rm ph}, l_{i}+1, l_{\rm ph}-1, \mu)}
      {\sin{\theta}\,d\theta} \cdot
      \Bigl\{
        J_{1}(l_{i},l_{f},l_{\rm ph}-1)\; -
        l_{i} \cdot J_{2}(l_{i},l_{f},l_{\rm ph}-1)
      \Bigr\}\; - \\
  & - &
    \sqrt{\displaystyle\frac{(l_{i}+1)\,l_{\rm ph}}{(2l_{i}+1)(2l_{\rm ph}+1)}} \cdot
    \displaystyle\frac{d\, I_{E}\,(l_{i}, l_{f}, l_{\rm ph}, l_{i}+1, l_{\rm ph}+1, \mu)}
      {\sin{\theta}\,d\theta} \cdot
      \Bigl\{
        J_{1}(l_{i},l_{f},l_{\rm ph}+1)\; - l_{i} \cdot J_{2}(l_{i},l_{f},l_{\rm ph}+1)
      \Bigr\},
\end{array}
\label{eq.2.4.11.2}
\end{equation}
and
\begin{equation}
\begin{array}{lcl}
\vspace{2mm}
  \displaystyle\frac{d\, \tilde{p}_{l_{\rm ph}\mu}^{M}} {\sin{\theta}\,d\theta} & = &
      \displaystyle\frac{d\, \tilde{I}\,(l_{i},l_{f},l_{\rm ph}, l_{\rm ph}, \mu)} {\sin{\theta}\,d\theta} \cdot
      \tilde{J}\, (l_{i},l_{f},l_{\rm ph}), \\
  \displaystyle\frac{d\, \tilde{p}_{l_{\rm ph}\mu}^{E}} {\sin{\theta}\,d\theta} & = &
    \sqrt{\displaystyle\frac{l_{\rm ph}+1}{2l_{\rm ph}+1}}
      \displaystyle\frac{d\, \tilde{I}\,(l_{i},l_{f},l_{\rm ph},l_{\rm ph}-1,\mu)} {\sin{\theta}\,d\theta} \cdot
      \tilde{J}\,(l_{i},l_{f},l_{\rm ph}-1) -
    \sqrt{\displaystyle\frac{l_{\rm ph}}{2l_{\rm ph}+1}}
      \displaystyle\frac{d\, \tilde{I}\,(l_{i},l_{f},l_{\rm ph},l_{\rm ph}+1,\mu)} {\sin{\theta}\,d\theta} \cdot
      \tilde{J}\,(l_{i},l_{f},l_{\rm ph}+1).
\end{array}
\label{eq.2.4.11.3}
\end{equation}
One can see that integration of such a defined functions over the $\theta$ angle inside region from 0 to $\pi$ gives the full matrix elements  $p_{l_{\rm ph}\mu}^{M}$ and $p_{l_{\rm ph}\mu}^{E}$ defined by eq.~(\ref{eq.2.4.6.4}),
and matrix elements $\tilde{p}_{l_{\rm ph}\mu}^{M}$ and  $\tilde{p}_{l_{\rm ph}\mu}^{E}$ defined by eq.~(\ref{eq.2.4.6.7}).

\subsection{Angular probability of emission of photon with impulse $\mathbf{k}_{\rm ph}$ and polarization $\mathbf{e}^{(\alpha)}$
\label{sec.2.6}}

I define the probability of transition of the system for time unit from the initial $i$-state into the final $f$-states, being in the given interval $d \nu_{f}$, with emission of photon with possible impulses inside the given interval $d \nu_{\rm ph}$, so
(see Ref.~\cite{Landau.v3.1989}, (42,5) \S~42, p.~189; Ref.~\cite{Berestetsky.1989}, \S~44, p.~191):
\begin{equation}
\begin{array}{l}
  \vspace{2mm}
  d W = \displaystyle\frac{|a_{fi}|^{2}}{T} \cdot d\nu =
    2\pi \:|F_{fi}|^{2} \: \delta (w_{f} - w_{i} + w_{\rm ph}) \cdot d\nu, \\
  \begin{array}{lll}
    d \nu = d\nu_{f} \cdot d\nu_{\rm ph}, &
    d \nu_{\rm ph} = \displaystyle\frac{d^{3} k_{\rm ph}}{(2\pi)^{3}} =
          \displaystyle\frac{w_{\rm ph}^{2} \, dw_{\rm ph} \,d\Omega_{\rm ph}}{(2\pi c)^{3}}, &
  w_{i} - w_{f} = \displaystyle\frac{E_{i} - E_{f}}{\hbar} = w_{fi},
  \end{array}
\end{array}
\label{eq.2.6.1}
\end{equation}
where $d\nu_{\rm ph}$ and $d\nu_{f}$ are intervals defined for photon and particle in the final $f$-state,
$d\Omega_{\rm ph} = d\,\cos{\theta_{\rm ph}} = \sin{\theta_{\rm ph}} \,d\theta_{\rm ph} \,d\varphi_{\rm ph}$, $k_{\rm ph}=w_{\rm ph}/c$.
However, we have to take into account that in multipolar expansion (\ref{eq.2.4.3.1}) for the vectorial potential of the electromagnetic field we oriented the frame so that axis $z$ be directed along the vector $\mathbf{k}_{\rm ph}$. So, we have to do not use $d \Omega_{\rm ph}$ in eq.~(\ref{eq.2.6.1}).
$F_{fi}$ is integral over space with summation by quantum numbers of the system in the final $f$-state. Such procedure is averaging by these characteristics and, so, $F_{fi}$ is independent on them. Interval $d\,\nu_{f}$ has only new characteristics and quantum numbers, by which integration and summation in $F_{fi}$ was not performed.
Integrating eq.~(\ref{eq.2.6.1}) over $dw_{fi}$ and substituting eq.~(\ref{eq.2.2.3}), we find:
\begin{equation}
\begin{array}{ll}
  d W = \displaystyle\frac{Z_{eff}^{2} \,e^{2}}{m^{2}}\:
        \displaystyle\frac{\hbar\, w_{\rm ph}}{2\pi \,c^{3}} \; \Bigl|p(k_{i}, k_{f})\Bigr|^{2}\; dw_{\rm ph}.
\end{array}
\label{eq.2.6.2}
\end{equation}
This is the probability of the photon emission with impulse $\mathbf{k}_{\rm ph}$ (and with averaging by polarization $\mathbf{e}^{(\alpha)}$) where the integration over angles of the particle motion after the photon emission has already fulfilled.

I define the following probability of emission of photon with momentum $\mathbf{k}_{\rm ph}$ when after such emission the particle moves (or tunnels) along direction $\mathbf{n}_{\rm r}^{f}$: \emph{differential probability concerning angle $\theta$ is such a function, definite integral of which over the angle $\theta$ with limits from 0 to $\pi$ equals to the total probability of the photon emission (\ref{eq.2.6.2})}.
Let us consider function:
\begin{equation}
\begin{array}{ccl}
  \displaystyle\frac{d^{2} W(\theta_{f})} {dw_{\rm ph}\; d\cos{\theta_{f}}} & = &
  \displaystyle\frac{Z_{\rm eff}^{2}\, \hbar\, e^{2}}{2\pi\, c^{3}}\: \displaystyle\frac{w_{\rm ph}}{m^{2}} \;
    \biggl\{p\,(k_{i},k_{f}) \displaystyle\frac{d\, p^{*}(k_{i},k_{f}, \theta_{f})}{d\cos{\theta_{f}}} + {\rm h. e.} \biggr\}.
\end{array}
\label{eq.2.6.3}
\end{equation}
This probability is inversely proportional to normalized volume $V$. With a purpose to have the probability independent on $V$, I divide eq.~(\ref{eq.2.6.3}) on flux $j$ of outgoing $\alpha$-particles, which is inversely proportional to this volume $V$ also. Using quantum field theory approach
(where $v(\mathbf{p}) = |\mathbf{p}| / p_{0}$ at $c=1$, see~\cite{Bogoliubov.1980}, \S~21.4, p.~174):
\begin{equation}
\begin{array}{cc}
  j = n_{i}\, v(\mathbf{p}_{i}), &
  v_{i} = |\mathbf{v}_{i}| = \displaystyle\frac{c^{2}\,|\mathbf{p}_{i}|} {E_{i}} =
          \displaystyle\frac{\hbar\,c^{2}\,k_{i}} {E_{i}},
\end{array}
\label{eq.2.6.4}
\end{equation}
where $n_{i}$ is average number of particles in time unit before photon emission (we have $n_{i}=1$ for the normalized wave function in the initial $i$-state), $v(\mathbf{p}_{i})$ is module of velocity of outgoing particle in the frame system where colliding center is not moved, I obtain the \emph{differential absolute probability}
(let us name $dW$ as the \emph{relative probability}):
\begin{equation}
\begin{array}{ccl}
  \vspace{2mm}
  \displaystyle\frac{d\,P (\varphi_{f}, \theta_{f})}{dw_{\rm ph}} & = &
  \displaystyle\frac{d^{2}\,W (\varphi_{f}, \theta_{f})}{dw_{\rm ph}\; d\cos{\theta_{f}}} \cdot
    \displaystyle\frac{E_{i}} {\hbar\, c^{2}\, k_{i}} = 
    \displaystyle\frac{Z_{eff}^{2} \,e^{2}}{2\pi\,c^{5}}\:
      \displaystyle\frac{w_{\rm ph}\,E_{i}}{m^{2}\,k_{i}} \;
      \biggl\{p\,(k_{i},k_{f}) \displaystyle\frac{d\, p^{*}(k_{i},k_{f}, \Omega_{f})}{d\,\cos{\theta_{f}}} + {\rm h. e.} \biggr\}.
\end{array}
\label{eq.2.6.5}
\end{equation}
Note that alternative theoretical way for calculations of the angular bremsstrahlung probabilities in $\alpha$-decays was developed in \cite{Jentschura.2008.PRC} based on different definition of the angular probability, different connection of the matrix element with the angle $\theta$ between fragment and photon emitted, application of some approximations.

Using formula (\ref{eq.2.2.4}), we rewrite eq.~(\ref{eq.2.6.5}) as
\begin{equation}
\begin{array}{ccl}
  \vspace{2mm}
  \displaystyle\frac{d\,P (\varphi_{f}, \theta_{f})}{dw_{\rm ph}} & = &
  \displaystyle\frac{d\,P_{\rm el} (\varphi_{f}, \theta_{f})}{dw_{\rm ph}} +
  \displaystyle\frac{d\,P_{\rm mag,1} (\varphi_{f}, \theta_{f})}{dw_{\rm ph}} +
  \displaystyle\frac{d\,P_{\rm mag,2} (\varphi_{f}, \theta_{f})}{dw_{\rm ph}} +
  \displaystyle\frac{d\,P_{\rm interference} (\varphi_{f}, \theta_{f})}{dw_{\rm ph}},
\end{array}
\label{eq.2.6.6}
\end{equation}
where
\begin{equation}
\begin{array}{ccl}
  \vspace{2mm}
  \displaystyle\frac{d\,P_{\rm el} (\varphi_{f}, \theta_{f})}{dw_{\rm ph}} & = &
    \displaystyle\frac{Z_{eff}^{2} \,e^{2}}{2\pi\,c^{5}}\:
      \displaystyle\frac{w_{\rm ph}\,E_{i}}{m^{2}\,k_{i}} \;
      \biggl\{p_{\rm el}\,(k_{i},k_{f})
      \displaystyle\frac{d\, p_{\rm el}^{*}(k_{i},k_{f}, \Omega_{f})}{d\,\cos{\theta_{f}}} + {\rm h. e.} \biggr\}, \\

  \vspace{2mm}
  \displaystyle\frac{d\,P_{\rm mag,1} (\varphi_{f}, \theta_{f})}{dw_{\rm ph}} & = &
    \displaystyle\frac{Z_{eff}^{2} \,e^{2}}{2\pi\,c^{5}}\:
      \displaystyle\frac{w_{\rm ph}\,E_{i}}{m^{2}\,k_{i}} \;
      \biggl\{p_{\rm mag,1}\,(k_{i},k_{f})
      \displaystyle\frac{d\, p_{\rm mag,1}^{*}(k_{i},k_{f}, \Omega_{f})}{d\,\cos{\theta_{f}}} + {\rm h. e.} \biggr\}, \\

  \vspace{2mm}
  \displaystyle\frac{d\,P_{\rm mag,2} (\varphi_{f}, \theta_{f})}{dw_{\rm ph}} & = &
    \displaystyle\frac{Z_{eff}^{2} \,e^{2}}{2\pi\,c^{5}}\:
      \displaystyle\frac{w_{\rm ph}\,E_{i}}{m^{2}\,k_{i}} \;
      \biggl\{p_{\rm mag,2}\,(k_{i},k_{f})
      \displaystyle\frac{d\, p_{\rm mag,2}^{*}(k_{i},k_{f}, \Omega_{f})}{d\,\cos{\theta_{f}}} + {\rm h. e.} \biggr\}, \\

  \vspace{2mm}
  \displaystyle\frac{d\,P_{\rm interference} (\varphi_{f}, \theta_{f})}{dw_{\rm ph}} & = &
    \displaystyle\frac{Z_{eff}^{2} \,e^{2}}{2\pi\,c^{5}}\:
      \displaystyle\frac{w_{\rm ph}\,E_{i}}{m^{2}\,k_{i}} \;
      \biggl\{
        p_{\rm el}\,(k_{i},k_{f})
        \displaystyle\frac{d\,(
          p_{\rm mag,1}^{*}(k_{i},k_{f},\Omega_{f}) +
          p_{\rm mag,2}^{*}(k_{i},k_{f}, \Omega_{f}))}{d\,\cos{\theta_{f}}} + \\

  & + & \quad
        p_{\rm mag,1}\,(k_{i},k_{f})
        \displaystyle\frac{d\,(
          p_{\rm el}^{*}(k_{i},k_{f},\Omega_{f}) +
          p_{\rm mag,2}^{*}(k_{i},k_{f}, \Omega_{f}))}{d\,\cos{\theta_{f}}} + \\
  & + & \quad
        p_{\rm mag,2}\,(k_{i},k_{f})
        \displaystyle\frac{d\,(
          p_{\rm el}^{*}(k_{i},k_{f},\Omega_{f}) +
          p_{\rm mag,1}^{*}(k_{i},k_{f}, \Omega_{f}))}{d\,\cos{\theta_{f}}} +

      {\rm h. e.} \biggr\}.
\end{array}
\label{eq.2.6.6}
\end{equation}
For clarity of further analysis, we call $d\,P_{\rm el}$ as \emph{electric component of emission} (or electric emission), $d\,P_{\rm mag,1}$ as \emph{magnetic component of emission} (or magnetic emission), $d\,P_{\rm mag,2}$ as \emph{correction of magnetic component of emission} (or correction of magnetic emission), $d\,P_{\rm interference}$ as \emph{interference component of emission}. Sometimes, we shall omit variables $\varphi_{f}$, $\theta_{f}$ in brackets of these functions.

For description of the bremsstrahlung which accompanies collisions of protons off nuclei, we shall consider
in this paper only normalized cross-section as
\begin{equation}
\begin{array}{ccl}
  \displaystyle\frac{d^{2}\,\sigma}{dw_{\rm ph}\,d\cos{\theta_{f}}} & = &
  N_{0}\, w_{\rm ph} \cdot
  \biggl\{p\,(k_{i},k_{f}) \displaystyle\frac{d\, p^{*}(k_{i},k_{f}, \Omega_{f})}{d\,\cos{\theta_{f}}} + {\rm h. e.} \biggr\},
\end{array}
\label{eq.2.6.7}
\end{equation}
where $N_{0}$ is normalization factor (determined by normalization of the calculated curve of the full bremsstrahlung spectrum on 1 point of experimental data), and in calculations of matrix elements we use boundary condition of elastic scattering for the wave function of the proton-nucleus system in the state before emission of photon.


\section{Results
\label{sec.3}}

Let us estimate the bremsstrahlung probability accompanying the proton-decay. I calculate the bremsstrahlung probability by eq.~(\ref{eq.2.6.5}). The potential of interaction between the proton and the daughter nucleus is defined in eqs.~(26)--(27) with parameters calculated by eqs.~(28)--(29) in  \cite{Maydanyuk.2011.JPG}.
The wave functions of the decaying system in the states before and after the photon emission are calculated
concerning such potential in the spherically symmetric approximation.
The boundary conditions and normalization are used in form of~(B.1)--(B.9) in  \cite{Maydanyuk.2011.JPG}.
To choose the convenient proton-emitters for calculations and analysis, one can use systematics presented in Ref.~\cite{Aberg_Nazarewicz.1997.PRC} (see Table II in the cited paper). In particular, in \cite{Maydanyuk.2011.JPG} the $^{157}{\rm Ta}$, $^{161}{\rm Re}$, $^{167}{\rm Ir}$ and $^{185}{\rm Bi}$ nuclei decaying from the $2s_{1/2}$ state (at $l_{i}=0$), the $^{109}_{53}{\rm I}_{56}$, $^{112}_{55}{\rm Cs}_{57}$ nuclei decaying from the $1d_{5/2}$ state and the $^{146}_{69}{\rm Tm}_{77}$, $^{151}_{71}{\rm Lu}_{80}$ nuclei decaying from the $0h_{11/2}$ state
(at $l_{i} \ne 0$) were selected.
In this paper I shall analyze only one nucleus $^{146}_{69}{\rm Tm}_{77}$ at $l_{i} \ne 0$ (as calculations for this nucleus are essentially more difficult than for nuclei at $l_{i}=0$), with a main emphasis to study new physical effects in frameworks of the proposed model (assuming that such studied effects should be similar for other nuclei). For the $^{146}_{69}{\rm Tm}_{77}$ nucleus we have $l_{i}=5$, $l_{f}=4$, $Q=1.140$~MeV \cite{Maydanyuk.2011.JPG}.


\subsection{Electrical, magnetic emissions and angular distributions
\label{sec.3.1}}

\begin{figure}[htbp]
\centerline{\includegraphics[width=85mm]{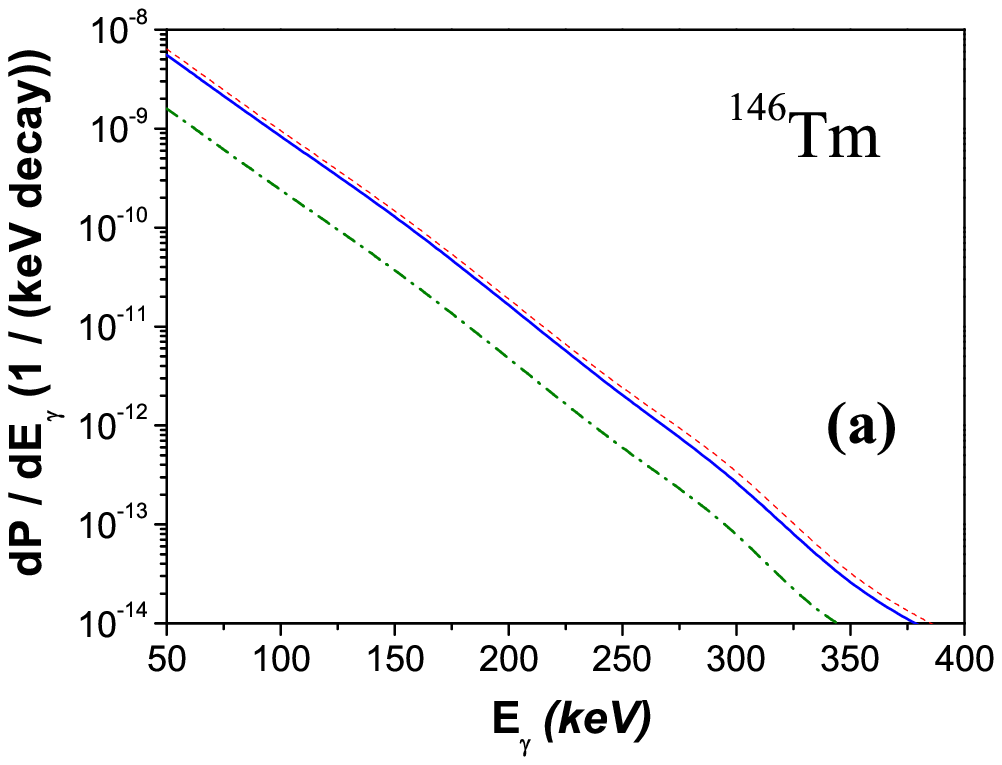}
\hspace{-1mm}\includegraphics[width=85mm]{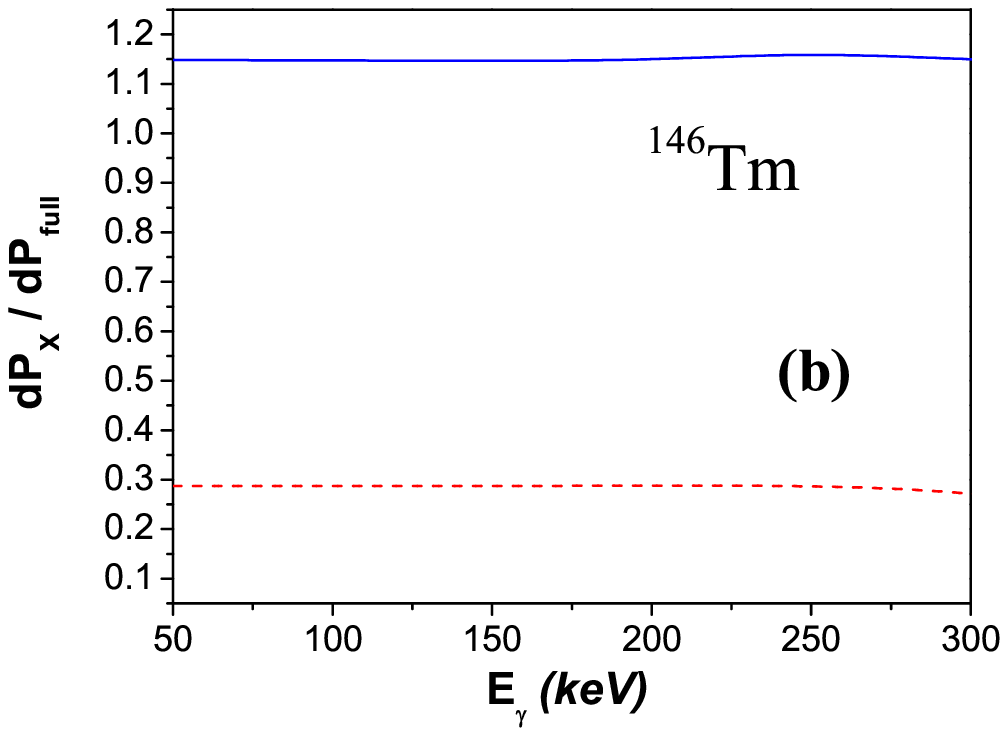}}
\vspace{2mm}
\caption{\small (Color online)
The full bremsstrahlung spectrum, electric and magnetic components of emission defined by eq.~(\ref{eq.2.6.6}) (at $\theta=90^{\circ}$):
(a) the full spectrum (full blue line), electric component $dP_{\rm el}$ (red dashed line) and magnetic component $dP_{\rm mag, 1}$ (green dash-dotted line),
(b) ratio of the components to the full spectrum (full blue line is for $dP_{\rm el} / dP_{\rm full}$, red dashed line for $dP_{\rm mag, 1} / dP_{\rm full}$).
One can see that the magnetic emission gives contribution about 28 percents inside energy region 50--300~keV.
\label{fig.1}}
\end{figure}

At first, let us clarify how much the magnetic emission is visible on the background of the full bremsstrahlung spectrum (to understand if there is a sense to study it, at all). The result of calculations of the bremsstrahlung probabilities during proton decay of $^{146}{\rm Tm}$ (at the chosen angle $\theta=90^{\circ}$ between the directions of the proton motion (with its possible tunneling) and the photon emission)
are presented in Fig.~\ref{fig.1}.
The electric and magnetic components are included also on these figures. One can see that the magnetic emission is smaller than electric one. But it gives contribution about 28 percents into the full spectrum (see Fig.~\ref{fig.1}~(b)), i.e. it is not so small to be neglected and it should be taken into account in further calculations of the bremsstrahlung spectra during nuclear decays with emission of charged fragments with non-zero spin. However, the magnetic component suppresses the full emission probability: according to Fig.~\ref{fig.1}~(b) (see blue solid line), inclusion of the magnetic component into calculations is determined by $P_{\rm el}/P_{\rm full} \simeq 1.14$, which is larger unity. This effect of suppressing of the total emission can be explained by a presence of not small destructive interference between the electric and magnetic components inside whole studied energy region. According to Fig.~\ref{fig.1}~(b), ratios of the electric and magnetic components to full spectrum are not changed in dependence on the energy of the emitted photon. As we find, the correction of the magnetic component $dP_{\rm mag, 2}$ is smaller than the electric and magnetic components by $10^{6}$ times (so we shall neglect by such a contribution in further analysis).

\begin{figure}[htbp]
\centerline{\includegraphics[width=85mm]{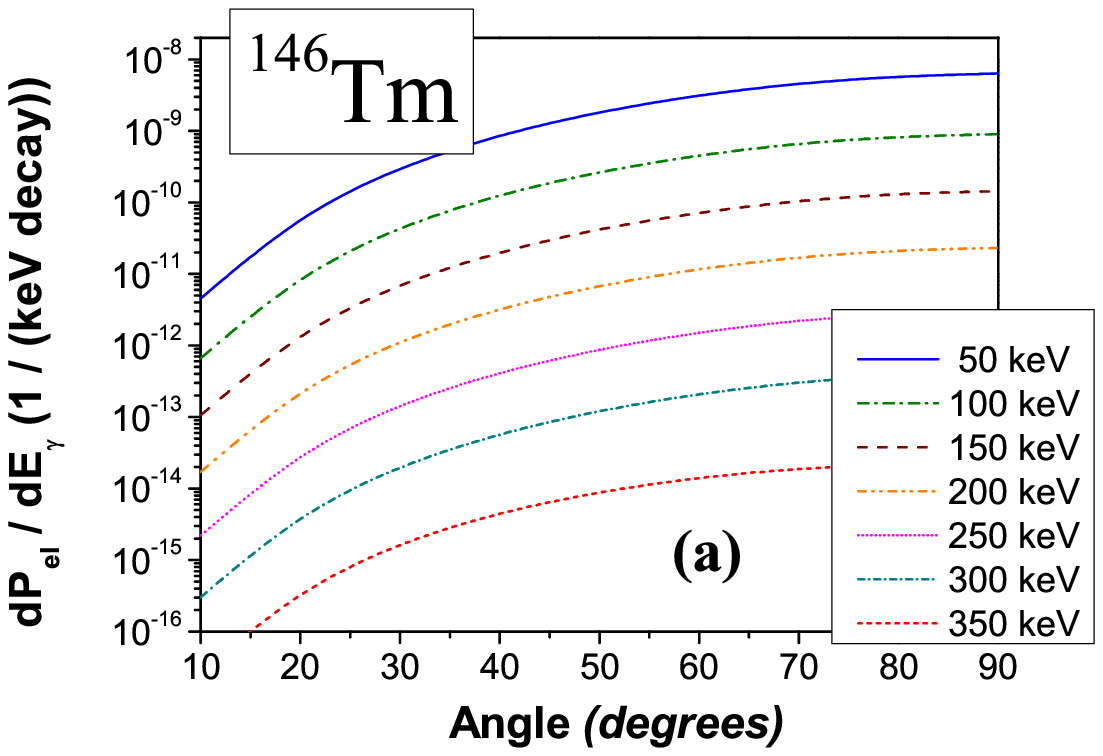}
\hspace{-1mm}\includegraphics[width=85mm]{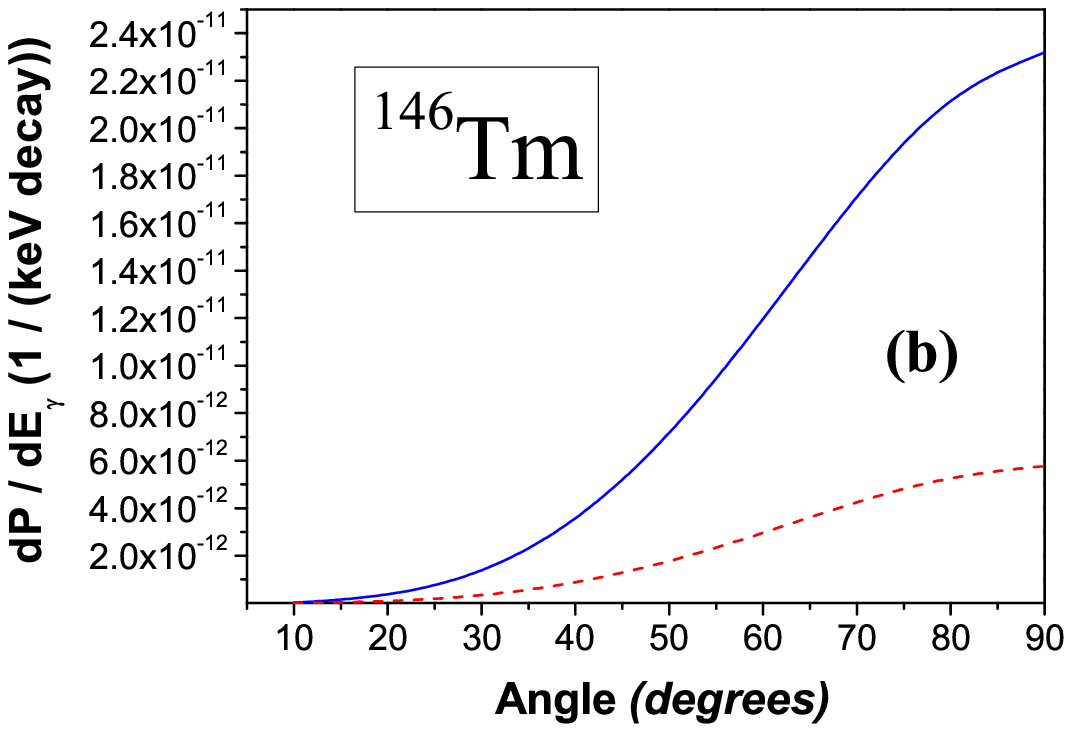}}
\vspace{2mm}
\caption{\small (Color online)
The angular distributions of the bremsstrahlung emission during proton decay of the $^{146}{\rm Tm}$ nucleus:
(a) the electric component of emission, $dP_{\rm el}$, calculated at different energies of the emitted photons;
(b) the electric component $dP_{\rm el}$ (full blue line) and magnetic component $dP_{\rm mag,1}$ (red dashed line) for the chosen photon energy 200~keV. One can see that both spectra increase proportionally (similarly) with increasing of the angle.
\label{fig.2}}
\end{figure}

Now we shall analyze how the magnetic emission is changed on the $\theta$ angle between the outgoing proton and emitted photon. In particular, let us find if there are some values of such angle, where the magnetic emission increases strongly relatively electric one. In Fig.~\ref{fig.2} the angular distributions of the electric and magnetic emissions
during the proton decay of $^{146}{\rm Tm}$ are shown.
One can see that the electric and magnetic components increase proportionally (similarly) with increasing of the $\theta$ angle. From Tabl.~\ref{table.1} it follows that there is no any angular value,
where the magnetic emission increases essentially relatively electric one.
\begin{table}
\begin{center}
\begin{tabular}{|c|c|c|c|} \hline
 & \multicolumn{3}{|c|}{Emission probability}
 \\ \cline{2-4}
  Angle & Electric & Magnetic & $dP_{\rm mag,1} / dP_{\rm el}$ \\
 $\theta$ & component $dP_{\rm el}$ & component $dP_{\rm mag,1}$ & \\ \hline
  $10^{\circ}$ & $1.704 \times 10^{-14}$ & $4.198 \times 10^{-15}$ & $0.24630$ \\
  $20^{\circ}$ & $2.580 \times 10^{-13}$ & $6.357 \times 10^{-14}$ & $0.24636$ \\
  $30^{\circ}$ & $1.192 \times 10^{-12}$ & $2.940 \times 10^{-13}$ & $0.24647$ \\
  $40^{\circ}$ & $3.329 \times 10^{-12}$ & $8.212 \times 10^{-13}$ & $0.24665$ \\
  $50^{\circ}$ & $6.952 \times 10^{-12}$ & $1.716 \times 10^{-12}$ & $0.24692$ \\
  $60^{\circ}$ & $1.188 \times 10^{-11}$ & $2.939 \times 10^{-12}$ & $0.24730$ \\
  $70^{\circ}$ & $1.727 \times 10^{-11}$ & $4.281 \times 10^{-12}$ & $0.24779$ \\
  $80^{\circ}$ & $2.158 \times 10^{-11}$ & $5.361 \times 10^{-12}$ & $0.24841$ \\
  $90^{\circ}$ & $2.319 \times 10^{-11}$ & $5.779 \times 10^{-12}$ & $0.24916$ \\
  \hline
\end{tabular}
\end{center}
\caption{Electric and magnetic components of emission in dependence on the $\theta$ angle between directions of outgoing proton and emitted photon at 200~keV of photon energy. One can see that ratio of magnetic emission on electric one is not changed practically inside whole angular region.
\label{table.1}}
\end{table}

\subsection{How are the electric and magnetic emissions changed in dependence on distance between the proton and the daughter nucleus?
\label{sec.3.2}}

Usually, authors of papers on the bremsstrahlung, which accompanies different types of collisions of particle between themselves and with nuclei, decays and fission of nuclei, calculate the spectra on the basis of integration over all space coordinates. In the relativistic models of collisions of nucleons off nucleons and nuclei (at intermediate energies) calculations are preformed in impulse representation mainly. Such approaches miss information on how much intensive emission is in dependence on distance between centers-of-masses of two studied objects. However, it is natural to think that photons are emitted with different intensity in dependence on such a distance. One can suppose that electric and magnetic photons are emitted by different ways.
We put such questions:

\vspace{-2mm}
\begin{enumerate}
\item
Can the magnetic emission be stronger than electric one in some space region?

\vspace{-2mm}
\item
How are the electric and magnetic emissions changed in dependence on distance between the proton and nucleus?

\vspace{-2mm}
\item
How much strong are the electric and magnetic emissions from the tunneling region? Is there principal difference between these types of emission from the tunneling region in comparison with the external one?
\end{enumerate}

In order to perform such an investigation, we shall define the probability of emission of the bremsstrahlung photons from the selected space region. In the presented formalism the dependence of emission on the distance is determined by the radial integrals $J_{1}(l_{i},l_{f},n)$, $J_{2}(l_{i},l_{f},n)$ and $J_{3}(l_{i},l_{f},n)$ in eq.~(\ref{eq.2.4.6.3}) and (\ref{eq.2.4.6.6}), where integration if performed over full space region. So, to obtain emission from arbitrary selected interval $r \in [r_{1}, r_{2}]$,
we shall consider the following integral:
\begin{equation}
\begin{array}{ccl}
  J_{m}(l_{i},l_{f},n; r_{1}, r_{2}) & = &
  \displaystyle\int\limits_{r_{1}}^{r_{2}} f_{m}(r)\; dr,
\end{array}
\label{eq.3.2.1}
\end{equation}
where $m=1,2,3$ and $f_{m}(r)$ is integrant function of the corresponding radial integral $J_{m}(l_{i},l_{f},n)$, defined in eq.~(\ref{eq.2.4.6.3}) or (\ref{eq.2.4.6.6}). In particular, $J_{m}(l_{i},l_{f},n; r_{1}, r_{2})$ transform to $J_{m}(l_{i},l_{f},n)$ at $r_{1} \to 0$ and $r_{2} \to +\infty$.
Now, for the emission from enough small interval $\Delta r$ near studied distance $r$ we obtain:
\begin{equation}
\begin{array}{ccl}
  J_{m}(l_{i},l_{f},n; r, r+\Delta r) & = &
  \displaystyle\int\limits_{r}^{r+\Delta r} f_{m}(r^{\prime})\; dr^{\prime}.
\end{array}
\label{eq.3.2.2}
\end{equation}
From here we define amplitude of emission in dependence on the distance $r$ on the basis of such a radial function:
\begin{equation}
\begin{array}{ccl}
  J_{m}(l_{i},l_{f},n; r) & = &
  \lim\limits_{\Delta r \to 0} \displaystyle\frac{J_{m}(l_{i},l_{f},n; r, r+\Delta r)}{\Delta r} =
  \lim\limits_{\Delta r \to 0} \displaystyle\frac{1}{\Delta r}\;
    \displaystyle\int\limits_{r}^{r+\Delta r} f_{m}(r^{\prime})\; dr^{\prime} =
  \lim\limits_{\Delta r \to 0} \displaystyle\frac{1}{\Delta r}\;
    f_{m}(r)\; \displaystyle\int\limits_{r}^{r+\Delta r} dr^{\prime} = \\
  & = &
  f_{m}(r)\; \lim\limits_{\Delta r \to 0} \displaystyle\frac{1}{\Delta r}\; \Delta r =
  f_{m}(r).
\end{array}
\label{eq.3.2.3}
\end{equation}
After this, matrix elements and probability of emission with included dependence on the distance $r$ can be defined as before, where we shall use $J_{m}(l_{i},l_{f},n; r)$ instead of the radial integrals $J_{m}(l_{i},l_{f},n)$. For denotation of new characteristics with dependence on the distance $r$ we shall include variable $r$ inside brackets.

\begin{figure}[htbp]
\centerline{\includegraphics[width=65mm]{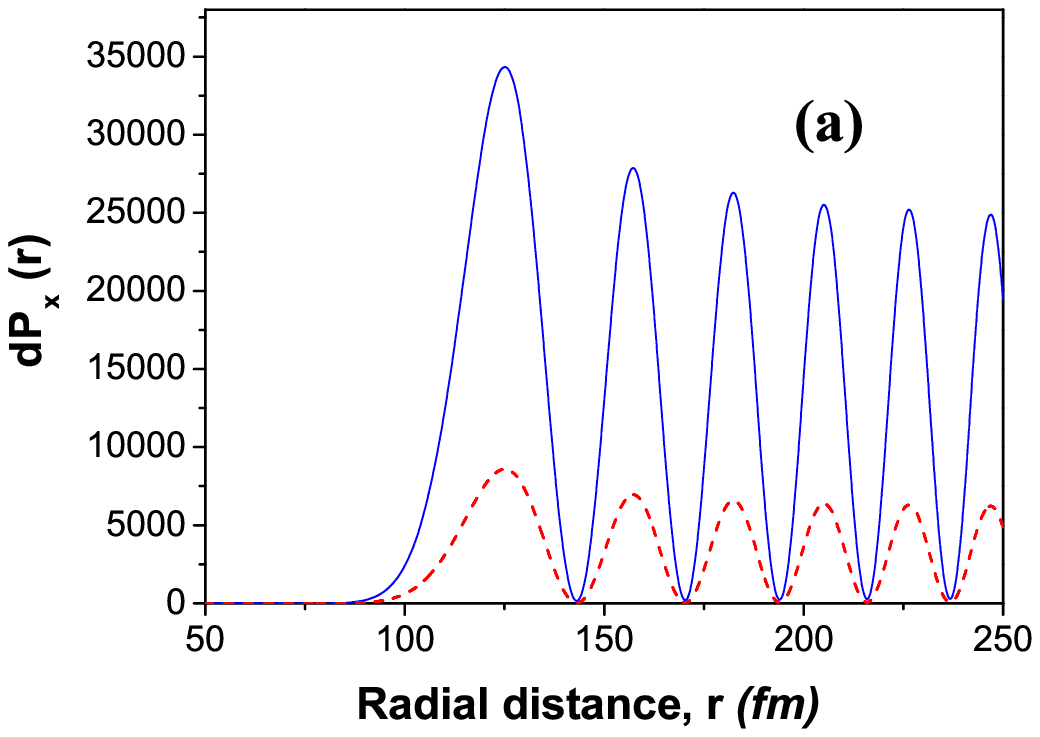}
\hspace{-5mm}\includegraphics[width=65mm]{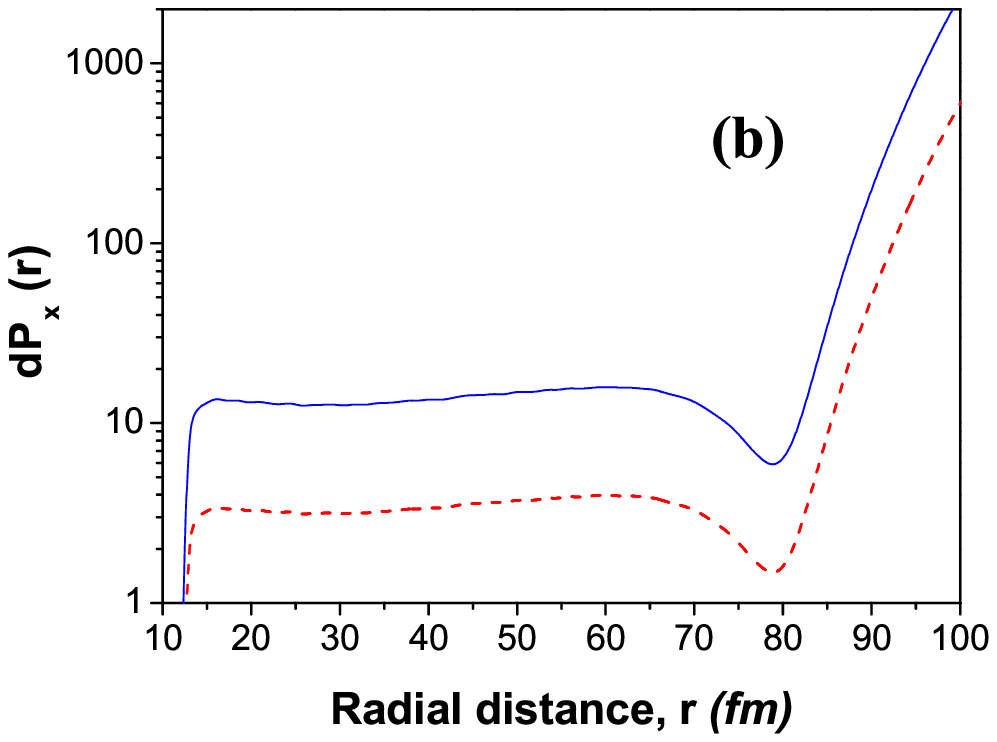}
\hspace{-5mm}\includegraphics[width=65mm]{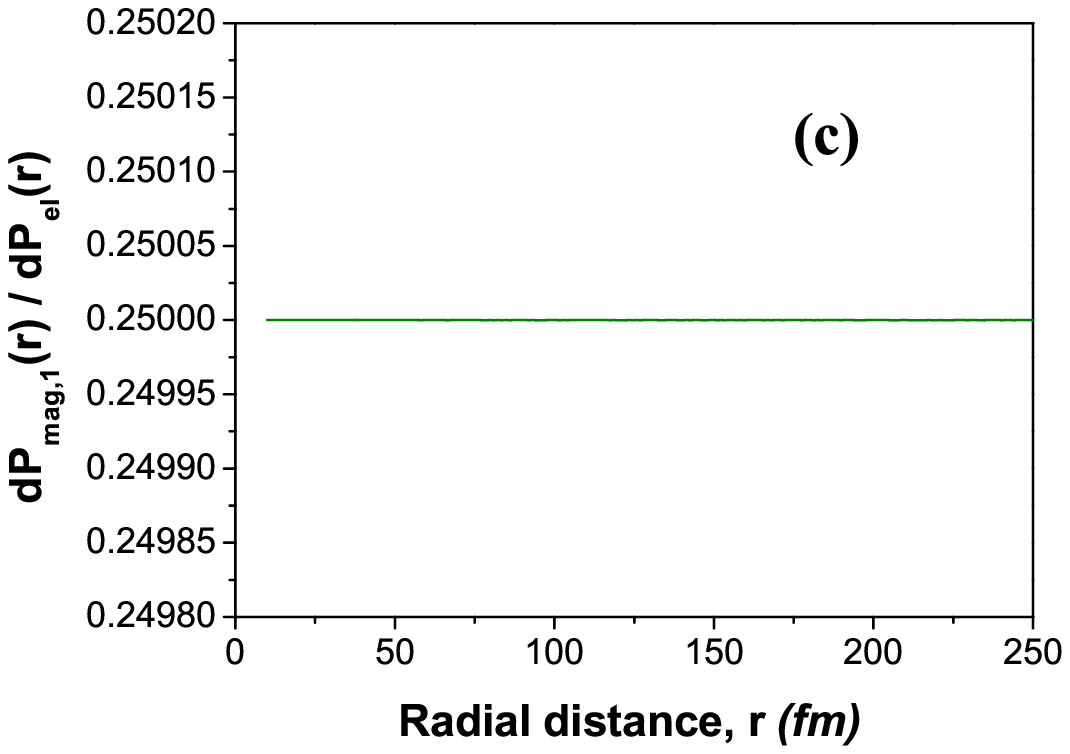}}
\vspace{-2mm}
\caption{\small (Color online)
The magnetic component $dP_{\rm mag,1}(r)$ and the electric component $dP_{\rm el}(r)$ in dependence on distance $r$ between centers-of-masses of the proton and daughter nucleus at 200 keV of the emitted photon energy (at $\theta=90^{\circ}$):
(a) the magnetic component $dP_{\rm mag,1}(r)$ (red dashed line) and the electric component $dP_{\rm el}(r)$ (full blue line) inside the space region up to 250~fm. One can see that both functions oscillate similarly in the external region outside the barrier, while they are essentially smaller inside the tunneling region;
(b) the magnetic component $dP_{\rm mag,1}(r)$ (red dashed line) and the electric component $dP_{\rm el}(r)$ (full blue line) inside the tunneling region (up to 80~fm). One can see that both functions have monotonous behavior (with possible one well, and without any oscillation) in this region. After crossing from the barrier region into the external one the first oscillation is appeared with further peak sharply increased (this demonstrates more intensive emission from the external region in comparison on the tunneling region). One can see also that after crossing from the barrier region into the internal one (near 12~fm) strong decreasing of both functions is appeared (with oscillations) --- this points on the extremely smaller bremsstrahlung emission from the space region of nucleus;
(c) ratio of the magnetic component to the electric one, $dP_{\rm mag,1}(r) / dP_{\rm el}(r)$ (full green line). One can see that this characteristic is not changed inside whole studied region of distances, it is the same in the tunneling and external regions.
\label{fig.3}}
\end{figure}

\begin{figure}[htbp]
\centerline{\includegraphics[width=65mm]{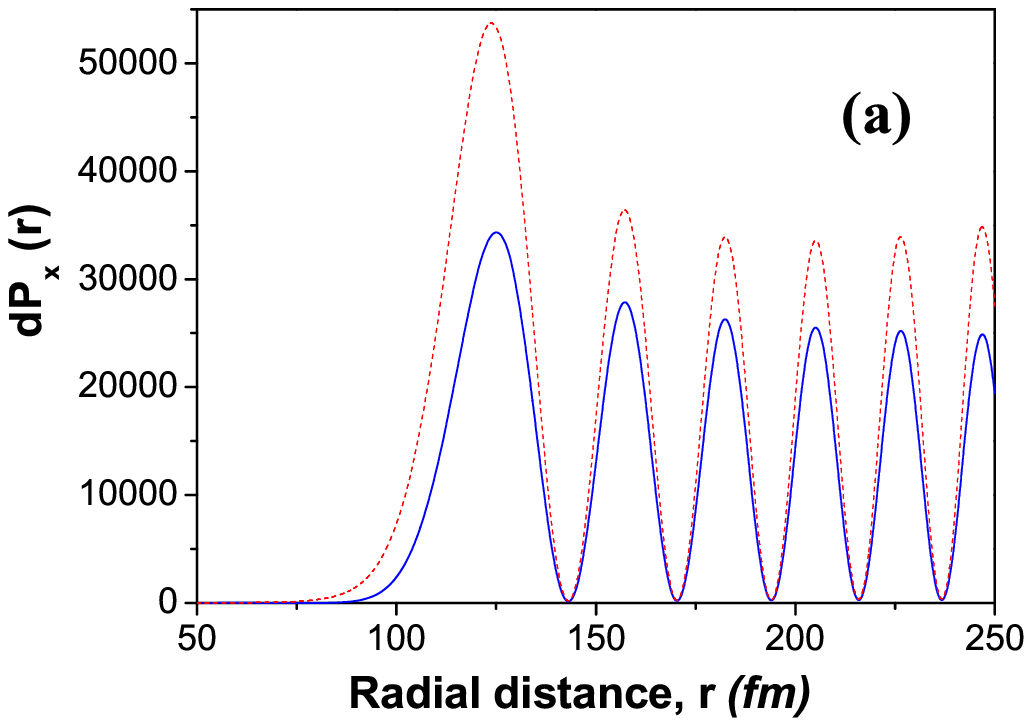}
\hspace{-5mm}\includegraphics[width=65mm]{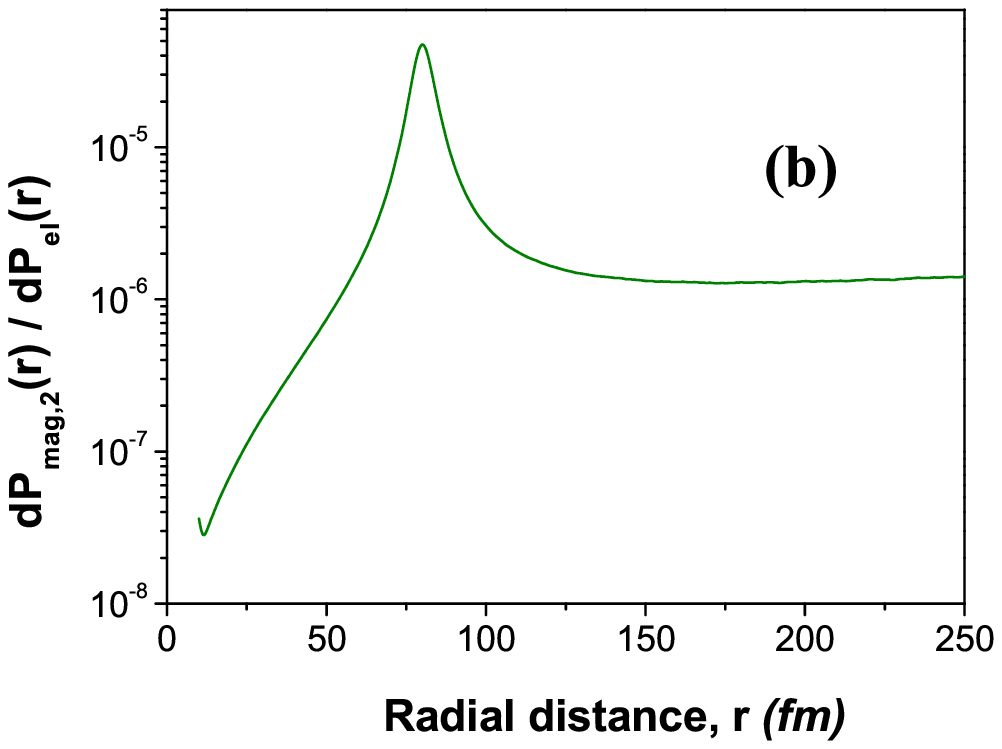}
\hspace{-5mm}\includegraphics[width=65mm]{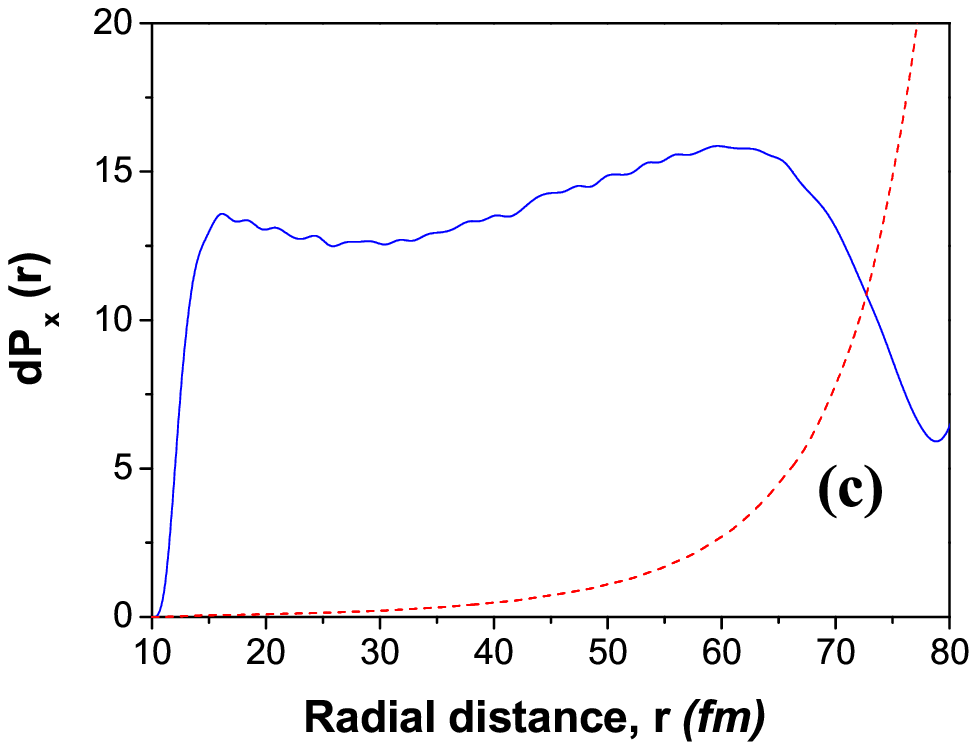}}
\vspace{-2mm}
\caption{\small (Color online)
Correction of the magnetic component of emission, $dP_{\rm mag,2}(r)$, in dependence on the distance $r$ between center-of-masses of the outgoing proton and the daughter nucleus at 200~keV of energy of the emitted photon (at $\theta=90^{\circ}$):
(a) the correction of the magnetic component $dP_{\rm mag,2}(r) \times 10^{6}$ (red dashed line) and the electric component $dP_{\rm el}(r)$ (full blue line) in region up to 250~fm. One can see that in the external region outside the barrier both functions oscillate similarly, while in the tunneling region (up to 80~fm) they are essentially smaller;
(b) ratio of the correction of the magnetic component to the electric one, $dP_{\rm mag, 2}(r) / dP_{\rm el}(r)$. One can see that there is a sharp peak close to 80~fm (that corresponds to the external turning point);
(c) the correction of the magnetic component $dP_{\rm mag, 2}(r) \times 10^{5}$ (red dashed line) and the electric component $dP_{\rm el}(r)$ (full blue line) in the tunneling region. One can see that in this region these two functions has principally different behavior. By this difference one can explain the presence of the peak in the previous figure~(b).
\label{fig.4}}
\end{figure}

The magnetic component $dP_{\rm mag,1} (r)$ on the background of the electric one $dP_{\rm el} (r)$ in dependence on the distance $r$ is shown in Fig.~\ref{fig.3}. One can see that behaviors of both functions are similar: they oscillate in the external region (having maxima and minima at similar space locations), while they have monotonous shapes with one possible well in the tunneling region. In general, the magnetic emission suppresses the full emission inside whole space region. The emission from the internal region up to the barrier is the smallest,
and from the external region --- the strongest.
Behavior of the correction of the magnetic component $dP_{\rm mag,2}(r)$ on the background of the electric one $dP_{\rm el}(r)$ in dependence on distance $r$ is shown in Fig.~\ref{fig.4}. In general, this function is essentially smaller. In the tunneling region it increases monotonously, in contrast to the electric and magnetic components (see Fig.~\ref{fig.4}~(c)). This causes a sharp peak of the function $dP_{\rm mag,2}(r) / dP_{\rm el}(r)$ close to the external boundary of the barrier (external turning point)
shown in Fig.~\ref{fig.4}~(b).
This peak could be interesting for further research, as it corresponds to the external space boundary of the barrier. But, unfortunately, this peak is extremely small (in comparison with the full spectrum) for any reasonable searches of its experimental measurements.


\subsection{Spectra of the emitted soft photons
\label{sec.3.3}}

From point of view of theory, it can be interesting to know what happens with the bremsstrahlung spectrum at limit of energy of the emitted photons to zero. In particular, let us analyze if this spectrum increases infinitely or tends to definite finite value, and which limit is in that case.

\begin{figure}[htbp]
\centerline{\includegraphics[width=65mm]{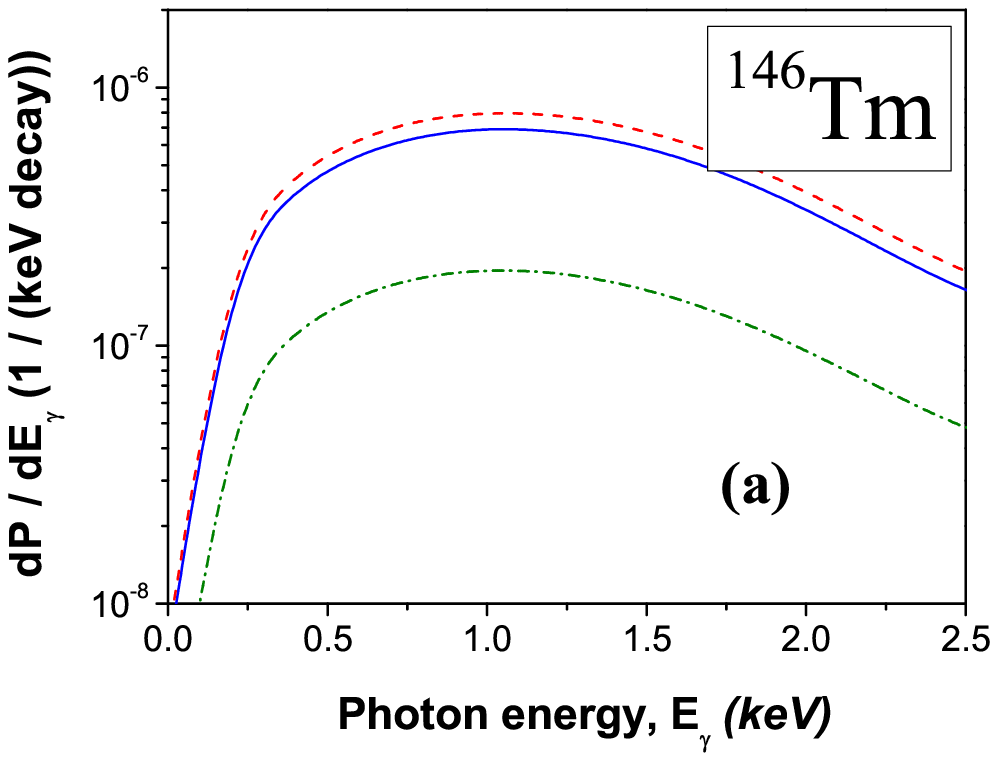}
\hspace{-5mm}\includegraphics[width=65mm]{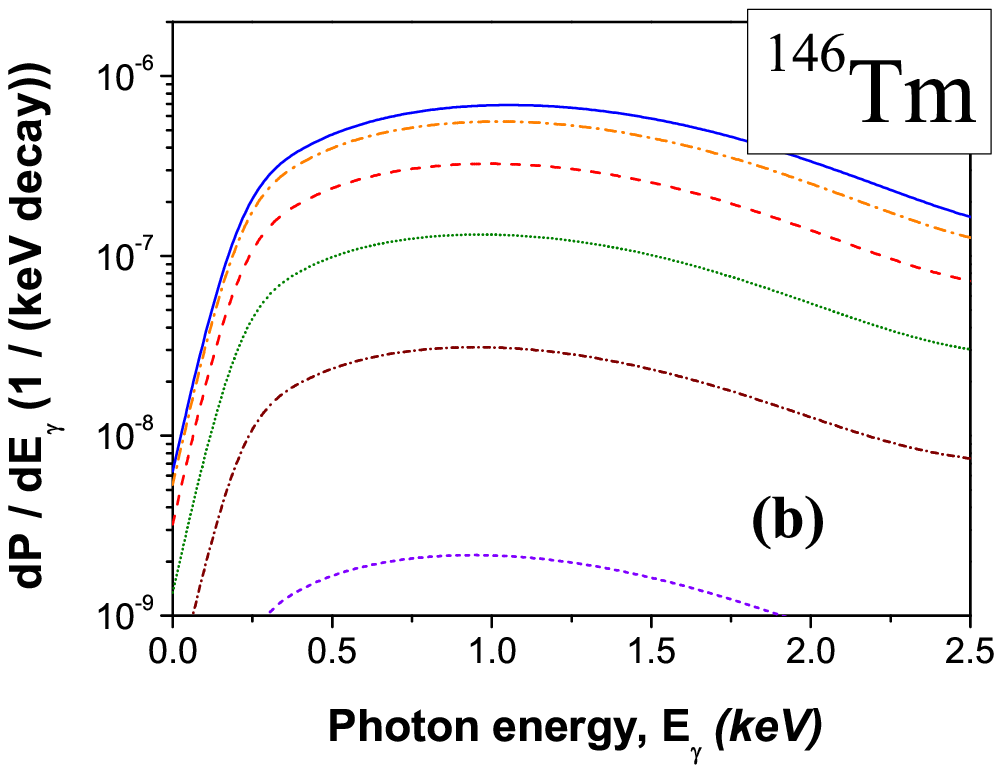}}
\vspace{-2mm}
\caption{\small (Color online)
The bremsstrahlung spectra for near-zero energy of the emitted photons (up to 2.5~keV):
(a) full spectrum (full blue line), electric component $dP_{\rm el}$ (red dashed line) and magnetic component $dP_{\rm mag,1}$ (green dash-dotted line) at $\theta=90^{\circ}$;
(b) full spectrum in dependence on the $\theta$ angle
(full blue line for для $\theta=90^{\circ}$,
orange dash-dotted line for $\theta=75^{\circ}$,
red dashed line for $\theta=60^{\circ}$,
olive short dotted line for $\theta=45^{\circ}$,
wine short dash-dotted line for $\theta=30^{\circ}$,
violent short dashed line for $\theta=15^{\circ}$).
\label{fig.5}}
\end{figure}

For low energies of photons (i.e. for soft photons) two prevailing approaches are known: the first approach is started from the early work \cite{Low.1958.PR} of Low and it is based on application of soft-photon theorem to all nuclear bremsstrahlung processes, the second one is based on application of the approximation of Feshbach and Yennie \cite{Feshbach.1962.NP}, which is more effective near resonances (see \cite{Pluiko.1987.PEPAN} for analysis). However, as it was noted in \cite{Pluiko.1987.PEPAN} (see p.~376), there is another way of development of the bremsstrahlung theory, i.e. potential one, to which our model can be referred. According to theory of QED, divergence in calculation of the matrix element is appeared at limit of photon energy to zero (infrared catastrophe, see p.~258--273 in~\cite{Ahiezer.1981}; p.~194--200 in~\cite{Nelipa.1977}; p.~194, 225, 231 in~\cite{Bogoliubov.1980}). However, we obtain the convergent integrals and the finite probability of the bremsstrahlung emission in our approach.
In particular, let us consider the first integral in eqs.~(\ref{eq.2.4.6.3}) for $n=0$ at limit $w_{\rm ph} \to 0$:
\begin{equation}
\begin{array}{ccl}
  \lim\limits_{w_{\rm ph} \to 0}
  J_{1}(l_{i},l_{f},n=0) & = &
  \lim\limits_{w_{\rm ph} \to 0}
  \hspace{-4mm}
  \displaystyle\int\limits^{R_{0}=1/k_{\rm ph}}_{0}
    \hspace{-3mm}
    \displaystyle\frac{dR_{i}(r, l_{i})}{dr}\: R^{*}_{f}(l_{f},r)\,
    j_{0}(k_{\rm ph}r)\; r^{2} dr +
  \lim\limits_{w_{\rm ph} \to 0}
  \hspace{-4mm}
  \displaystyle\int\limits^{+\infty}_{R_{0}=1/k_{\rm ph}}
    \hspace{-3mm}
    \displaystyle\frac{dR_{i}(r, l_{i})}{dr}\: R^{*}_{f}(l_{f},r)\,
    j_{0}(k_{\rm ph}r)\; r^{2} dr.
\end{array}
\label{eq.3.3.1}
\end{equation}
At $w_{\rm ph} \to 0$ we have $j_{0}(k_{\rm ph}r) = \sin(k_{\rm ph}r)/(k_{\rm ph}r) \to 1$ ($k_{\rm ph}=w_{\rm ph}/c$). So, one can see that the first item converges (according to the chosen boundary conditions, $\chi_{f}(r) = 0$ at $r = 0$, where $R_{f}(r) = \chi_{f}(r)/r$ \cite{Maydanyuk.2011.JPG}). The second item does not include small energies of photon ($k_{\rm ph}>1/R_{0}$) and, therefore, it is standard integral in our calculations of the spectra of not-soft photons, i.e. it converges also. The same result can be obtained at arbitrary chosen $n$ and for $J_{2}(l_{i},l_{f},n)$, $\tilde{J}\,(l_{i},l_{f},n)$. On such a basis, according to eqs.~(\ref{eq.2.4.4.8}), (\ref{eq.2.4.6.4}) and (\ref{eq.2.4.6.7}), all matrix elements $p_{\rm el}$, $p_{\rm mag,1}$ and $p_{\rm mag, 2}$ (and the angular matrix elements) converge at arbitrary values of quantum numbers $l_{i}$, $l_{f}$.
According to eq.~(\ref{eq.2.6.5}), we obtain:
\begin{equation}
\begin{array}{lcl}
  \lim\limits_{w_{\rm ph} \to 0}
    \displaystyle\frac{d\,P (\varphi_{f}, \theta_{f})}{dw_{\rm ph}} & = &
  \lim\limits_{w_{\rm ph} \to 0}
    \displaystyle\frac{Z_{eff}^{2} \,e^{2}}{2\pi\,c^{5}}\:
      \displaystyle\frac{w_{\rm ph}\,E_{i}}{m^{2}\,k_{i}} \;
      \biggl\{p\,(k_{i},k_{f}) \displaystyle\frac{d\, p^{*}(k_{i},k_{f}, \Omega_{f})}{d\,\cos{\theta_{f}}} + {\rm h. e.} \biggr\} = 0.
\end{array}
\label{eq.3.3.2}
\end{equation}
Our calculations at energy of the emitted photons up to 2.5~keV are shown in Fig.~\ref{fig.5}.
One can see that at decreasing of the photon energy the bremsstrahlung probability increases slowly up to finite maximum, and then it decreases monotonously. According to our estimations, the probability has finite maximum at energy of the emitted photon smaller 1.5~keV.
So, there is no the infrared catastrophe in our approach\footnote{It is interesting to note that such a proposed definition of probability, eq.~(\ref{eq.2.6.5}), allows to describe enough well experimental data of the bremsstrahlung emission during $\alpha$-decay without any normalization of the calculated spectra on experiment (see Fig.~1 in~\cite{Maydanyuk.2009.TONPPJ}).}.


\subsection{Spectra in collisions of protons off nuclei at intermediate energies
\label{sec.3.4}}

In finishing, I shall shortly demonstrate applicability of the proposed model and calculations for description of experimental spectra of the bremsstrahlung during collisions of protons off nuclei at intermediate incident energies of protons. I calculate the normalized cross-sections by eq.~(\ref{eq.2.6.7}), use the same form of the proton-nucleus potential and parameters (defined as for the problem of proton-decay studied above)\footnote{A key problem in obtaining the reliable bremsstrahlung spectra is difficulty to achieve stability in calculations of the matrix elements. Also I suppose that this is the main reason why a main idea of the proposed potential approach was not developed essentially for calculations of the bremsstrahlung spectra at intermediate energies which accompany different kinds of nuclear processes. In order to achieve the stability, I applied the approach presented in Appendix in~\cite{Maydanyuk.2010.PRC} inside the radial region from $R_{\rm as}$ to $R_{\rm max}$. For simplicity of analysis, I used the same values for these two parameters: $R_{\rm as} = 0.\,9 \times (R_{R}+7a_{R})$, $R_{\rm max}$ is chosen so large when the spectrum is not changed after its variations, $R_{R}$ and $a_{R}$ are potential parameters defined in eqs.~(29) of \cite{Maydanyuk.2011.JPG}.}.

\begin{figure}[htbp]
\centerline{\includegraphics[width=67mm]{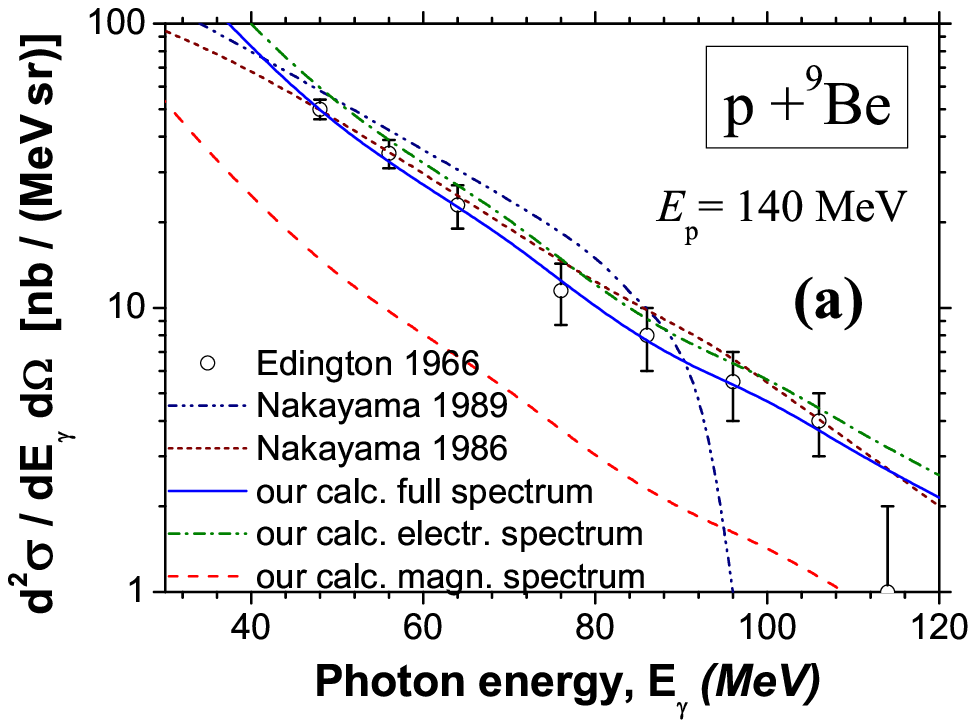}
\hspace{-8mm}\includegraphics[width=67mm]{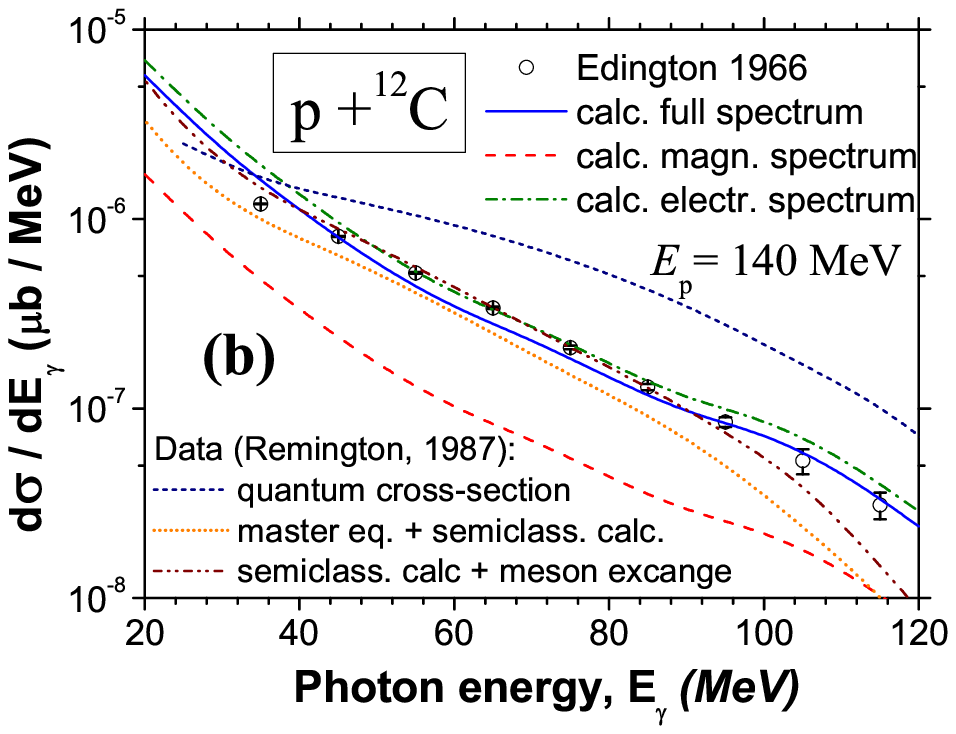}
\hspace{-8mm}\includegraphics[width=67mm]{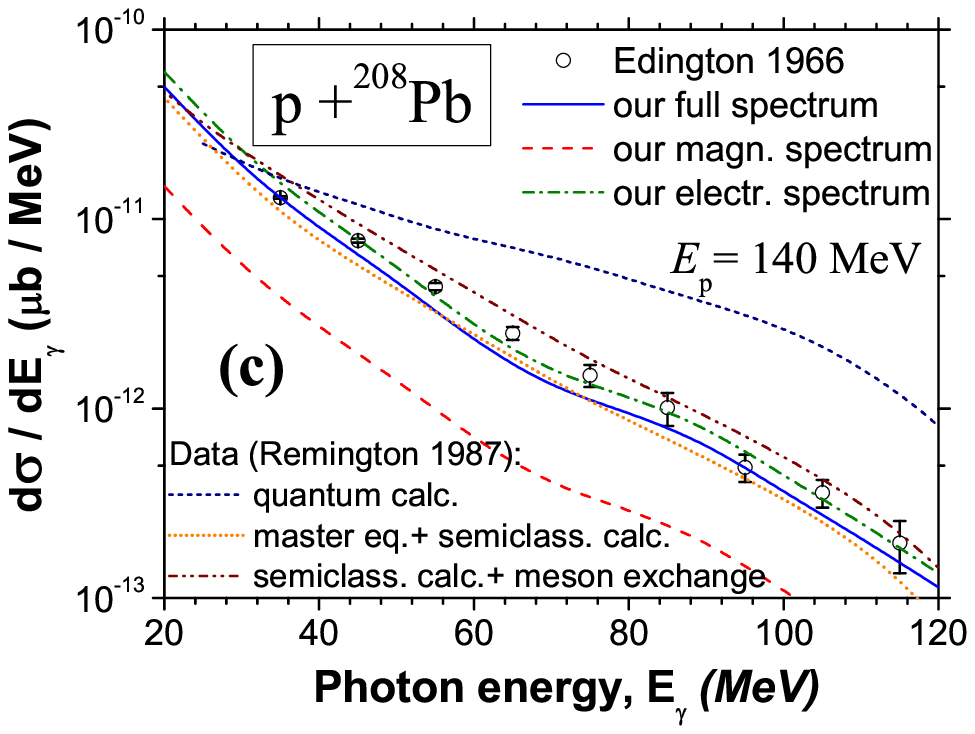}}
\vspace{-2mm}
\caption{\small (Color online)
The proton nucleus bremsstrahlung probability rates in the laboratory system at the incident energy $T_{\rm lab}=140$~MeV (in our calculations we use photon emission angle $\theta=90^{\circ}$):
(a) Comparison for $p+^{9}{\rm Be}$ between the calculations by our model (blue solid line is for full spectrum, green dash-dotted line for electric contribution, red dashed line for magnetic contribution), calculations from (Nakayama 1986: \cite{Nakayama.1986.PRC}, wine short-dashed line), calculations from (Nakayama 1989: \cite{Nakayama.1989.PRC}, navy dash double-dotted line) and experimental data (Edington 1966: \cite{Edington.1966.NP});
(b,~c) Comparison for $p+^{12}{\rm C}$ and $p+^{208}{\rm Pb}$ between the calculations by our model (blue solid line is for full spectrum, green dash-dotted line for electric contribution, red dashed line for magnetic contribution), calculations by Remington, Blann and Bertsch in (Remington 1987: \cite{Remington.1987.PRC}, wine dash double-dotted line is for calculations by master equation using the semiclassical bremsstrahlung cross sections, orange short dotted line for semiclassical cross sections multiplied by 2 for meson exchange, and navy short dashed line for quantum bremsstrahlung cross sections), and experimental data (Edington 1966: \cite{Edington.1966.NP}).
\label{fig.6}}
\end{figure}
In Fig.~\ref{fig.6}~(a) one can see that our approach can describe enough well experimental data for $p+^{9}{\rm Be}$ in energy region from 20~MeV to 120~MeV in comparison with results obtained by Nakayama and Bertsch in \cite{Nakayama.1986.PRC} and calculations performed by Nakayama in \cite{Nakayama.1989.PRC}.
In next figure~(b) we compare our calculations for $p+^{12}{\rm C}$ with experimental data~\cite{Edington.1966.NP} and results obtained by Remington, Blann and Bertsch in~\cite{Remington.1987.PRC}. Such comparison shows: in energy region of the emitted photons up to 90 MeV our full spectrum (see solid blue line) is enough close to experimental data and calculations obtained using master equation and semiclassical bremsstrahlung formula (see wine dash double-dotted line), the semiclassical cross-sections multiplied by factor 2 for meson exchange (see orange short dotted line) in~\cite{Remington.1987.PRC}. But for hard photons with energy from 90 to 120 MeV we achieve better agreement with experimental data than results of \cite{Remington.1987.PRC}. Comparison of our results with quantum calculations performed in \cite{Remington.1987.PRC} (see navy short dashed line in that figure) indicates on absolute applicability (and availability) of the quantum approach in description of the emitted photons of high energy in collisions of protons off nuclei. At the same time, such an approach allows to deeper study quantum properties (such as non-locality, for example) of the considered colliding process. In last figure~(c) similar comparison is performed for $p+^{208}{\rm Pb}$. On all figure we add our calculations for the magnetic and electric bremsstrahlung emission.

\begin{figure}[htbp]
\centerline{\includegraphics[width=67mm]{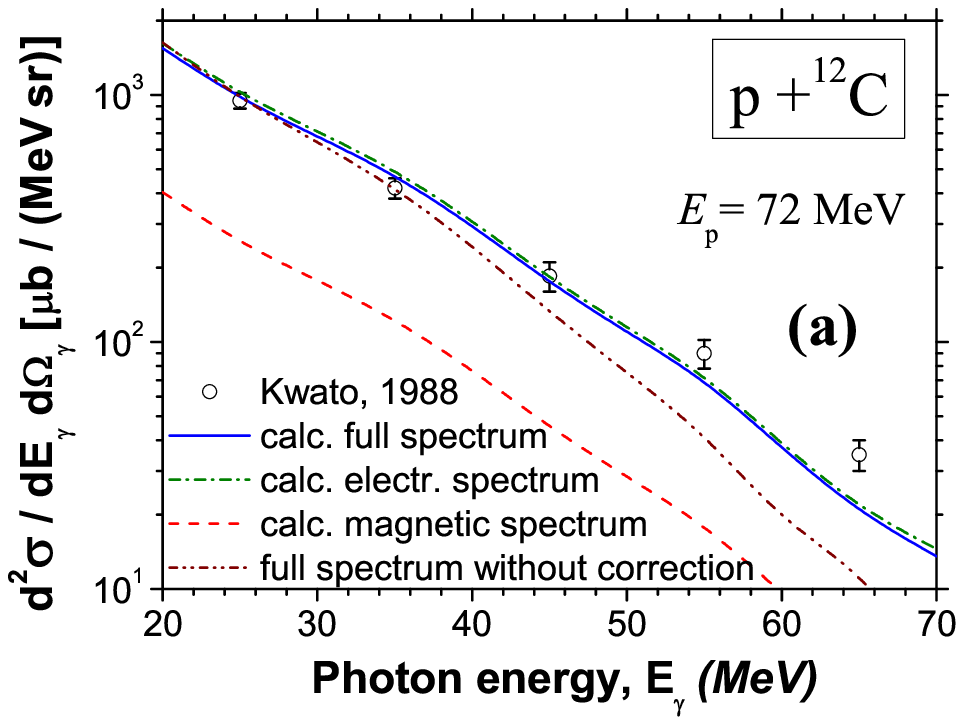}
\hspace{-8mm}\includegraphics[width=67mm]{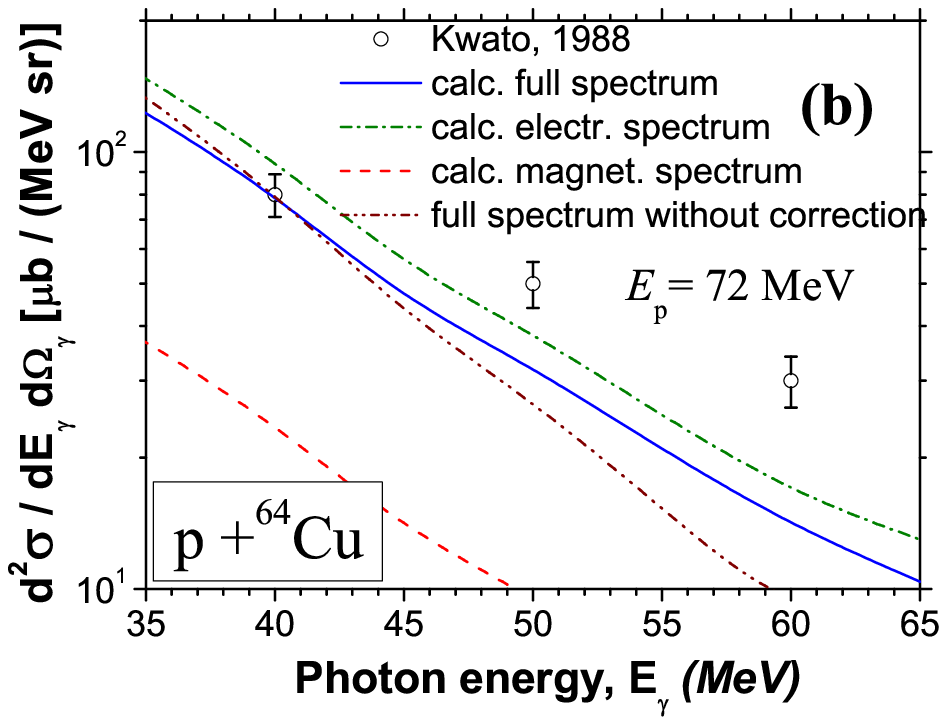}
\hspace{-8mm}\includegraphics[width=67mm]{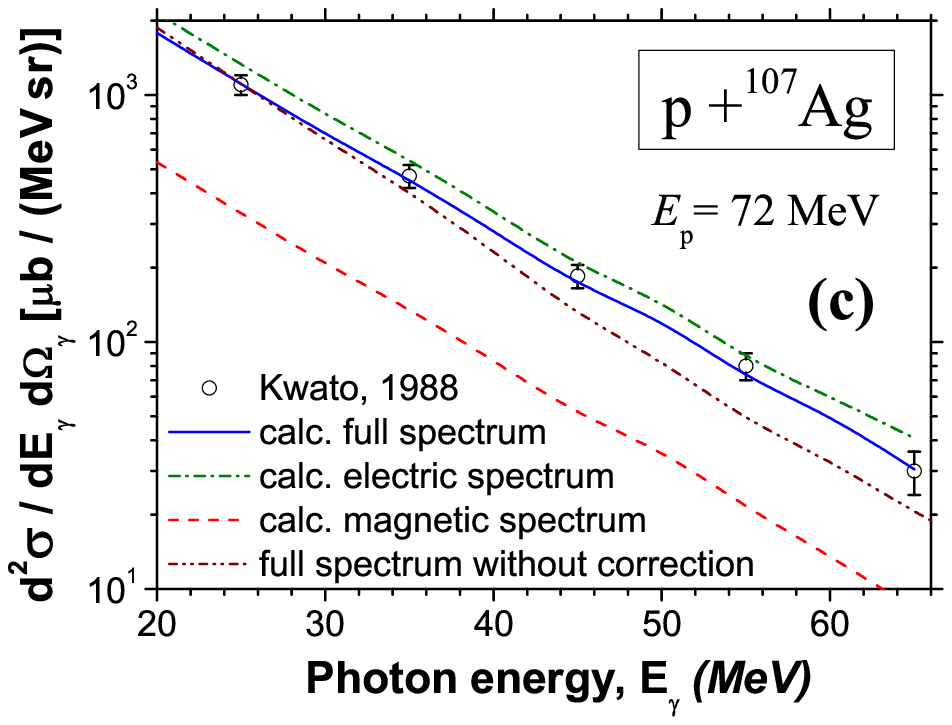}}
\vspace{-2mm}
\caption{\small (Color online)
The proton nucleus bremsstrahlung probability rate in the laboratory system at the incident energy $T_{\rm lab}=72$~MeV and photon emission angle $\theta=90^{\circ}$:
Comparison for $p+^{12}{\rm C}$ (a), $p+^{64}{\rm Cu}$ (b) and $p+^{107}{\rm Ag}$ (c) between the full cross-section calculated by eq.~(\ref{eq.2.6.7}) (wine dash double-dotted line),
the corrected cross-section obtained by eq.~(\ref{eq.2.6.7}) with division on $k_{f}$ (blue solid line) and experimental data (Kwato 1988: \cite{Kwato_Njock.1988.PLB}).
We add the electric component (green dash-doted line) and magnetic component (red dashed line) to all figures.
\label{fig.7}}
\end{figure}

In Fig.~\ref{fig.7} we present our calculations of the bremsstrahlung cross-sections for collisions $p+^{9}{\rm C}$, $p+^{64}{\rm Cu}$ and $p+^{107}{\rm Ag}$ in comparison with experimental data~\cite{Kwato_Njock.1988.PLB} at the incident proton energy $T_{\rm lab}=72$~MeV. Here, we show the full spectrum calculated by eq.~(\ref{eq.2.6.7}) and the corrected spectrum obtained by eq.~(\ref{eq.2.6.7}) with division on $k_{f}$ (according to formula (13) of cross-section defined in \cite{Kopitin.1997.YF}). Comparison with quantum calculations performed by Kopitin, Dolgopolov, Churakova and Kornev in \cite{Kopitin.1997.YF} (see Fig.~1 in the cited paper) shows more stable calculations in our approach.
In addition, this answers on assumption putted in \cite{Kopitin.1997.YF} that quantum approach (with included nuclear component of potential) is absolutely able
to describe well experimental data of the bremsstrahlung during proton-nucleus collisions.

\begin{figure}[htbp]
\centerline{\includegraphics[width=67mm]{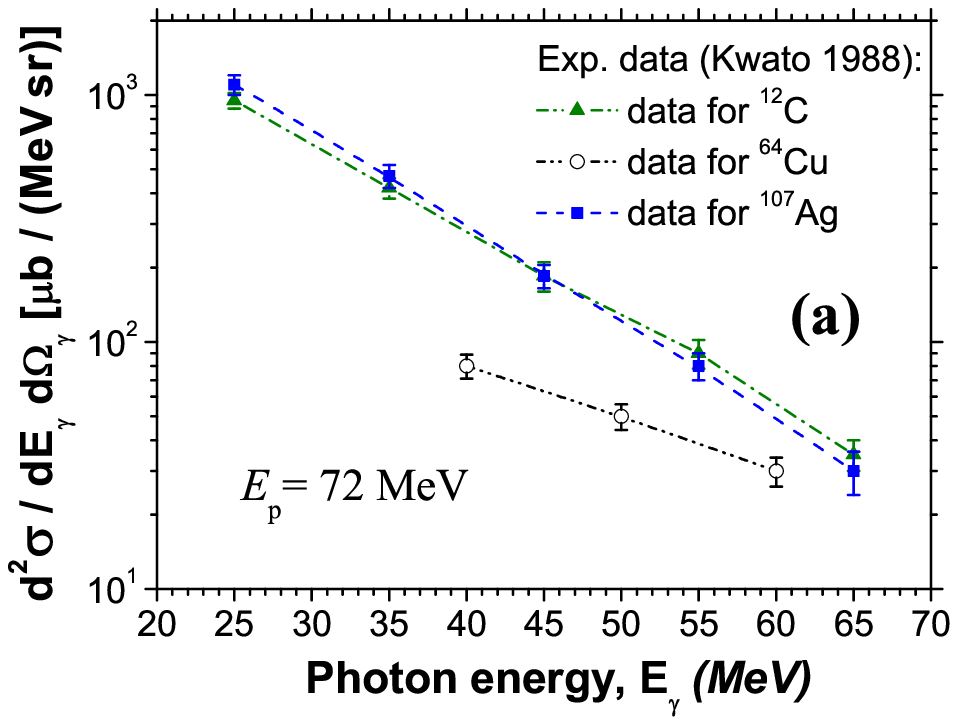}
\hspace{-8mm}\includegraphics[width=67mm]{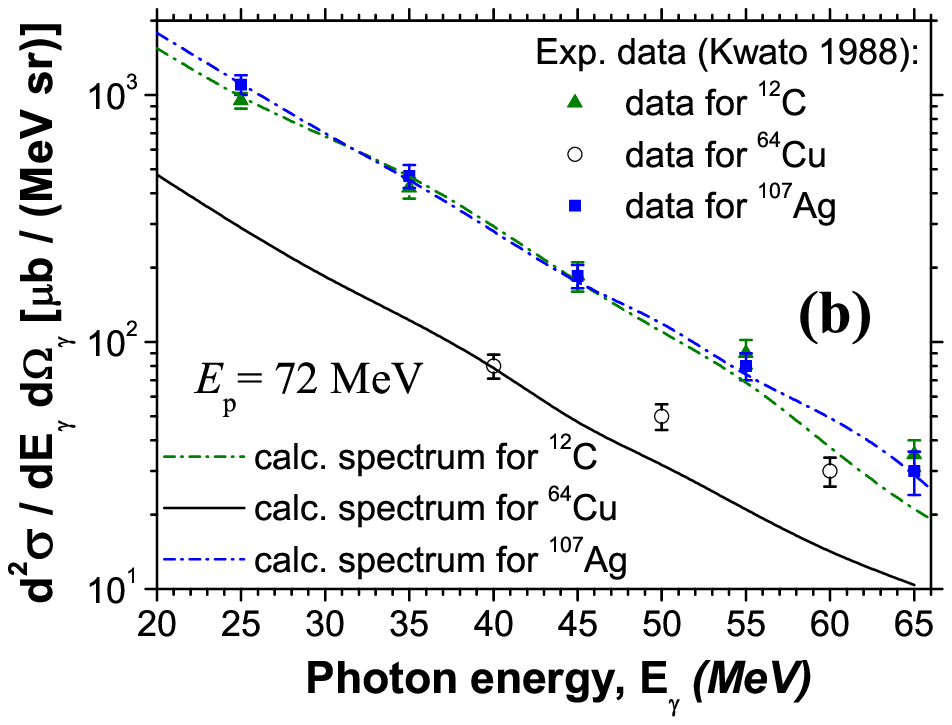}}
\vspace{-2mm}
\caption{\small (Color online)
The experimental data (Kwato 1988: \cite{Kwato_Njock.1988.PLB}) for $p+^{12}{\rm C}$, $p+^{64}{\rm Cu}$ and $p+^{107}{\rm Ag}$ at $T_{\rm lab}=72$~MeV:
(a) One can see that data for $^{64}{\rm Cu}$ are located lower than the data for $^{12}{\rm C}$ and $^{107}{\rm Ag}$. At the same time, the data for $^{64}{\rm Cu}$ are decreased more slowly with increasing of the photon energy than the data for $^{12}{\rm C}$ and $^{107}{\rm Ag}$.
(b) The comparison between experimental data reinforced by calculations of the full cross-sections (blue solid lines in Fig.~\ref{fig.7}, $\theta=90^{\circ}$): inclusion of the calculated curves, describing general tendency of the spectra, only reinforces difference in behavior between experimental data, indicated on the previous figure (a).
\label{fig.8}}
\end{figure}

Comparing results of calculations obtained for $p + ^{12}{\rm C}$, $p + ^{64}{\rm Cu}$ and $p + ^{107}{\rm Ag}$ at the incident energy $T_{\rm lab}=72$~MeV, one can find worse agreement between theory and experiment for the $^{64}{\rm Cu}$ nucleus. This situation looks to be enough strange, as all measurements were made by the same group of experimentalists (and it is difficult to expect that the cross section for the $^{64}{\rm Cu}$ nucleus is the experimental error). For evidence, let us consider all these experimental data in one figure (see Fig.~\ref{fig.8}). One can see that data for $^{64}{\rm Cu}$ are located lower than the data for $^{12}{\rm C}$ and $^{107}{\rm Ag}$. At the same time, the data for $^{64}{\rm Cu}$ are decreased more slowly with increasing of the photon energy than the data for $^{12}{\rm C}$ and $^{107}{\rm Ag}$. In particular, one can expect that further continuation of all these data for higher photon energies have to lead to their intersection at one point (or these are evident deviations from monotonous decreasing tends of the spectra), that has never been observed before.

Such a picture disagrees with early observed general tendency of the bremsstrahlung spectra in nuclear processes. This logics explains why the spectrum is decreased more strongly (with increasing of the photon energies), if this spectrum is lower. Such a tendency is based on correspondence between shape of the barrier with tunneling length of the emitted fragment: the energy of proton is lower, the length of tunneling is larger, the total emission of photons is less intensive (because of it is less intensive for photons emitted from tunneling region, than from above-barrier regions). We demonstrated this tendency on the example of two spectra in the $\alpha$-decay of the $^{214}{\rm Po}$ and $^{226}{\rm Ra}$ nuclei (see Fig.~3 and explanations in~\cite{Giardina.2008.MPLA}).

But, if this difference between experimental data is supported by future measurements, then such a result would be very interesting. This will be a direct indication on influence on the spectra of some other hidden characteristics of the proton-nucleus system, which are not included into current calculations. This will the indicate on a presence of new aspects in the bremsstrahlung spectra. One can assume that structure of the proton-nucleus system, dynamics of its nucleons,
other early not studied
properties can be important in such new developments\footnote{For example, in the problem of the bremsstrahlung emission accompanying ternary fission of $^{252}{\rm Cf}$ (where this nucleus is separated on the $\alpha$-particle and two heavy fragments) we shown that dynamics of relative motion of all participated fragments, and geometry of nuclear separation have strong influence on the bremsstrahlung spectrum~\cite{Maydanyuk.2011.JPCS}.}.


\subsection{Role of the multipolar components in the angular analysis
\label{sec.3.5}}

The first calculations of the multipolar components of the bremsstrahlung emission of higher order in tasks of nuclear decays were obtained by Tkalya in~\cite{Tkalya.1999.JETP,Tkalya.1999.PHRVA}. Studying emission of the bremsstrahlung photons during $\alpha$-decay of the $^{226}{\rm Ra}$, $^{210}{\rm Po}$ and $^{214}{\rm Po}$ nuclei, he shown that the multipolar term $E2$ is essentially smaller in comparison with $E1$ (see Fig.~1 in~\cite{Tkalya.1999.PHRVA}, ratio between contributions $P^{E1}/P^{E2}$ is about 50--1000 for the photon energy range up to 900~keV).
There are also estimations obtained by Kurgalin, Chuvilsky and Churakova for the multipolar term $E2$ of the emitted photons in $\alpha$-decay of $^{210}{\rm Po}$ \cite{Kurgalin.2004.VVGU}: according to their calculations, contribution of the $E2$ multipole is smaller than $E1$ by 50--500 times for the photon energies up to 800~keV. We studied this question also and found the multipolar terms $E2$ and $M2$ to be very small.
Authors of paper \cite{Jentschura.2008.PRC} investigated the dipole and quadrupole contributions in the semiclassical consideration to the bremsstrahlung probability in $\alpha$-decay,
studied interference between such contributions\footnote{%
Expansion in \cite{Jentschura.2008.PRC} and the multipolar expansion in the given paper have different basis and sense. In \cite{Jentschura.2008.PRC} dipole and quadrupole contributions are defined as the first term (at $l_{f}=1$) and the second term (at $l_{f}=2$) of expansion of wave function $\varphi_{f}(\mathbf{r})$ of the $\alpha$-nucleus system in the state after emission of photon (see eqs.~(B1)--(B4) in~\cite{Jentschura.2008.PRC}), at representation of the effective charge for two-charged nuclear system (see eqs.~(A1)--(A4) in~\cite{Jentschura.2008.PRC}). The multipolar approach in this paper is based on the standard multipolar expansion of the wave function of photon
(\ref{eq.2.4.3.5}).}.
I have not found any information about other attempts to estimate the $E2$ multipolar term and the multipoles of higher order, which could be obtained up to now.
By such reasons, calculations of the bremsstrahlung spectra in the multipolar approach usually are performed on the basis of the first multipolar term, which gives the prevailing contribution into the full spectrum (usually 4--5 first digits of the calculated probability are stable in our approach, as minimum).

Also it is more difficult to obtain reliable estimations of the multipolar terms of higher order because of essentially smaller convergence of their calculations. This is real practical difficulty (which can be alienated from many researchers trying to obtain the multipolar terms of higher order).
Indications on difficulty of such problems and perspective of their solution I find in papers of authors, who calculated the bremsstrahlung spectra in different nuclear tasks with realistic potentials (for example, see \cite{Pluiko.1987.PEPAN,Kamanin.1989.PEPAN,Kopitin.1997.YF}).

\vspace{1mm}
In order to understand more clearly, how the angular bremsstrahlung probability is changed in dependence on quantum numbers $l_{i}$, $l_{f}$ and $l_{\rm ph}$ (which defines the multipolar term), we rewrite formulas separating components which describe this angular dependence.
This information is completely included in the differential matrix elements:
\begin{equation}
\begin{array}{lcl}
\vspace{0mm}
  \displaystyle\frac{d\, p_{l_{\rm ph}\mu}^{M}}{\sin{\theta}\,d\theta} & = &
  \delta_{\mu, m_{i}-m_{f}}\;
  P_{l_{f}}^{|m_{f}|}\;
  \displaystyle\sum\limits_{\mu^{\prime} = \pm 1}
  \Bigl\{
    \delta_{l_{i} \ne 0}\:
    c_{1}^{\mu^{\prime}} \:
      P_{l_{i}-1}^{|m_{i} - \mu^{\prime}|}\; -
    c_{2}^{\mu^{\prime}}\:
      P_{l_{i}+1}^{|m_{i} - \mu^{\prime}|}\;
  \Bigr\} \cdot
  P_{l_{\rm ph}}^{|\mu - \mu^{\prime}|}, \\

\vspace{0mm}
  \displaystyle\frac{d\, p_{l_{\rm ph}\mu}^{E}}{\sin{\theta}\,d\theta} & = &
  \delta_{\mu, m_{i}-m_{f}}\;
  P_{l_{f}}^{|m_{f}|}\,
  \displaystyle\sum\limits_{\mu^{\prime} = \pm 1}\,
  \biggl\{
    \Bigl[
      \delta_{l_{i} \ne 0}\;
      c_{3}^{\mu^{\prime}}\:
      P_{l_{i}-1}^{|m_{i} - \mu^{\prime}|} +
    c_{5}^{\mu^{\prime}}\:
      P_{l_{i}+1}^{|m_{i} - \mu^{\prime}|}\,
      \Bigr]\:
      P_{l_{\rm ph}-1}^{|\mu - \mu^{\prime}|} - \\
\vspace{1mm}
  & - &
    \Bigl[
      \delta_{l_{i} \ne 0}\; c_{4}^{\mu^{\prime}}\:
      P_{l_{i}-1}^{|m_{i} - \mu^{\prime}|} +
      c_{6}^{\mu^{\prime}}\:
      P_{l_{i}+1}^{|m_{i} - \mu^{\prime}|}\,
    \Bigr]\:
    P_{l_{\rm ph}+1}^{|\mu - \mu^{\prime}|}
  \biggr\},
\end{array}
\label{eq.3.5.1}
\end{equation}
\begin{equation}
\begin{array}{lcllcl}
\vspace{1mm}
  \displaystyle\frac{d\, \tilde{p}_{l_{\rm ph}\mu}^{M}} {\sin{\theta}\,d\theta} & = &
  \delta_{m_{i}, m_{f}}\,
    c_{7} \cdot
    P_{l_{i}}^{|m_{i}|}\:
    P_{l_{f}}^{|m_{f}|}\:
    P_{l_{\rm ph}}^{0}, \\

  \displaystyle\frac{d\, \tilde{p}_{l_{\rm ph}\mu}^{E}} {\sin{\theta}\,d\theta} & = &
  \delta_{m_{i}, m_{f}}\,
  P_{l_{i}}^{|m_{i}|}\:
  P_{l_{f}}^{|m_{i}|}\:
  \Bigl\{
    c_{8}\; P_{l_{\rm ph}-1}^{0} -
    c_{9}\; P_{l_{\rm ph}+1}^{0}
  \Bigr\},
\end{array}
\label{eq.3.5.2}
\end{equation}
where
\begin{equation}
\begin{array}{lcl}
\vspace{2mm}
  c_{1}^{\mu^{\prime}} & = &
  \sqrt{\displaystyle\frac{l_{i}}{2l_{i}+1}}\;
  C_{l_{i} l_{f} l_{ph} l_{i}-1, l_{ph}}^{m_{i} m_{f} \mu^{\prime}} \cdot
  \Bigl[
    J_{1}(l_{i},l_{f},l_{\rm ph}) +
    (l_{i}+1) \cdot J_{2}(l_{i},l_{f},l_{\rm ph})
  \Bigr], \\

  c_{2}^{\mu^{\prime}} & = &
  \sqrt{\displaystyle\frac{l_{i}+1}{2l_{i}+1}}\;
  C_{l_{i} l_{f} l_{ph} l_{i}+1, l_{ph}}^{m_{i} m_{f} \mu^{\prime}} \cdot
  \Bigl[J_{1}(l_{i},l_{f},l_{\rm ph}) - l_{i} \cdot J_{2}(l_{i},l_{f},l_{\rm ph}) \Bigr],
\end{array}
\label{eq.3.5.3}
\end{equation}
\begin{equation}
\begin{array}{lcl}
\vspace{0mm}
  c_{3}^{\mu^{\prime}} & = &
    \sqrt{\displaystyle\frac{l_{i}\,(l_{\rm ph}+1)}{(2l_{i}+1)(2l_{\rm ph}+1)}}\;
    C_{l_{i} l_{f} l_{ph} l_{i}-1, l_{\rm ph}-1}^{m_{i} m_{f} \mu^{\prime}} \cdot
    \Bigl[ J_{1}(l_{i},l_{f},l_{\rm ph}-1)\; + (l_{i}+1) \cdot J_{2}(l_{i},l_{f},l_{\rm ph}-1) \Bigr], \\

\vspace{0mm}
  c_{4}^{\mu^{\prime}} & = &
    \sqrt{\displaystyle\frac{l_{i}\,l_{\rm ph}}{(2l_{i}+1)(2l_{\rm ph}+1)}}\;
    C_{l_{i} l_{f} l_{ph} l_{i}-1, l_{\rm ph}+1}^{m_{i} m_{f} \mu^{\prime}} \cdot
    \Bigl[ J_{1}(l_{i},l_{f},l_{\rm ph}+1)\; + (l_{i}+1) \cdot J_{2}(l_{i},l_{f},l_{\rm ph}+1) \Bigr], \\

\vspace{0mm}
  c_{5}^{\mu^{\prime}} & = &
    \sqrt{\displaystyle\frac{(l_{i}+1)(l_{\rm ph}+1)}{(2l_{i}+1)(2l_{\rm ph}+1)}}\;
    C_{l_{i} l_{f} l_{ph} l_{i}+1, l_{\rm ph}-1}^{m_{i} m_{f} \mu^{\prime}} \cdot
    \Bigl[ J_{1}(l_{i},l_{f},l_{\rm ph}-1)\; - l_{i} \cdot J_{2}(l_{i},l_{f},l_{\rm ph}-1) \Bigr], \\

  c_{6}^{\mu^{\prime}} & = &
    \sqrt{\displaystyle\frac{(l_{i}+1)\,l_{\rm ph}}{(2l_{i}+1)(2l_{\rm ph}+1)}}\;
    C_{l_{i} l_{f} l_{ph} l_{i}+1, l_{\rm ph}+1}^{m_{i} m_{f} \mu^{\prime}} \cdot
    \Bigl[ J_{1}(l_{i},l_{f},l_{\rm ph}+1)\; - l_{i} \cdot J_{2}(l_{i},l_{f},l_{\rm ph}+1) \Bigr],
\end{array}
\label{eq.3.5.4}
\end{equation}
\begin{equation}
\begin{array}{lcl}
\vspace{2mm}
  c_{7} & = &
  C_{l_{i} l_{f} l_{\rm ph} l_{\rm ph}}^{m_{i} \mu} \cdot
  \tilde{J}\, (l_{i},l_{f},l_{\rm ph}), \\
\vspace{2mm}
  c_{8} & = &
    \sqrt{\displaystyle\frac{l_{\rm ph}+1}{2l_{\rm ph}+1}}\;
    C_{l_{i} l_{f} l_{\rm ph}, l_{\rm ph}-1}^{m_{i} \mu} \cdot
    \tilde{J}\,(l_{i},l_{f},l_{\rm ph}-1), \\
  c_{9} & = &
    \sqrt{\displaystyle\frac{l_{\rm ph}}{2l_{\rm ph}+1}}\;
    C_{l_{i} l_{f} l_{\rm ph}, l_{\rm ph}+1}^{m_{i} \mu} \cdot
    \tilde{J}\,(l_{i},l_{f},l_{\rm ph}+1).
\end{array}
\label{eq.3.5.5}
\end{equation}
Here, $c_{1}^{\mu^{\prime}}$ ... $c_{6}^{\mu^{\prime}}$ and $c_{7}$ ... $c_{9}$ are not dependent on the $\theta$ angle. Function $\delta_{l_{i} \ne 0}$ is defined as $\delta_{l_{i} \ne 0} = 0$ at $l_{i}=0$ and $\delta_{l_{i} \ne 0} = 1$ at $l_{i} \ne 0$. Formulas for the first some values of $l_{i}$ and $l_{f}$ are presented in Appendix~\ref{sec.app.3}. On the basis of these formulas we conclude the following.

\vspace{1mm}
\begin{enumerate}
\item
Numbers $l_{i}$ and $l_{f}$ determine basic shape of the angular distribution of the bremsstrahlung probability, number $l_{\rm ph}$ determines oscillations in this shape:

\begin{enumerate}
\item
Number of oscillations of this shape is minimal at $l_{\rm ph}=1$ and increases at increasing of $l_{\rm ph}$.

\item
$c_{1}^{\mu^{\prime}}$ \ldots $c_{6}^{\mu^{\prime}}$ and $c_{7}$ \ldots $c_{9}$ are weights of oscillations at each chosen $l_{\rm ph}$. As integrals $J_{1}$, $J_{2}$ are decreased at increasing of $l_{\rm ph}$ (at fixed $w_{\rm ph}$), so each matrix element with next value of $l_{\rm ph}$ gives own new contribution into base shape of the probability distribution with smaller intensity, but larger number of oscillations.
\end{enumerate}

\item
If polynomials $P_{l_{i}\pm 1}^{|m_{i} - \mu^{\prime}|}$ at some chosen $l_{i}$ or polynomials $P_{l_{f}}^{|m_{f}|}$ at chosen $l_{f}$ in eqs.~(\ref{eq.3.5.1}) (polynomials $P_{l_{i}}^{|m_{i}|}$ at some chosen $l_{i}$ or polynomials $P_{l_{f}}^{|m_{f}|}$ at the chosen $l_{f}$ in eqs.~(\ref{eq.3.5.2})) equal to zero for some values of the $\theta$ angle, than the differential matrix elements in eqs.~(\ref{eq.3.5.1}) (in eqs.~(\ref{eq.3.5.2})) equal to zero at any value of $l_{\rm ph}$ for this $\theta$ angle.
\end{enumerate}

\begin{figure}[htbp]
\centerline{\includegraphics[width=65mm]{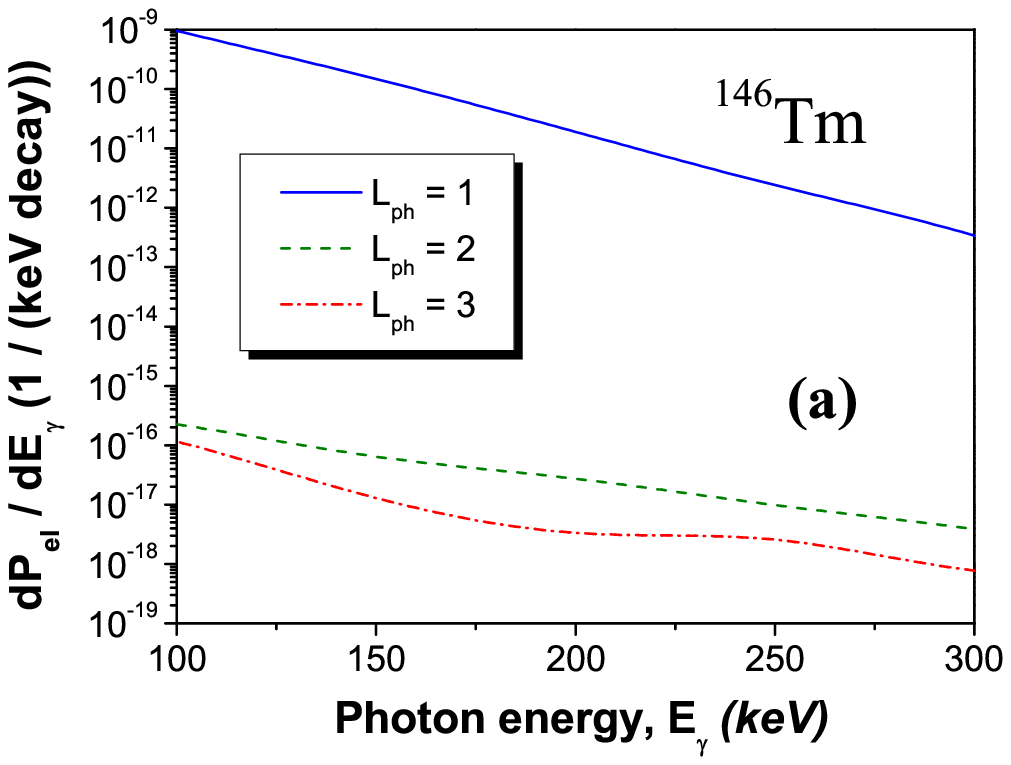}
\hspace{-5mm}\includegraphics[width=65mm]{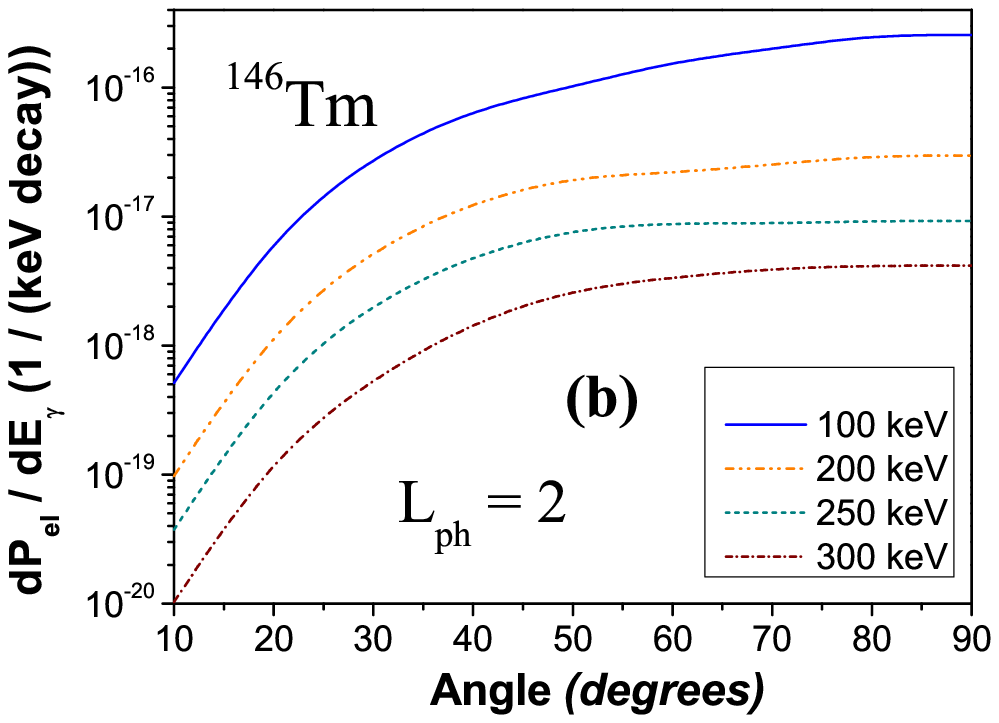}
\hspace{-5mm}\includegraphics[width=65mm]{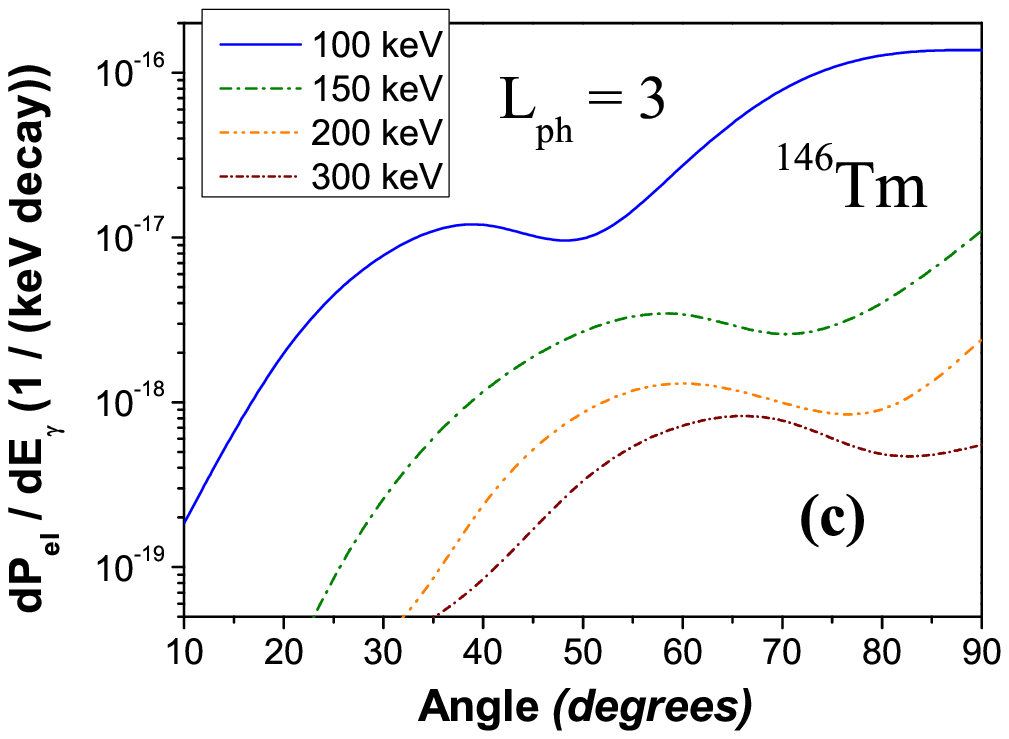}}
\vspace{-2mm}
\caption{\small (Color online)
Contributions of the electric component $dP_{\rm el}$ of the bremsstrahlung emission for proton decay of the $^{146}{\rm Tm}$ nucleus for the first three multipoles ($l_{\rm ph} = 1,2,3$).
(a) The spectra at $\theta = 90^{\circ}$: one can see that the first contribution at $l_{\rm ph}=1$ (blue solid line) is essentially larger in comparison with contributions at $l_{\rm ph}=2$ (green dashed line) and $l_{\rm ph}=3$ (red dash-dotted line), i.~e. the first multipolar contribution is prevailing inside whole energy region of the emitted photons.
(b) The multipolar contribution at $l_{\rm ph}=2$ in dependence on the $\theta$ angle:
one additional extremum can appear in each curve inside the angular region from 0 up to $90^{\circ}$, but it is practically smoothed (at current computer accuracy of calculations). However, at small values of $\theta$ each curve is increased more sharply in comparison with the angular spectra at $l_{\rm ph}=1$ (see Fig.~\ref{fig.2}~(a)).
(c) The multipolar contribution at $l_{\rm ph}=3$ in dependence on the $\theta$ angle:
appearance of else one new extremum in each curve forms one new oscillation. There is displacement of maximum and minimum of each spectrum in direction of larger values of $\theta$ with increasing of energy of the emitted photons.
\label{fig.9}}
\end{figure}

The angular contributions of the electric component $dP_{\rm el}$ of the bremsstrahlung emission during the proton decay of the $^{146}{\rm Tm}$ nucleus for the first three multipoles are presented in Fig.~\ref{fig.9}. In figure (a) one can see that the second and third multipolar contributions (at $l_{\rm ph}=2$ and $l_{\rm ph}=3$, $\theta=90^{\circ}$) are smaller on 5--7 orders of magnitude in comparison with the first one (at $l_{\rm ph}=1$, $\theta=90^{\circ}$). The angular distributions of these multipolar contributions are shown in next figures (b,~c) for $l_{\rm ph}=2$ and $l_{\rm ph}=3$. In particular, one can see that for smaller values of the $\theta$ angle the emission is more intensive at increasing of the multipolar order $l_{\rm ph}$ (at the same fixed $l_{i}$ and $l_{f}$ for $^{146}{\rm Tm}$).

\section{Conclusions
\label{sec.conclusions}}

The new model of the bremsstrahlung emission which accompanies proton decay and collisions of protons off nuclei in the energy region from the lowest up to intermediate, has been developed. This model includes spin formalism, potential approach for description of interaction between protons and nuclei, and operator of emission includes the component of magnetic emission (defined on the basis of Pauli equation). In the problem of the bremsstrahlung during the proton decay in the first time a role of the magnetic emission is studied using such a model. For such investigations the $^{146}{\rm Tm}$ nucleus is chosen.
We obtain the following:

\begin{enumerate}
\item
Inside energy region from 50 up to 300~keV the magnetic emission gives contribution about 28 percents into the full spectrum (see Fig.~\ref{fig.1}), i.e. it is not so small and should be taken into account in further estimations of spectra of the bremsstrahlung emission during nuclear decays with emission of the charged fragments with non-zero spin. However, the magnetic component suppresses the full emission probability: inclusion of the magnetic component into calculations is determined by $P_{\rm el}/P_{\rm full} \simeq 1.14$, which is larger unity. This effect of suppressing of the total emission can be explained by a presence of not small destructive interference between the electric and magnetic components inside whole studied energy region. Ratios of the electric and magnetic components to full spectrum are not changed in dependence on the energy of the emitted photon. The correction of the magnetic component $dP_{\rm mag, 2}$ is smaller than the electric and magnetic components by $10^{6}$ times.

\item
With increasing of the $\theta$ angle between directions of the outgoing proton and emitted photon the electric and magnetic components increase proportionally (see Fig.~\ref{fig.2}), but ratio between them is not changed (see Tabl.~\ref{table.1}). So, there is no some angular value, where the magnetic emission increases essentially relatively electric one.

\item
The magnetic component $dP_{\rm mag,1}(r)$ is dependent on distance $r$ between centers-of-masses of the proton and daughter nucleus similarly as the electric component $dP_{\rm mag,1}(r)$ (ratio between such two components is not changed inside region from 5~fm up to 250~fm). In the external region both components oscillate (having maxima and minima at similar space locations), while in the tunneling region they have monotonous shapes with one possible well (see Fig.~\ref{fig.3}). In general, the magnetic emission suppresses the full emission inside whole space region. The emission from the internal region up to the barrier is the smallest, and from the external region --- the strongest.

\item
The correction of the magnetic component $dP_{\rm mag,2}(r)$ is essentially smaller than the electric one $dP_{\rm el}(r)$ in dependence on distance $r$ (see Fig.~\ref{fig.4}). In the tunneling region it increases monotonously, in contrast to the electric and magnetic components. This causes a sharp peak of the function $dP_{\rm mag,2}(r) / dP_{\rm el}(r)$ close to the external boundary of the barrier (near 80~fm).

\item
At decreasing of the photon energy up to zero, the bremsstrahlung probability increases slowly up to finite maximum (at energy of the emitted photon less 1.5~keV), and then it monotonously decreases to zero (see Fig.~~\ref{fig.5}). The angular distribution of the probabilities of the bremsstrahlung emission at such small energies looks like the angular distributions inside the energy region from 50 to 350~keV studied above.
We show that there is no the infrared catastrophe in our approach.
\end{enumerate}
It is demonstrated that the model is able to describe enough well experimental data of the bremsstrahlung emission
which accompanies collisions of protons off
the $^{9}{\rm C}$, $^{64}{\rm Cu}$ and $^{107}{\rm Ag}$ nuclei at the incident energy $T_{\rm lab}=72$~MeV (at the photon energy up to 60~MeV),
the $^{9}{\rm Be}$, $^{12}{\rm C}$ and $^{208}{\rm Pb}$ nuclei at the incident energy $T_{\rm lab}=140$~MeV (at the photon energy up to 120~MeV).


\appendix
\section{Linear and circular polarizations of the photon emitted
\label{sec.2.4.2}}

Rewrite vectors of \emph{linear} polarization $\mathbf{e}^{(\alpha)}$ through \emph{vectors of circular polarization} $\mathbf{\xi}_{\mu}$ with opposite directions of rotation (see Ref.~\cite{Eisenberg.1973}, (2.39), p.~42):
\begin{equation}
\begin{array}{ccc}
  \mathbf{\xi}_{-1} = \displaystyle\frac{1}{\sqrt{2}}\,
                      \bigl(\mathbf{e}^{(1)} - i\mathbf{e}^{(2)}\bigr), &
  \mathbf{\xi}_{+1} = -\displaystyle\frac{1}{\sqrt{2}}\,
                      \bigl(\mathbf{e}^{(1)} + i\mathbf{e}^{(2)}\bigr), &
  \mathbf{\xi}_{0} = \mathbf{e}^{(3)},
\end{array}
\label{eq.2.4.2.1}
\end{equation}
where
\begin{equation}
\begin{array}{ccc}
  h_{\pm} = \mp \displaystyle\frac{1 \pm i}{\sqrt{2}}, &
  h_{-1} + h_{+1} = -i\sqrt{2}, &
  \sum\limits_{\alpha = 1,2} \mathbf{e}^{(\alpha),*} =
    h_{-1} \mathbf{\xi}_{-1}^{*} + h_{+1} \mathbf{\xi}_{+1}^{*}.
\end{array}
\label{eq.2.4.2.2}
\end{equation}
We have (in Coulomb gauge at $\mathbf{e}^{(3)}=0$):
\begin{equation}
\begin{array}{cc}
  \mathbf{e}^{(1)} = \displaystyle\frac{1}{\sqrt{2}}\, \bigl(\xibf_{-1} - \xibf_{+1}\bigr), &
  \mathbf{e}^{(2)} = \displaystyle\frac{i}{\sqrt{2}}\, \bigl(\xibf_{-1} + \xibf_{+1}\bigr),
\end{array}
\label{eq.2.4.2.3}
\end{equation}
\begin{equation}
\begin{array}{ccc}
  \displaystyle\sum\limits_{\mu = \pm 1} \xi_{\mu}^{*} \cdot \xi_{\mu} =
  \displaystyle\frac{1}{2}\,
    \bigl(\mathbf{e}^{(1)} - i\mathbf{e}^{(2)}\bigr)\, \bigl(\mathbf{e}^{(1)} - i\mathbf{e}^{(2)}\bigr)^{*} +
  \displaystyle\frac{1}{2}\,
    \bigl(\mathbf{e}^{(1)} + i\mathbf{e}^{(2)}\bigr)\, \bigl(\mathbf{e}^{(1)} + i\mathbf{e}^{(2)}\bigr)^{*} = 2.
\end{array}
\label{eq.2.4.2.4}
\end{equation}
We shall find also multiplications of vectors $\xibf_{\pm 1}$. From eq.~(\ref{eq.2.4.2.1}) we obtain:
\begin{equation}
\begin{array}{cc}
  \xibf_{-1}^{*} = - \xibf_{+1}, &
  \xibf_{+1}^{*} = - \xibf_{-1}.
\end{array}
\label{eq.2.4.2.5} 
\end{equation}
From here we find:
\begin{equation}
\begin{array}{lcl}
  \vspace{1mm}
  \Bigl[\xibf_{-1} \times \xibf_{+1}\Bigr] & = &
  \Bigl[ \displaystyle\frac{1}{\sqrt{2}}\, \bigl(\mathbf{e}^{(1)} - i\mathbf{e}^{(2)}\bigr) \times
    \displaystyle\frac{-1}{\sqrt{2}}\, \bigl(\mathbf{e}^{(1)} + i\mathbf{e}^{(2)}\bigr) \Bigr] =
  -\,\displaystyle\frac{1}{2}\, \Bigl[ \bigl(\mathbf{e}^{(1)} - i\mathbf{e}^{(2)}\bigr) \times
    \bigl(\mathbf{e}^{(1)} + i\mathbf{e}^{(2)}\bigr) \Bigr] = \\
  \vspace{2mm}
  & = &
  -\,\displaystyle\frac{1}{2}\,
  \Bigl\{ i\, \bigl[\mathbf{e}^{(1)} \times \mathbf{e}^{(2)}\bigr] -
          i\, \bigl[\mathbf{e}^{(2)} \times \mathbf{e}^{(1)}\bigr] \Bigr\} =
  - i\, \bigl[\mathbf{e}^{(1)} \times \mathbf{e}^{(2)}\bigr] =
  - i\, \mathbf{e}_{\rm z},
\end{array}
\label{eq.2.4.2.6} 
\end{equation}

\vspace{-2mm}
\begin{equation}
\begin{array}{lcllcl}
  \vspace{2mm}
  \Bigl[\xibf_{-1}^{*} \times \xibf_{+1}\Bigr] & = & -\, \Bigl[\xibf_{+1} \times \xibf_{+1}\Bigr] = 0, &

  \hspace{10mm}
  \Bigl[\xibf_{-1}^{*} \times \xibf_{-1}\Bigr] & = & -\, \Bigl[\xibf_{+1} \times \xibf_{-1}\Bigr] =
  i\, \mathbf{e}_{\rm z}, \\

  \Bigl[\xibf_{+1}^{*} \times \xibf_{-1}\Bigr] & = & -\, \Bigl[\xibf_{-1} \times \xibf_{-1}\Bigr] = 0, &

  \hspace{10mm}
  \Bigl[\xibf_{+1}^{*} \times \xibf_{+1}\Bigr] & = & -\, \Bigl[\xibf_{-1} \times \xibf_{+1}\Bigr] =
  -i\, \mathbf{e}_{\rm z}.
\end{array}
\label{eq.2.4.2.7} 
\end{equation}

\section{Angular integrals $I_{E}$, $I_{M}$ and $\tilde{I}$
\label{sec.app.2}}

We shall calculate the integrals in eqs.~(\ref{eq.2.4.6.3}) and (\ref{eq.2.4.6.6}):
\begin{equation}
\begin{array}{lcl}
  I_{M}\, (l_{i}, l_{f}, l_{\rm ph}, l_{1}, \mu) & = &
    \displaystyle\int
      Y_{l_{f}m_{f}}^{*}(\mathbf{n}_{\rm r})\,
      \mathbf{T}_{l_{i}\, l_{1},\, m_{i}}(\mathbf{n}_{\rm r})\,
      \mathbf{T}_{l_{\rm ph}\,l_{\rm ph},\, \mu}^{*}(\mathbf{n}_{\rm r})\; d\Omega, \\

  I_{E}\, (l_{i}, l_{f}, l_{\rm ph}, l_{1}, l_{2}, \mu) & = &
    \displaystyle\int
      Y_{l_{f}m_{f}}^{*}(\mathbf{n}_{\rm r})\,
      \mathbf{T}_{l_{i} l_{1},\, m_{i}}(\mathbf{n}_{\rm r})\,
      \mathbf{T}_{l_{\rm ph} l_{2},\, \mu}^{*}(\mathbf{n}_{\rm r})\; d\Omega, \\

  \tilde{I}\, (l_{i}, l_{f}, l_{\rm ph}, n, \mu) & = &
    \xibf_{\mu} \displaystyle\int
      Y_{l_{f}m_{f}}^{*}(\mathbf{n}_{\rm r})\,
      Y_{l_{i}m_{i}}(\mathbf{n}_{\rm r})\,
      \mathbf{T}_{l_{\rm ph} n,\, \mu}^{*}(\mathbf{n}_{\rm r})\; d\Omega.
\end{array}
\label{eq.app.2.1}
\end{equation}
Substituting the function $\mathbf{T}_{jl,m}(\mathbf{n}_{\rm r})$ defined by eq.~(\ref{eq.2.4.3.3}),
we obtain (at ${\mathbf \xi}_{0} = 0$):
\begin{equation}
\begin{array}{lcl}
  I_{M}\, (l_{i}, l_{f}, l_{\rm ph}, l_{1}, \mu) & = &
    \displaystyle\sum\limits_{\mu^{\prime} = \pm 1}
      (l_{1}, 1, l_{i} \,\big| \,m_{i}-\mu^{\prime}, \mu^{\prime}, m_{i})\;
      (l_{\rm ph}, 1, l_{\rm ph} \,\big|\, \mu-\mu^{\prime}, \mu^{\prime}, \mu)\; \times \\
  & \times &
    \displaystyle\int
      Y_{l_{f}m}^{*}(\mathbf{n}_{\rm r}) \cdot
      Y_{l_{1},\, m_{i}-\mu^{\prime}}(\mathbf{n}_{\rm r}) \cdot
      Y_{l_{\rm ph},\, \mu-\mu^{\prime}}^{*} (\mathbf{n}_{\rm r})\; d\Omega, \\

  I_{E}\, (l_{i}, l_{f}, l_{\rm ph}, l_{1}, l_{2}, \mu) & = &
    \displaystyle\sum\limits_{\mu^{\prime} = \pm 1}
      (l_{1}, 1, l_{i} \,\big| \,m_{i}-\mu^{\prime}, \mu^{\prime}, m_{i})\;
      (l_{2}, 1, l_{\rm ph} \,\big|\, \mu-\mu^{\prime}, \mu^{\prime}, \mu)\; \times \\
  & \times &
    \displaystyle\int
      Y_{l_{f}m}^{*}(\mathbf{n}_{\rm r}) \cdot
      Y_{l_{1},\, m_{i}-\mu^{\prime}} (\mathbf{n}_{\rm r}) \cdot
      Y_{l_{2},\, \mu-\mu^{\prime}}^{*} (\mathbf{n}_{\rm r})\; d\Omega.
\end{array}
\label{eq.app.2.2}
\end{equation}
\begin{equation}
\begin{array}{ccl}
  \tilde{I}\, (l_{i}, l_{f}, l_{\rm ph}, n, \mu) & = &
    (n, 1, l_{\rm ph} \,\big| \,0, \mu, \mu) \times
    \displaystyle\int
      Y_{l_{f}m_{f}}^{*}(\mathbf{n}_{\rm r})\,
      Y_{l_{i}m_{i}}(\mathbf{n}_{\rm r})\,
      Y_{n0}^{*}(\mathbf{n}_{\rm r})\; d\Omega.
\end{array}
\label{eq.app.2.3}
\end{equation}
Here, we have taken orthogonality of vectors $\xi_{\pm 1}$ into account.
In these formulas we shall find angular integral:
\begin{equation}
\begin{array}{l}
\vspace{1mm}
  \displaystyle\int \:
    Y_{l_{f}m_{f}}^{*}({\mathbf n}_{\rm r})\,
    Y_{l_{1},\, m_{i}-\mu^{\prime}}(\mathbf{n}_{\rm r})\,
    Y_{n,\, \mu-\mu^{\prime}}^{*}(\mathbf{n}_{\rm r})\;
    d\Omega =

    (-1)^{l_{f} + n + m_{i} - \mu^{\prime}}\;
    i^{l_{f}+l_{1}+n + |m_{f}| + |m_{i} - \mu^{\prime}| + |m_{i} - m_{f}-\mu^{\prime}|}\; \times \\

  \;\times
    \sqrt{
      \displaystyle\frac{(2l_{f}+1)\, (2l_{1}+1)\, (2n+1)}{16\pi}
      \displaystyle\frac{(l_{f}-|m_{f}|)!}{(l_{f}+|m_{f}|)!}\;
      \displaystyle\frac{(l_{1}-|m_{i}-\mu^{\prime}|)!} {(l_{1}+|m_{i}-\mu^{\prime}|)!}\;
      \displaystyle\frac{(n-|m_{i} - m_{f}-\mu^{\prime}|)!}{(n+|m_{i} - m_{f} -\mu^{\prime}|)!}}\;
    \times \\

  \;\times
    \displaystyle\int\limits_{0}^{\pi}\:
      P_{l_{f}}^{|m_{f}|}(\cos{\theta})\;
      P_{l_{1}}^{|m_{i} - \mu^{\prime}|}(\cos{\theta})\;
      P_{n}^{|m_{i} - m_{f} - \mu^{\prime}|} (\cos{\theta}) \cdot
      \sin{\theta}\, d\theta,
\end{array}
\label{eq.app.2.4}
\end{equation}
where $P_{l}^{m}(\cos{\theta})$ are associated Legandre's polynomials, and we obtain conditions:
\begin{equation}
\begin{array}{llllll}
  \mbox{for integrals } I_{M}, I_{E}: &
  \mu = m_{i} - m_{f}, &
  n \ge |\mu - \mu^{\prime}| = |m_{i} - m_{f} + \mu^{\prime}|, & \mu = \pm 1, \\

  \mbox{for integral } \tilde{I}: &
  m_{i} = m_{f}. & &
\end{array}
\label{eq.app.2.5}
\end{equation}
Using formula (\ref{eq.app.2.4}), we calculate integrals (\ref{eq.app.2.2}) and (\ref{eq.app.2.3}):
\begin{equation}
\begin{array}{lcl}
  I_{M}\, (l_{i}, l_{f}, l_{\rm ph}, l_{1}, \mu) & = &
    \delta_{\mu, m_{i}-m_{f}}\;
    \displaystyle\sum\limits_{\mu^{\prime} = \pm 1}
      C_{l_{i} l_{f} l_{ph} l_{1} l_{ph}}^{m_{i} m_{f} \mu^{\prime}}
      \displaystyle\int\limits_{0}^{\pi}\:
        f_{l_{1} l_{f} l_{\rm ph}}^{m_{i} m_{f} \mu^{\prime}}(\theta)\; \sin{\theta}\,d\theta, \\

  I_{E}\, (l_{i}, l_{f}, l_{\rm ph}, l_{1}, l_{2}, \mu) & = &
    \delta_{\mu, m_{i}-m_{f}}\;
    \displaystyle\sum\limits_{\mu^{\prime} = \pm 1}
      C_{l_{i} l_{f} l_{ph} l_{1} l_{2}}^{m_{i} m_{f} \mu^{\prime}}
      \displaystyle\int\limits_{0}^{\pi} \:
      f_{l_{1} l_{f} l_{2}}^{m_{i} m_{f} \mu^{\prime}}(\theta)\; \sin{\theta}\,d\theta, \\

  \tilde{I}\, (l_{i}, l_{f}, l_{\rm ph}, n, \mu) & = &
    C_{l_{i} l_{f} l_{\rm ph} n}^{m_{i} \mu}
    \displaystyle\int\limits_{0}^{\pi}\:
    f_{l_{i} l_{f} n}^{m_{i} m_{i} 0}(\theta)\; \sin{\theta}\,d\theta,
\end{array}
\label{eq.app.2.6}
\end{equation}

\vspace{-5mm}
\noindent
where
\begin{equation}
\begin{array}{lcl}
\vspace{1mm}
  C_{l_{i} l_{f} l_{ph} l_{1} l_{2}}^{m_{i} m_{f} \mu^{\prime}} & = &
    (-1)^{l_{f} + l_{2} + m_{i} - \mu^{\prime}}\;
    i^{l_{f}+l_{1}+ l_{2} + |m_{f}| + |m_{i} - \mu^{\prime}| + |m_{i} - m_{f}-\mu^{\prime}|}\; \times \\
  \vspace{2mm}
  & \times &
    (l_{1}, 1, l_{i} \,\big| \,m_{i}-\mu^{\prime}, \mu^{\prime}, m_{i})\;
    (l_{2}, 1, l_{\rm ph} \,\big|\, m_{i}-m_{f} -\mu^{\prime}, \mu^{\prime}, m_{i}-m_{f})\; \times \\
  & \times &
    \sqrt{
      \displaystyle\frac{(2l_{f}+1)\, (2l_{1}+1)\, (2l_{2}+1)}{16\pi}
      \displaystyle\frac{(l_{f}-|m_{f}|)!}{(l_{f}+|m_{f}|)!}\;
      \displaystyle\frac{(l_{1}-|m_{i}-\mu^{\prime}|)!}{(l_{1}+|m_{i}-\mu^{\prime}|)!}\;
      \displaystyle\frac{(l_{2}-|m_{i}-m_{f}-\mu^{\prime}|)!} {(l_{2} +|m_{i} - m_{f} -\mu^{\prime}|)!}},
\end{array}
\label{eq.app.2.7}
\end{equation}

\vspace{-4mm}
\begin{equation}
\begin{array}{lcl}
\vspace{1mm}
  C_{l_{i} l_{f} l_{\rm ph} n}^{m_{i} \mu} & = &
  (-1)^{l_{f} + n + m_{i} + |m_{i}|}\;
  i^{l_{f}+l_{i} + n} \cdot
  (n, 1, l_{\rm ph} \,\big| \,0, \mu, \mu) \cdot
    \sqrt{
      \displaystyle\frac{(2l_{f}+1)\, (2l_{i}+1)\, (2n+1)}{16\pi}
      \displaystyle\frac{(l_{f}-|m_{i}|)!}{(l_{f}+|m_{i}|)!}\;
      \displaystyle\frac{(l_{i}-|m_{i}|)!}{(l_{i}+|m_{i}|)!} },
\end{array}
\label{eq.app.2.8}
\end{equation}

\vspace{-5mm}
\begin{equation}
  f_{l_{1} l_{f} l_{2}}^{m_{i} m_{f} \mu^{\prime}}(\theta) =
    P_{l_{1}}^{|m_{i} - \mu^{\prime}|}(\cos{\theta})\;
    P_{l_{f}}^{|m_{f}|}(\cos{\theta})\;
    P_{l_{2}}^{|m_{i} - m_{f} - \mu^{\prime}|} (\cos{\theta}).
\label{eq.app.2.9}
\end{equation}
We define differential functions on the integrals (\ref{eq.app.2.6}) with angular dependence as
\begin{equation}
\begin{array}{lcl}
  \displaystyle\frac{d\, I_{M}\, (l_{i}, l_{f}, l_{\rm ph}, l_{1}, \mu)} {\sin{\theta}\,d\theta} & = &
  \delta_{\mu, m_{i}-m_{f}}\;
  \displaystyle\sum\limits_{\mu^{\prime} = \pm 1}
    C_{l_{i} l_{f} l_{ph} l_{1} l_{ph}}^{m_{i} m_{f} \mu^{\prime}} \cdot
    f_{l_{1} l_{f} l_{\rm ph}}^{m_{i} m_{f} \mu^{\prime}}(\theta), \\

  \displaystyle\frac{d\, I_{E}\, (l_{i}, l_{f}, l_{\rm ph}, l_{1}, l_{2}, \mu)} {\sin{\theta}\,d\theta} & = &
  \delta_{\mu, m_{i}-m_{f}}\;
  \displaystyle\sum\limits_{\mu^{\prime} = \pm 1}
    C_{l_{i} l_{f} l_{ph} l_{1} l_{2}}^{m_{i} m_{f} \mu^{\prime}} \cdot
    f_{l_{1} l_{f} l_{2}}^{m_{i} m_{f} \mu^{\prime}}(\theta), \\

  \displaystyle\frac{d\, \tilde{I}\, (l_{i}, l_{f}, l_{\rm ph}, n, \mu)} {\sin{\theta}\,d\theta} & = &
    \delta_{m_{i} m_{f}}\,
    C_{l_{i} l_{f} l_{\rm ph} n}^{m_{i} \mu}
    f_{l_{i} l_{f} n}^{m_{i} m_{i} 0}(\theta).
\end{array}
\label{eq.app.2.10}
\end{equation}

\section{Differential matrix elements for the fist $l_{i}$ and $l_{f}$
\label{sec.app.3}}

We write calculations for the fist some values of $l_{i}$ and $l_{f}$, at arbitrary $l_{\rm ph}$:

1. $l_{i}=0$, $l_{f}=0$:
%
\begin{equation}
\begin{array}{lcllcl}
  \displaystyle\frac{dp_{l_{\rm ph}\mu}^{M}}{\sin{\theta}\,d\theta} =
  \displaystyle\frac{dp_{l_{\rm ph}\mu}^{E}}{\sin{\theta}\,d\theta} = 0, & \quad

  \displaystyle\frac{\tilde{dp}_{l_{\rm ph}\mu}^{M}}{\sin{\theta}\,d\theta} =
  c_{7} \; P_{l_{\rm ph}}^{0}, & \quad

  \displaystyle\frac{\tilde{dp}_{l_{\rm ph}\mu}^{E}}{\sin{\theta}\,d\theta} =
    c_{8}\; P_{l_{\rm ph}-1}^{0} -
    c_{9}\; P_{l_{\rm ph}+1}^{0}, & \quad
  m_{i} = m_{f} = 0.
\end{array}
\label{eq.app.3.1}
\end{equation}

2. $l_{i}=0$, $l_{f}=1$:
\begin{equation}
\begin{array}{lcll}
  \vspace{0mm}
  \displaystyle\frac{d\,p_{l_{\rm ph}\mu}^{M}}{\sin{\theta}\,d\theta} & = &
  - \sin^{2}\theta\;
  \displaystyle\sum\limits_{\mu^{\prime} = \pm 1}
    c_{2}^{\mu^{\prime}} \cdot
    P_{l_{\rm ph}}^{|\mu - \mu^{\prime}|}, &
  m_{i}=0,\; m_{f}=\pm 1, \\

\vspace{0mm}
  \displaystyle\frac{d\, p_{l_{\rm ph}\mu}^{E}}{\sin{\theta}\,d\theta} & = &
  \sin^{2}\theta\;
  \displaystyle\sum\limits_{\mu^{\prime} = \pm 1}\,
  \Bigl\{
    c_{5}^{\mu^{\prime}}\: P_{l_{\rm ph}-1}^{|\mu - \mu^{\prime}|} -
    c_{6}^{\mu^{\prime}}\: P_{l_{\rm ph}+1}^{|\mu - \mu^{\prime}|}
  \Bigr\}, &
  m_{i}=0,\; m_{f}=\pm 1, \\

\vspace{2mm}
  \displaystyle\frac{d\, \tilde{p}_{l_{\rm ph}\mu}^{M}}{\sin{\theta}\,d\theta} & = &
  c_{7} \cdot \cos\theta\; P_{l_{\rm ph}}^{0}, &
  m_{i}=0,\; m_{f}=0, \\

  \displaystyle\frac{d\, \tilde{p}_{l_{\rm ph}\mu}^{E}}{\sin{\theta}\,d\theta} & = &
  \cos\theta\;
  \Bigl\{
    c_{8}\; P_{l_{\rm ph}-1}^{0} -
    c_{9}\; P_{l_{\rm ph}+1}^{0}
  \Bigr\}, &
  m_{i}=0,\; m_{f}=0.
\end{array}
\label{eq.app.3.2}
\end{equation}

3. $l_{i}=0$, $l_{f}=2$:
\begin{equation}
\begin{array}{lcll}
\vspace{0mm}
  \displaystyle\frac{d\, p_{l_{\rm ph}\mu}^{M}}{\sin{\theta}\,d\theta} & = &
  -\, 3\, \sin^{2}\theta\, \cos\theta\;
  \displaystyle\sum\limits_{\mu^{\prime} = \pm 1}
    c_{2}^{\mu^{\prime}} \cdot
    P_{l_{\rm ph}}^{|\mu - \mu^{\prime}|}, &
  m_{i}=0,\; m_{f}=\pm 1, \\

\vspace{0mm}
  \displaystyle\frac{d\, p_{l_{\rm ph}\mu}^{E}}{\sin{\theta}\,d\theta} & = &
  3\,\sin^{2}\theta\,\cos\theta\;
  \displaystyle\sum\limits_{\mu^{\prime} = \pm 1}\,
  \Bigl\{
    c_{5}^{\mu^{\prime}}\: P_{l_{\rm ph}-1}^{|\mu - \mu^{\prime}|} -
    c_{6}^{\mu^{\prime}}\: P_{l_{\rm ph}+1}^{|\mu - \mu^{\prime}|}
  \Bigr\}, &
  m_{i}=0,\; m_{f}=\pm 1, \\

\vspace{0mm}
  \displaystyle\frac{d\, \tilde{p}_{l_{\rm ph}\mu}^{M}} {\sin{\theta}\,d\theta} & = &
  \displaystyle\frac{c_{7}}{2}\: (3\,\cos^{2}\theta - 1)\: P_{l_{\rm ph}}^{0}, &
  m_{i}=0,\; m_{f}=0, \\

  \displaystyle\frac{d\, \tilde{p}_{l_{\rm ph}\mu}^{E}} {\sin{\theta}\,d\theta} & = &
  \displaystyle\frac{1}{2}\: (3\,\cos^{2}\theta - 1)\:
  \Bigl\{
    c_{8}\; P_{l_{\rm ph}-1}^{0} -
    c_{9}\; P_{l_{\rm ph}+1}^{0}
  \Bigr\}, &
  m_{i}=0,\; m_{f}=0. \\
\end{array}
\label{eq.app.3.3}
\end{equation}

4. $l_{i}=1$, $l_{f}=1$:
\begin{equation}
\begin{array}{lcll}
\vspace{1mm}
  \displaystyle\frac{d\, p_{l_{\rm ph}\mu}^{M}}{\sin{\theta}\,d\theta} & = &
  \cos\theta \cdot
  \displaystyle\sum\limits_{\mu^{\prime} = \pm 1}
  \biggl\{
    -\, \delta_{m_{i},0}\; 3\, c_{2}^{\mu^{\prime}}\, \sin^{2}\theta +
    \delta_{m_{i},\pm 1}\,
    \Bigl[
      \delta_{m_{i} \mu^{\prime}}\, c_{1}^{m_{i}} -
      c_{2}^{\mu^{\prime}}\: P_{2}^{|m_{i} - \mu^{\prime}|}\;
    \Bigr] \biggr\} \cdot
  P_{l_{\rm ph}}^{|\mu - \mu^{\prime}|}, &
  |m_{i}-m_{f}| = 1, \\

\vspace{1mm}
  \displaystyle\frac{d\, p_{l_{\rm ph}\mu}^{E}}{\sin{\theta}\,d\theta} & = &
  \cos\theta \cdot
  \displaystyle\sum\limits_{\mu^{\prime} = \pm 1}\,
  \biggl\{
    \delta_{m_{i}, 0}\; 3\,\sin^{2}\theta\, c_{5}^{\mu^{\prime}}\; +
    \delta_{m_{i}, \pm 1}\;
      \Bigl[
        \delta_{m_{i} \mu^{\prime}}\, c_{3}^{m_{i}} +
        c_{5}^{\mu^{\prime}}\: P_{2}^{|m_{i} - \mu^{\prime}|}\,
      \Bigr]\:
  \biggr\}\;
  P_{l_{\rm ph}-1}^{|\mu - \mu^{\prime}|} \quad - \\
\vspace{-1mm}
  & - &
  \cos\theta \cdot
  \displaystyle\sum\limits_{\mu^{\prime} = \pm 1}\,
  \biggl\{
    \delta_{m_{i}, 0}\; 3\,\sin^{2}\theta\, c_{6}^{\mu^{\prime}} +
    \delta_{m_{i}, \pm 1}\;
    \Bigl[
      \delta_{m_{i} \mu^{\prime}}\, c_{4}^{m_{i}} +
      c_{6}^{\mu^{\prime}}\: P_{2}^{|m_{i} - \mu^{\prime}|}
    \Bigr]\:
  \biggr\}\:
  P_{l_{\rm ph}+1}^{|\mu - \mu^{\prime}|}, &
  |m_{i}-m_{f}| = 1, \\
\end{array}
\label{eq.app.3.4}
\end{equation}
\begin{equation}
\begin{array}{lcll}
\vspace{1mm}
  \displaystyle\frac{d\, \tilde{p}_{l_{\rm ph}\mu}^{M}} {\sin{\theta}\,d\theta} & = &
  c_{7} \cdot (P_{1}^{|m_{i}|})^{2}\, P_{l_{\rm ph}}^{0}, &
  m_{i}=m_{f}=0,\pm 1, \\

  \displaystyle\frac{d\, \tilde{p}_{l_{\rm ph}\mu}^{E}} {\sin{\theta}\,d\theta} & = &
  \bigl(P_{1}^{|m_{i}|}\bigr)^{2}
  \Bigl\{
    c_{8}\; P_{l_{\rm ph}-1}^{0} -
    c_{9}\; P_{l_{\rm ph}+1}^{0}
  \Bigr\}, &
  m_{i}=m_{f}=0,\pm 1.
\end{array}
\label{eq.app.3.5}
\end{equation}



\end{document}